\documentclass[prl, twocolumn, longbibliography]{revtex4-2}
\usepackage{graphicx}
\usepackage{amsmath}
\usepackage{amssymb}
\usepackage{color}
\usepackage{appendix}
\usepackage{braket}
\usepackage{bbm}
\usepackage{subfigure}
\usepackage{mathrsfs}

\usepackage[colorlinks,urlcolor=blue,citecolor=blue,linkcolor=blue]{hyperref}
\usepackage{lipsum}
\usepackage{diagbox}
\usepackage{multirow}

\begin{document}
\title{Braiding Topology of Non-Hermitian Open-Boundary Bands}
\author{Yongxu Fu}
\email{yongxufu@pku.edu.cn}
\affiliation{International Center for Quantum Materials, School of Physics, Peking University, Beijing 100871, China}
\author{Yi Zhang}
\email{frankzhangyi@gmail.com}
\affiliation{International Center for Quantum Materials, School of Physics, Peking University, Beijing 100871, China}

\begin{abstract}
There has been much recent interest and progress on topological structures of the non-Hermitian Bloch bands. Here, we study the topological structures of non-Bloch bands of non-Hermitian multiband quantum systems under open boundary conditions, which has received limited attention in prior studies. Using a continuity criterion and an efficient sub-generalized Brillouin zone (sub-GBZ) algorithm, we establish a homotopic characterization---braiding topology, e.g., characterized by the band's total vorticity---for open-boundary bands and sub-GBZs. Such topological identification is robust without topological transition and emergent degenerate points, such as exceptional points. We further analyze the transition's impact on bands and spectral flows, including interesting properties unique to open boundaries, and numerically demonstrate our conclusions with tight-binding model examples. We unveil a crucial insight that open-boundary bands interchange their portions after encountering certain exceptional points. Our results enrich the foundational understanding of topological characterizations for generic non-Hermitian quantum systems. 
\end{abstract}
\maketitle

{\em Introduction.---}
The Bloch band theory is a well-established cornerstone in solid-state physics \cite{quinn2018,Phillips_2012,negele2018book,mahan2000book,Altland_Simons_2010, fradkin2013book,nagaosa2013book}. In the past few years, research on non-Hermitian quantum systems has made the extension of such band theory a major focus \cite{ashida2020,bergholtzrev2021,lin2023skineffect,okuma2023review,kunst2018,gong2018,kawabataprx,origin2020,slager2020,tai2023,yang2024complex}. For instance, a non-Bloch band theory has been developed for one-dimensional (1D) non-Hermitian quantum systems under open boundary condition (OBC) \cite{torres2018anomalous,yao2018,yokomizo2019,lee2019an,zhang2020,yang2020,fu2022,fu2023ana,wu2022,hu2024geo} and rapidly verified in various experimental platforms \cite{xiao2020exp,stegmaier2021exp,xiao2021ob,wand2021exp,wang2022morphing,liang2022exp,helbig2020exp,zhang2023exp,xiao2023observation}. Superseding the conventional Brillouin zone (BZ), the non-Bloch band theory's generalized Brillouin zone (GBZ) gives rise to the non-Hermitian skin effect and breaks down the bulk-boundary correspondence \cite{hasan2010topo,qi2011topo,bernevig2013book,qiu2016class}. Further generalizations have also been made to higher dimensions \cite{kawabata2019second,kawabatahigher,okugawa2020,fu2021,zhu2022hybrid,zhang2022universalnon,li2022gain,palacios2021,zou2021observation,wang2022amoeba,hu2023nonhermitian,xiong2023graph} and scale-free localization scenarios~\cite{li2021impurity,guo2023scale,libo2023scale,molignini2023anomalous,fu2023scalefree,xie2024scalefree}. 

Topological concepts such as the braid groups and the knots \cite{kassel2008braid,kauffman2013knots} have had widespread and fruitful applications in various physics topics. The nodal-line knots of topological semimetals \cite{chen2017nodal,yan2017nodal,bi2017nodal,lee2020imaging} are soon followed by the exceptional-line knots in non-Hermitian systems \cite{ding2018eps,yang2019hopf,bergholtz2019knot,marcus2019hyperbolic,yang2020jones,yang2020observation,bergholtz2021eps,zhang2021tidal,patil2022measuring,ding2022nonhermitian,hu2022knot,zhang2023symmetry,guo2023nonabelian,li2024braidingeps}. Braid and knot structures also emerge in the three-dimensional (3D) space spanned by the complex energy and the wave vector of the Bloch band theory \cite{wojcik2020homotopy,li2021homotopy,hu2021knots,wang2021topological,wojcik2022band,cao2023probing,li2023knots,rui2023braidings,yang2023homotopy,pang2024synthetic,chen2024machine,wang2024onedimensional}, applicable only for periodic boundary conditions (PBC). However, despite recent work on similar topology for non-Bloch bands under OBC \cite{li2023nonblochbraiding}, the general definition and properties remain unclear, especially for generic non-Hermitian systems with multiple non-Bloch bands with respective GBZs dubbed the sub-GBZs \cite{yang2020,fu2022,fu2023ana}. 

In this paper, we propose a continuity criterion and a homotopic formalism of the braid and knot topology for the non-Bloch bands of non-Hermitian quantum systems under OBC in the thermodynamic limit, even in the presence of multiple bands and sub-GBZs. Consequently, we can characterize such robust topology and the intermediate topological transitions, denoted by the emergence of exceptional points (EPs), by the non-Bloch bands' vorticity. We have also established an efficient and accurate numerical algorithm, demonstrated examples of non-Hermitian multiband models, and pointed out their intriguing characteristics unique to OBCs. We reveal that open-boundary bands interchange their parts as the system crosses particular EPs. Additionally, compared to PBC bands, the braiding of non-Hermitian OBC bands is significant due to its more diverse transition options---the commonly two-visit OBC eigenvalues offer multiple choices for exchange partners and flow directions at the transition. Our results offer an essential facet and thus pave the way toward our full understanding of the topology of generic non-Hermitian quantum systems under OBC.

{\em Continuity criterion of bulk energy bands under OBC.---} 
Consider a generic 1D non-interacting non-Hermitian tight-binding Hamiltonian:
\begin{align}
\label{eqgenric}
\hat{\mathcal{H}}=\sum_{j}\sum_{m=-\mathcal{R}_{1}}^{\mathcal{R}_{2}}\mathbf{c}^{\dagger}_{j}\mathcal{T}_{m}\mathbf{c}_{j+m},
\end{align}
with $N$ lattice sites (unit cells) and $n$ internal degrees of freedom on each site $j$---the dimension of the hopping matrices $\mathcal{T}_{m}$. Without loss of generality, we set the hopping ranges $\mathcal{R}_{1}=\mathcal{R}_{2}\equiv\mathcal{R}$. The direct diagonalization of such a non-Hermitian $\hat{\mathcal{H}}$ suffers from precision issues, especially for large system sizes \cite{supp}. Instead, the energy bands of $\hat{\mathcal{H}}$ may descend from the roots of the characteristic equation with respect to the reference energy $E$:
\begin{align}
    \label{eqcharaceq}
    f(\beta,E)\equiv\det \left[\mathcal{H}(\beta)-E\times\mathbbm{1}_{n\times n}\right]=0,
\end{align}
where $\mathcal{H}(\beta)=\sum_{m=-\mathcal{R}}^{\mathcal{R}}\mathcal{T}_{m}\beta^{m}$ is a matrix-valued Laurent polynomial of the complex variable $\beta$. In the presence of translation symmetry and PBC, the Bloch band theory dictates $\beta=e^{ik}$, where the wave vector (lattice momentum) $k\in [-\pi, \pi]$ is defined within the (first) BZ \footnote{For simplicity, we have set the lattice constant to 1}. For OBC, on the other hand, we need to adapt to the non-Bloch band theory \cite{yao2018,yokomizo2019}, where $\beta$'s satisfy the GBZ condition---the $2M\equiv 2n\mathcal{R}$ sorted roots of the characteristic Eq. (\ref{eqcharaceq}),  $\left|\beta_{1}\right|\leq \left|\beta_{2}\right|\leq\ldots \leq\left|\beta_{2M}\right|$, obey $\left|\beta_{M}\right|=\left|\beta_{M+1}\right|$. The bulk energy bands follow such BZs or GBZs accordingly. Throughout this paper, we refer to $k$ as the wave vector in the BZ of the Bloch band theory under PBC, while $\theta$ represents the complex argument of $\beta$ in the GBZ of the non-Bloch band theory under OBC.

Noteworthily, each OBC band $\varepsilon_{i}, i=1,2,\ldots,n$, is related to an ideally unique continuous sub-GBZ $\mathcal{C}_{i}$ for a generic non-Hermitian system in the thermodynamic limit \cite{yang2020,fu2022,fu2023ana}, disregarding the potential emergence of pseudo-gaps with a very low density of states \cite{li2022685}. However, due to the multiplicity of the roots of the characteristic Eq.~(\ref{eqcharaceq}), subtle complications may arise when we associate the OBC bands with the sub-GBZs. For instance, while the auxiliary GBZ (aGBZ) can give all of the sub-GBZs \cite{yang2020}, isolating the multivalued energy bands for each sub-GBZ remains uncontrolled in practice and thus intractable for generic non-Hermitian Hamiltonians except for certain simple cases \cite{fu2023ana}. Here, we overcome such ambiguity by enforcing the continuity criterion that a map from the sub-GBZ $\mathcal{C}_{i}$ to the OBC band $\varepsilon_{i}$ must be continuous. In other words, we require that infinitesimal variations on $\mathcal{C}_{i}$ only induce infinitesimal variations in $\varepsilon_{i}$ \footnote{The continuity condition is exact in the thermodynamic limit; in practice, however, we must set a numerical precision threshold for the $\varepsilon_{i}$ variations in finite-size calculations.}, which keep the bands' clear distinctions despite of coincidental roots thus intersecting sub-GBZs. 

Here, we propose an algorithm to enforce such a continuity criterion. First, we solve the resultant equation: 
\begin{align}
    \label{eqresl}
    \mathcal{R}_{E}\left[f(\beta,E),f(\beta e^{i\Theta},E)\right]=0,
\end{align}
over the range $\Theta\in [-\pi,\pi]$. Here, $\mathcal{R}_{E}$ represents the resultant of $f(\beta,E)$ and $f(\beta e^{i\Theta},E)$ relative to $E$ \cite{Gelfand1994,Cox2005}. For a given $\Theta$, the characteristic equations $f(\beta, E)=0$ and $f(\beta e^{i\Theta}, E)=0$ share a common energy $E$ if and only if the resultant Eq. (\ref{eqresl}) holds. Thus, we can obtain all $\beta_{p+1}=\beta_{p} e^{\pm i\Theta}$ solutions satisfying the characteristic Eq. (\ref{eqcharaceq}), so that $\left|\beta_{1}\right|\leq\ldots\leq\left|\beta_{p}\right|=\left|\beta_{p+1}\right|\leq\ldots \leq\left|\beta_{2M}\right|$, $p=1,2,\ldots,2M-1$, for a specific $\Theta$. Next, we select the $\beta$ solutions with $p=M$, which coincide with the GBZ condition, and map out the GBZ as we vary $\Theta$. Finally, after obtaining the set of GBZ with the corresponding ensemble of OBC bands, we apply the continuity criterion: as the argument $\theta$ of $\beta$ increases, both $\beta_i$ and $\varepsilon_{i}$ within each band $i$ evolve continuously. In practice, it helps to separate this ensemble of energies into individual bands $\varepsilon_{i}$ one by one, following the order of $\beta$'s argument at a specific numerical precision. Each $\{\beta_i\}$ for $\theta\in[-\pi, \pi]$ constitutes a sub-GBZ $\mathcal{C}_{i}$. Such an algorithm can effectively and unambiguously determine the multiple OBC bands and the associated sub-GBZs, except for the degenerate points (DPs) where multiple $\beta_i$ coincide and signal a topological transition; see detailed elucidation of the algorithm along with the examples and applications later and in the supplemental materials \cite{supp}. 

We note that although the mapping from $\mathcal{C}_{i}$ to $\varepsilon_{i}$ is continuous (and vice versa), it is not homeomorphic or injective but rather surjective. Namely, an $E$ value may correspond to two (or more) points on $\mathcal{C}_{i}$, dubbed as a $2$-bifurcation (or $\mathfrak{n}$-bifurcation) point \cite{wu2022,fu2023ana}, and passed through $2$ (or $\mathfrak{n}$) times as we move across $\mathcal{C}_{i}$. We note that the $\mathfrak{n}$-bifurcation point is usually within a single band and not a degeneracy between the bands. The degeneracy between two or more bands will grant interesting physical consequences, such as topological transitions and DPs, as we discuss next.

{\em Band braiding topology with sub-GBZs.---}
The continuity criterion yields continuous OBC bands $\varepsilon_{i}(\beta)$ with $\beta$ lying on the related sub-GBZ $\mathcal{C}_{i}$. For simplicity, we only consider cases where all sub-GBZs are closed, i.e., homeomorphic to the $1$-sphere $S^{1}$, and $r=\left|\beta\right|$ is a single-valued function of $\theta$ on each $\mathcal{C}_{i}$ in the main text. We also choose the same convention of $\theta$, i.e., identical starting points $\theta_0$, for all sub-GBZs' arguments so that $\beta$ remains consistent and continuous before and after a topological transition, where parts of the OBC bands may switch their partners. We discuss more complex cases with intertwined sub-GBZs in the supplemental materials \cite{supp}. 

The braiding topology is only available to multiband systems. We define two arbitrary bands $i\neq j$ as isolated if $\varepsilon_{i}(\beta_{1})\neq \varepsilon_{j}(\beta_{2})$ for any $\beta_{1,2}\in \mathcal{C}_{i,j}$, separated if they are not isolated yet $\varepsilon_{i}(\beta)\neq \varepsilon_{j}(\beta)$ for any coincidental $\beta\in \mathcal{C}_{i,j}$, and degenerate if $\varepsilon_{i}=\varepsilon_{j}$ at at least one degenerate point (DP) $\beta_i=\beta_j$ on $\mathcal{C}_{i,j}$. We wish to analyze the equivalence classes of the OBC bands $\varepsilon_{i}(\beta)$ in the absence of multiband degeneracies. However, as long as $\varepsilon_{i}=\varepsilon_{j}$ and $\theta_{i}=\theta_{j}$, we obtain $|\beta_{i}|=|\beta_{j}|$ as dictated by the GBZ condition, and then $\beta_{i}=\beta_{j}$; therefore, it suffices to relate the transition to $\varepsilon_{i}(\theta)$ instead of $\varepsilon_{i}(\beta)$ and focus instead on the continuous mapping from $\theta \in S^{1}$ to the OBC energies $\varepsilon_{i}\in \mathbbm{C}$. The subsequent homotopy characterization resembles that of a non-Hermitian Bloch Hamiltonian $\mathcal{H}(k)$ with the lattice momentum $k\in[-\pi,\pi]$ within the BZ \cite{wojcik2020homotopy,li2021homotopy,hu2021knots,yang2023homotopy} in our place of $\theta$. 

Strictly speaking, the relevant non-based map is from $S^{1}$ to $\mathcal{X}_{n}=\left(\mathrm{Conf}_{n}\times F_{n}\right)/\mathcal{S}_{n}$, where $(\varepsilon_{1},\ldots,\varepsilon_{n})\in \mathrm{Conf}_{n}$ is the ordered $n$ tuples of the complex energies, the quotient space $F_{n}=\mathrm{GL}_{n}(\mathbbm{C})/\mathrm{GL}_{1}^{n}(\mathbbm{C})$ describes the eigenvectors of $\mathcal{H}(\beta_i)$ with respect to the complex energies, and $\mathcal{S}_{n}$ is the permutation group. The homotopy equivalence classes $\left[S^{1},\mathcal{X}_{n}\right]$ can be simplified as the conjugacy classes of the braid group $B_{n}=\pi_{1}\left(\mathrm{Conf}_{n}/\mathcal{S}_{n}\right)$ due to $\pi_{1}\left(F_{n}\right)=0$, and further reduced to the pure braid group $PB_{n}$, a subgroup of $B_{n}$, for the closed sub-GBZs and $S^{1}$ manifold here. Notably, it is unnecessary to specify the starting point of $S^{1}$ in the homotopy characterization, as different choices correspond to braids within the same conjugacy class. More complicated scenarios with intertwined sub-GBZs and $B_{n}$ braid group classifications are in the supplemental materials \cite{supp}. 

Intuitively, the target $B_{n}$ conjugacy classes are equivalent to the geometric knots \footnote{We have unified our references of links and knots as knots for simplicity.} of strings ($i=1,2,\ldots,n$) in the $\left(\mathrm{Re}\left(\varepsilon\right),\mathrm{Im} \left(\varepsilon\right),\theta\right)$ 3D space, and the target $PB_{n}$ to the geometric knots of closed loops \cite{hu2021knots,kassel2008braid,kauffman2013knots} in the 3D space periodic in $\theta \in\left[-\pi,\pi\right]$. We may thus characterize the braiding topology of the OBC bands with the total vorticity over the bands: 
\begin{align}
\label{eqvortotal}
\nu=\frac{1}{2}\sum_{i\neq j}\nu_{ij},
\end{align}
relevant to the braid crossings \cite{supp}, where $\nu_{ij}$ is the vorticity between two bands $\varepsilon_{i}(\theta)$ and $\varepsilon_{j}(\theta)$ \footnote{The vorticity expression is always well-defined: the derivative of $\varepsilon_{i}-\varepsilon_{j}$ may be discontinuous at cusps or bifurcation points of the spectrum, yet such singularity originates from the absolute value instead of the argument.}: 
\begin{align}
\label{eqvortwobands}
\nu_{ij}=\frac{1}{2\pi}\oint_{S^{1}}\frac{d}{d\theta}\mathrm{arg}\left[\varepsilon_{i}(\theta)-\varepsilon_{j}(\theta)\right]d\theta,
\end{align}
in a similar way to the non-Hermitian Bloch bands. We note that the vorticity is not a complete characterization of the braiding topology, and a comprehensive description requires braid words \cite{hu2021knots,kassel2008braid,kauffman2013knots} or knot invariants (polynomials), e.g., the Jones polynomials \cite{yang2020jones,witten1989,frohlich1989,OZSVATH200458}, to encode the full information of the braiding topology. Topologically different braids may manifest the same vorticity \cite{hu2022knot}; nevertheless, braids with distinct vortices must be topologically different, which offers us an elementary way to distinguish the different braiding topology. For more complex scenarios, knot polynomials might be necessary, but the overall braiding topology and phase transition presented in this paper remain valid.

{\em Topological transition of OBC band braiding.---}
The DPs are commonly EPs, which are stable, or unstable degenerate points (UDPs), which rely on fine-tuning and are unstable and evolve into several EPs upon generic perturbations \cite{hu2021knots,shen2018,heiss_2012,lee2016anom,yang2021doubling, supp}. Isolated bands are always topologically trivial, while separated bands may host nontrivial braiding topology, e.g., finite vorticity, which remains robust and protected without going through DPs. Correspondingly, the degenerate bands denote the critical point of a topological transition, where the vorticity may undergo a change.

\begin{figure}
    \centering
    \includegraphics[width=1 \linewidth]{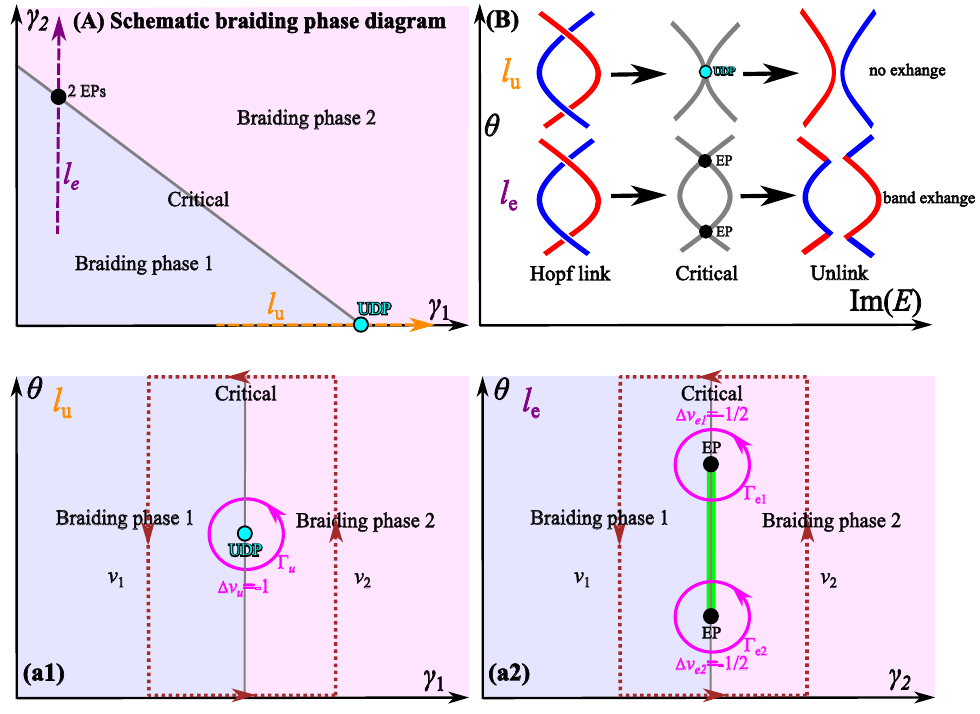}
    \includegraphics[width=1 \linewidth]{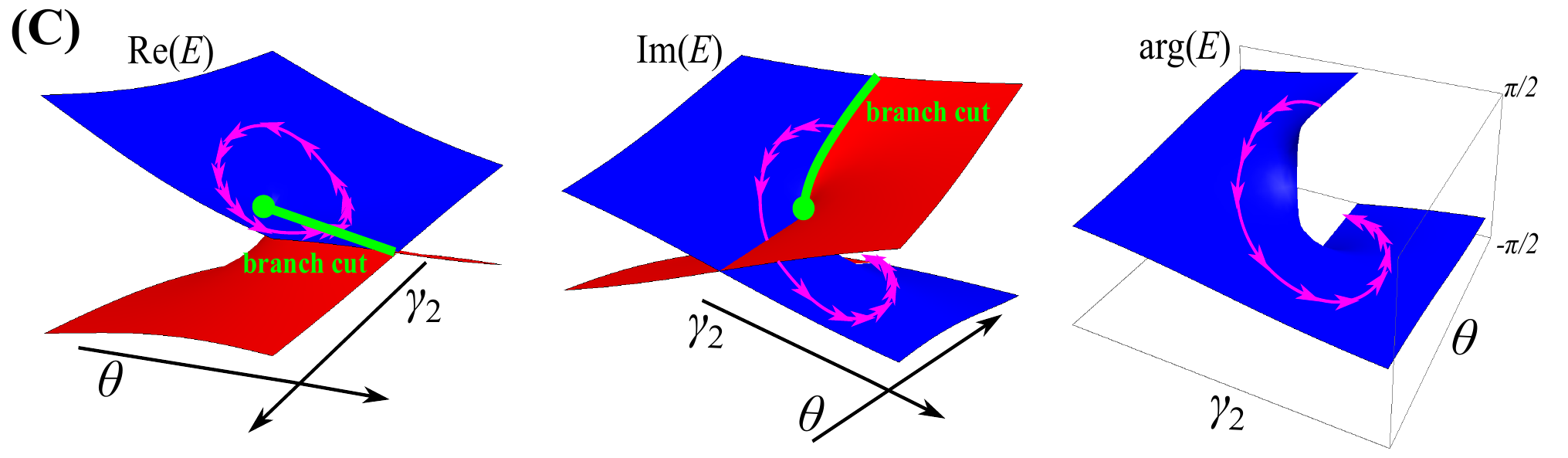}
    \caption{(A) A schematic phase diagram shows non-Hermitian models with different OBC band braidings are separated by intermediate topological transitions via UDP (orange arrowed line $l_{u}$) or EPs (purple arrowed line $l_{e}$). (B) The topological transition of OBC bands $\varepsilon_{1,2}$ from a Hopf link to an unlink may either go through a UDP directly or a pair of EPs, which exchange the bands between the EPs. The resulting change in total vorticity from $\nu_1$ to $\nu_2$ (brown dashed lines) corresponds to the contours around all DPs, (a1) a UDP with $\Delta \nu_u=-1$ or (a2) two EPs with $\Delta \nu_e=-1/2$ each and a branch cut in between (thick green line). (C) The Riemann surface structure shows the exchange between the two bands (blue and red surfaces) and the vorticity change as the contour (magenta arrowed curve) circles a single EP. }
    \label{figphasetran}
\end{figure}

For instance, we illustrate the different OBC band braidings and the topological transition in between on a two-parameter ($\gamma_{1,2}$) phase diagram in Fig. \ref{figphasetran}(A), where a two-fold degeneracy occurs on the critical line. We note that a higher-fold degeneracy can split into several robust two-fold degeneracies upon perturbations. Without loss of generality, we consider two types of transition: the degeneracy is introduced by stable EPs [along the $l_{e}$ line in Fig. \ref{figphasetran}(A)] or UDPs [along the $l_{u}$ line in Fig. \ref{figphasetran}(A)]. Without extra symmetries \cite{shen2018,heiss_2012,lee2016anom,yang2021doubling}, a UDP splits into two EPs upon a small perturbation, e.g., finite $\gamma_{2}$. 

Upon crossing each EP, two bands interchange and introduce a vorticity change of $\Delta \nu_{e}=\pm 1/2$. To see that, consider the effective Hamiltonian upon the two relevant bands close to the stable EP: $\mathcal{H}_{eff}=h_{x}\sigma_{x}+h_{y}\sigma_{y}+h_{z}\sigma_{z}$, where $\sigma_{x,y,z}$ are the Pauli matrices and $h_{x,y,z}$ are small parameters depending on $\theta$ and $\gamma_{1,2}$. The two bands $E_{\pm}=\pm\left(h_{x}^{2}+h_{y}^{2}+h_{z}^{2}\right)^{1/2}$ follow a multivalued function with a branch cut originating from the EP at $h_{x}^{2}+h_{y}^{2}+h_{z}^{2}=0$; see an example Riemann surface structure in the vicinity of the EP as a function of $\theta$ and a model parameter, e.g., $\gamma_{2}$, in Fig. \ref{figphasetran}(C). Any closed contour around the EP will traverse the branch cut and switch one band with the other, introducing a vorticity change of $\Delta\nu_e=\pm 1/2$ and indicating such an EP's stability against perturbations unless annihilated or merged with another EP \cite{supp}. 

After the topological transition, the two bands separate once again, but there may be a resulting change of total vorticity and an exchange of partners between parts of the bands. As we sum up $\theta$ for the total vorticity, the difference between the neighboring models [brown contours in Figs. \ref{figphasetran}(a1) and \ref{figphasetran}(a2)] receives contributions from all intermediate DPs (magenta circles): $\nu_{2}-\nu_{1}=\sum \Delta \nu$. For instance, two EPs with $\Delta\nu=-1/2$ allow a difference of $-1$ between the total vorticity of the models on two sides, e.g., a Hopf link and an unlink in Fig. \ref{figphasetran}(B), upon tuning the model parameter $\gamma_{2}$. In addition, after the two bands $\varepsilon_{1,2}$ touch at the two EPs, they separate and enter the unlink region, yet with their parts between the two EPs interchanged; see the lower panel in Fig. \ref{figphasetran}(B). 

On the other hand, topological transitions through EPs are not the only option, as a change of total vorticity can also happen via UDPs with $|\Delta \nu_{u}| \ge 1$, which may separate into multiple EPs. For example, the UDP with vorticity $\Delta \nu_{u}=-1$ [Fig. \ref{figphasetran}(a1)] can split into two separate EPs with $\Delta \nu_{e}=-1/2$ each and a branch cut in between [Fig. \ref{figphasetran}(a2)] upon perturbations. Similarly, we can visualize such a UDP as two $\Delta \nu_e=-1/2$ EPs merged together, leaving behind no branch cut or band exchange. Correspondingly, during the transition from the Hopf link to the unlink through a UDP [upper panel in Fig. \ref{figphasetran}(B)], the two bands $\varepsilon_{1}$ and $\varepsilon_{2}$ merely touch at the UDP and directly move across without any band exchanges, yielding an unlink afterward. The stable two-fold degeneracies constitute the fundamental ingredient of the braiding transition. Therefore, the braiding transition in non-Hermitian models with more than two bands can be understood as a combination of multiple two-fold degeneracies. The change in topological invariants, such as the total vorticity, adheres to the additive principle over all two-fold degeneracies. These arguments about the braiding phase transition generalize straightforwardly to topological transitions with arbitrary DPs and OBC band braidings.

\begin{figure*}
    \centering
    \includegraphics[width=0.24 \linewidth]{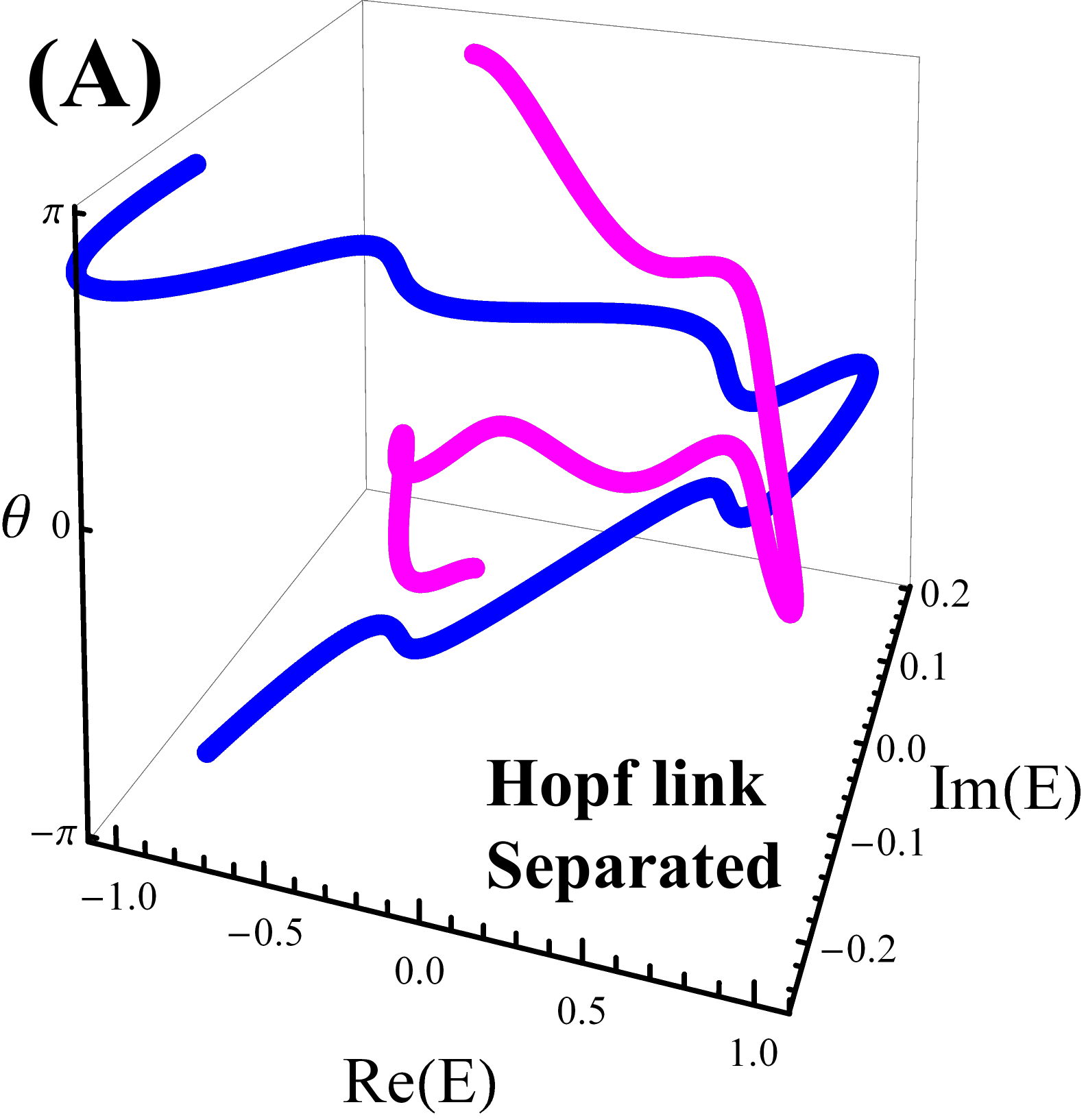}
    \includegraphics[width=0.24 \linewidth]{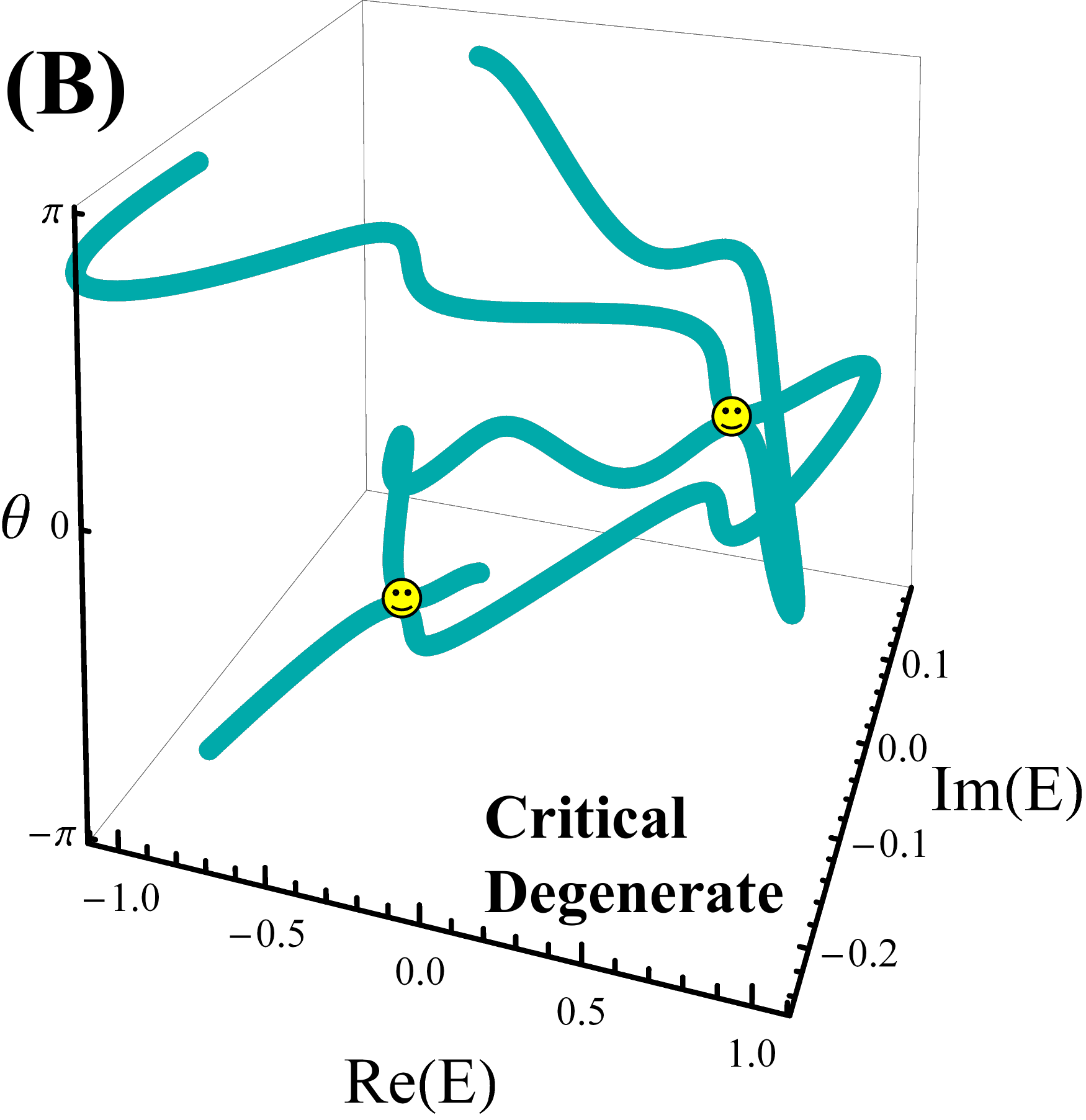}
    \includegraphics[width=0.24 \linewidth]{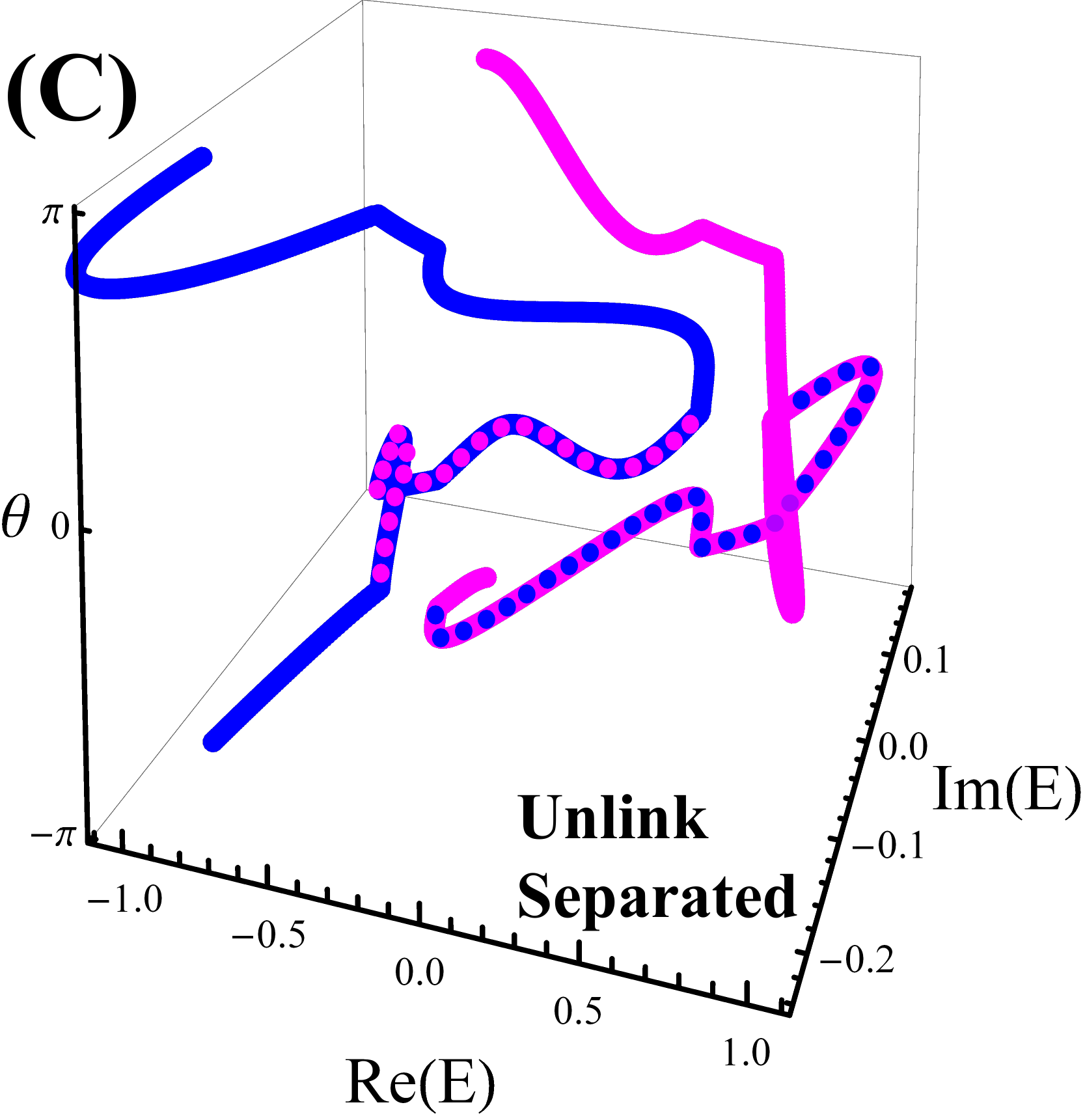}
    \includegraphics[width=0.24 \linewidth]{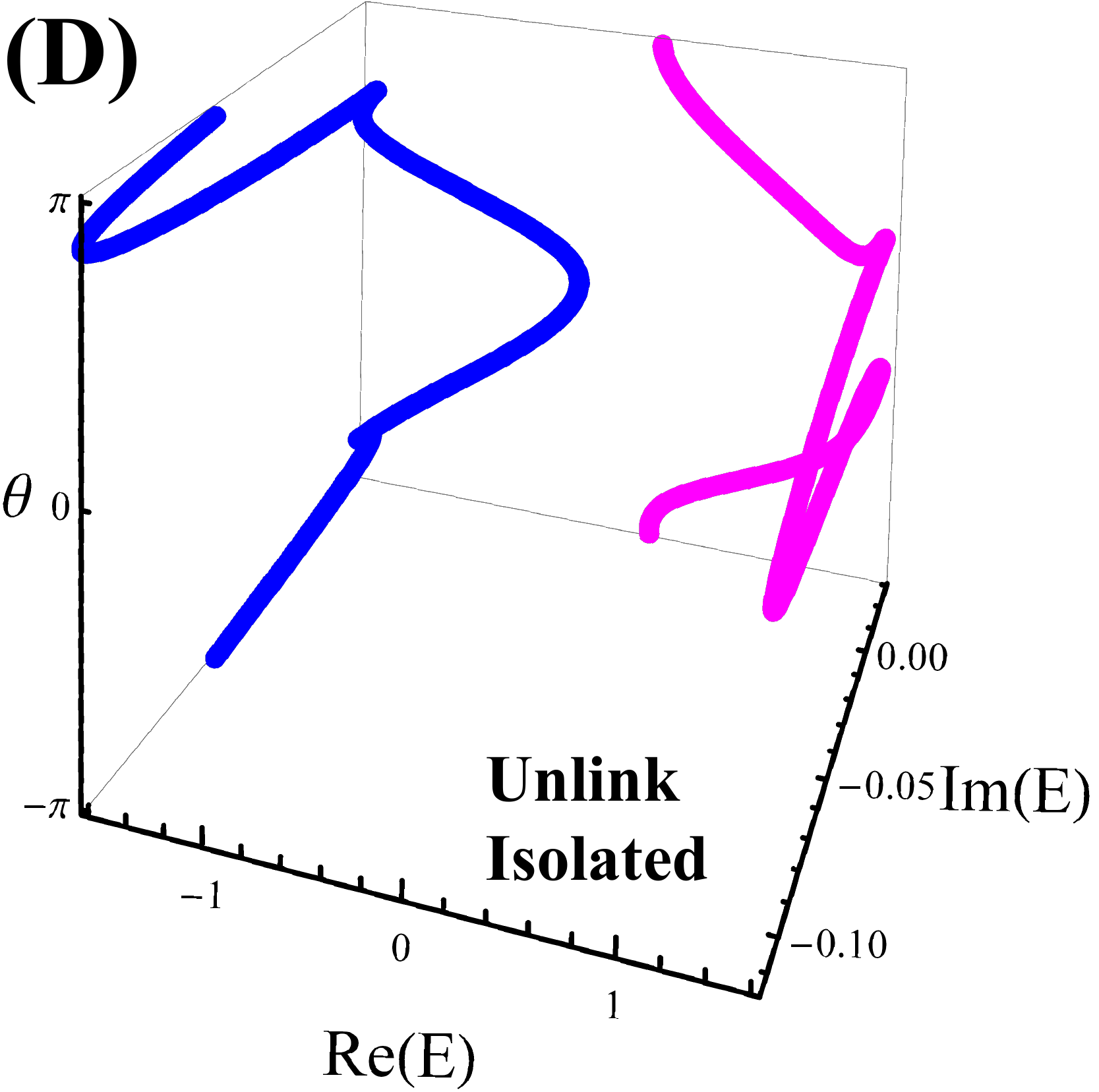}\\
    \includegraphics[width=0.252 \linewidth]{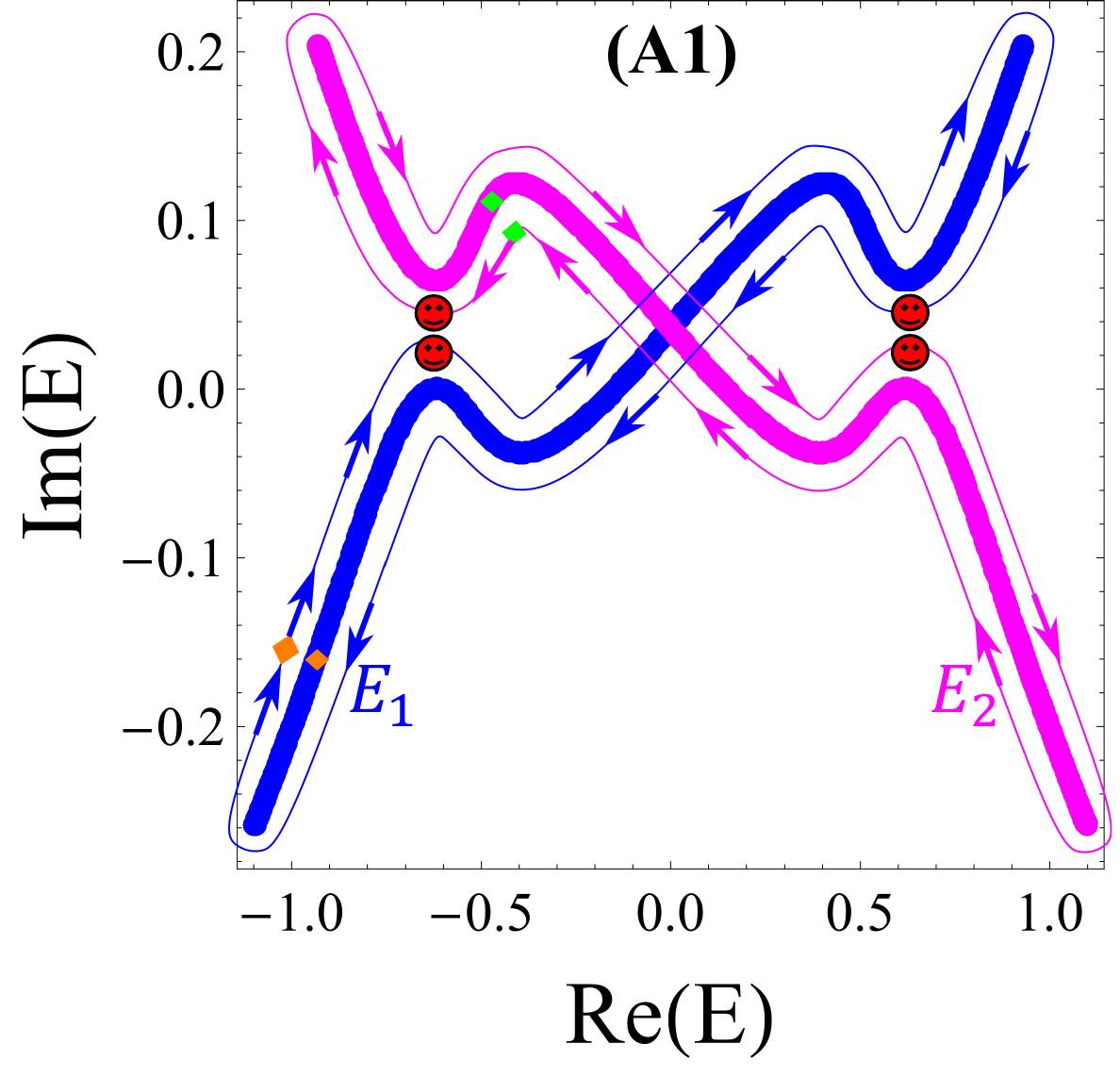}
    \includegraphics[width=0.225 \linewidth]{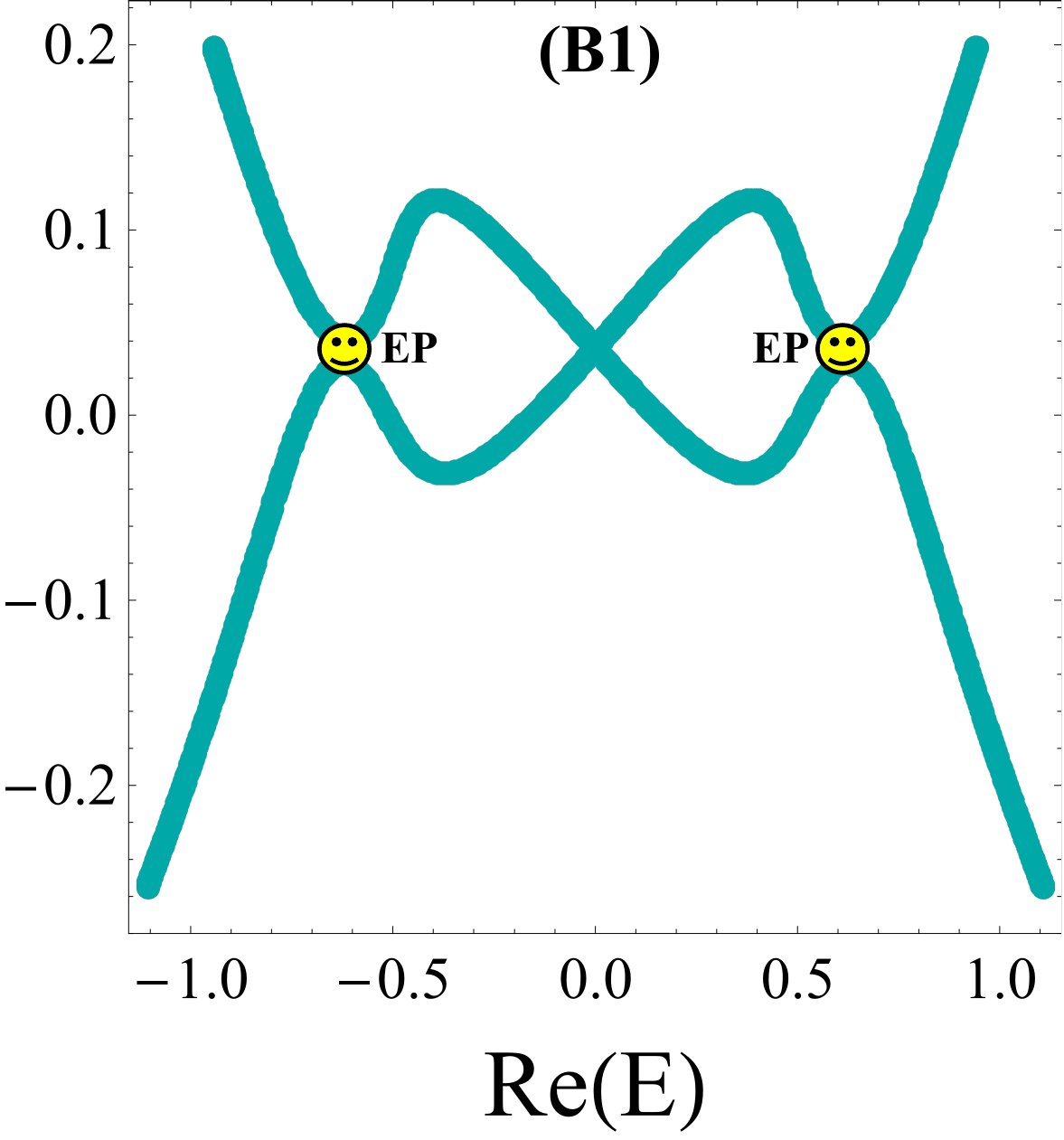}
    \includegraphics[width=0.225 \linewidth]{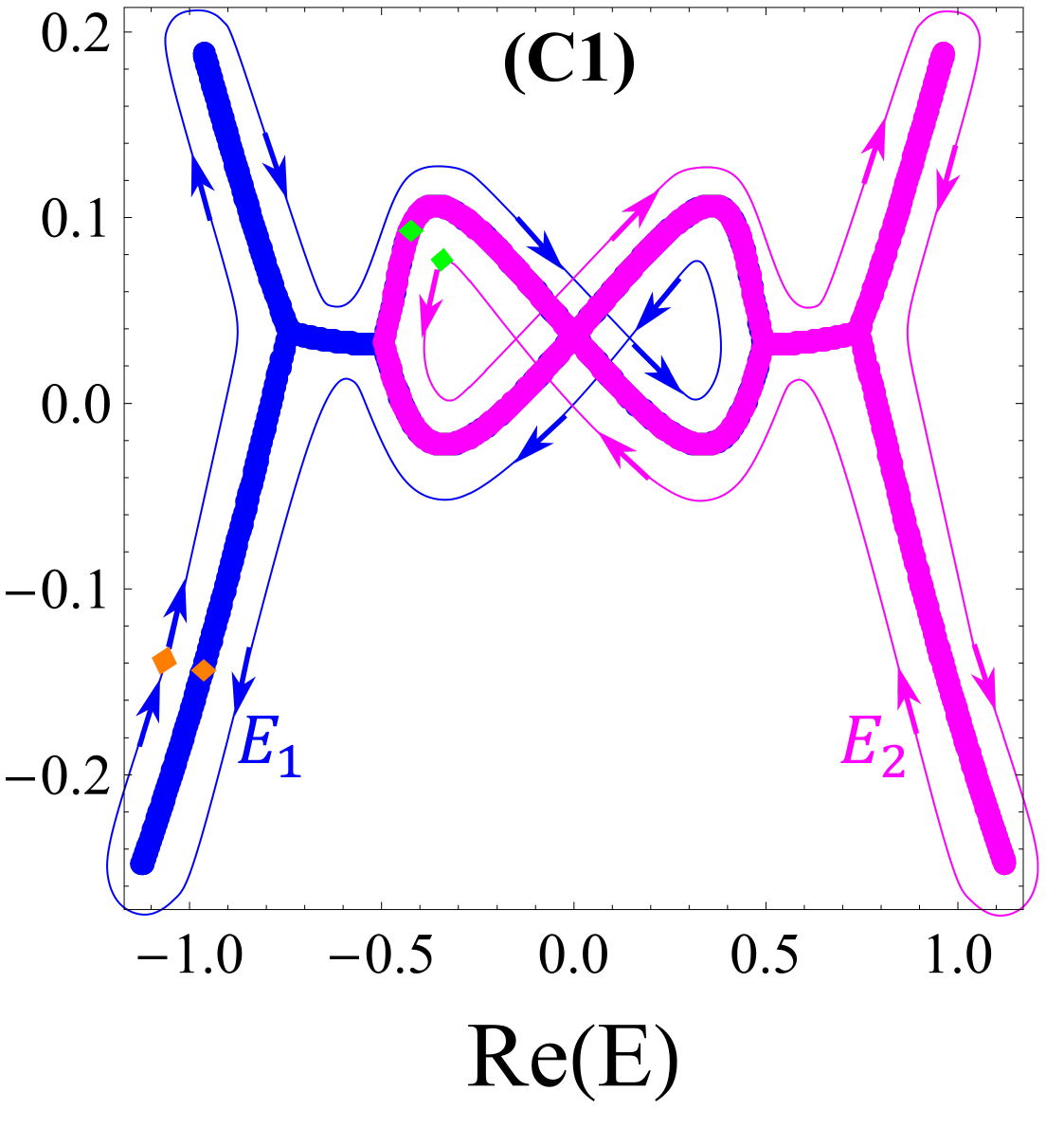}
    \includegraphics[width=0.232 \linewidth]{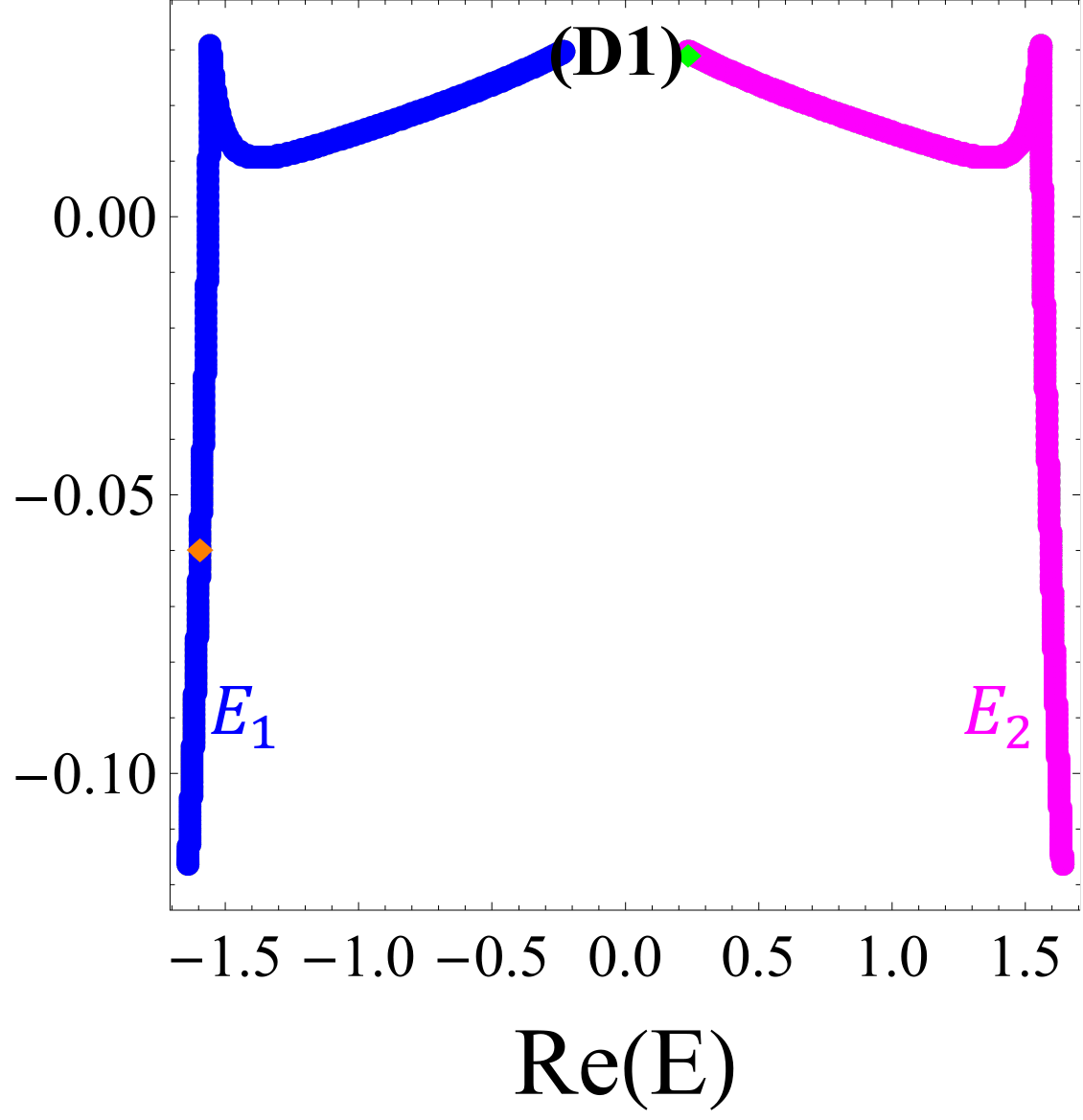}\\
    \includegraphics[width=0.255 \linewidth]{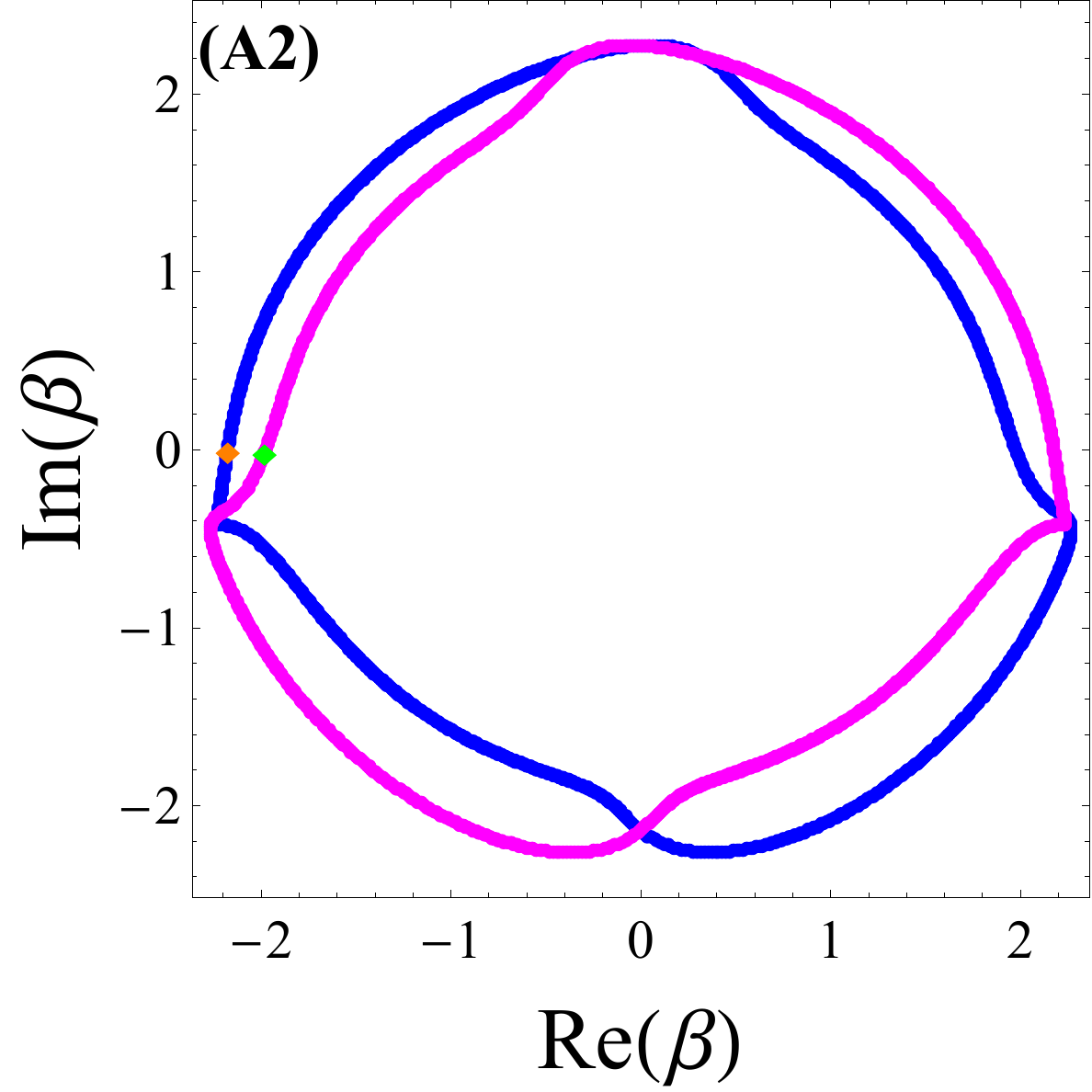}
    \includegraphics[width=0.234 \linewidth]{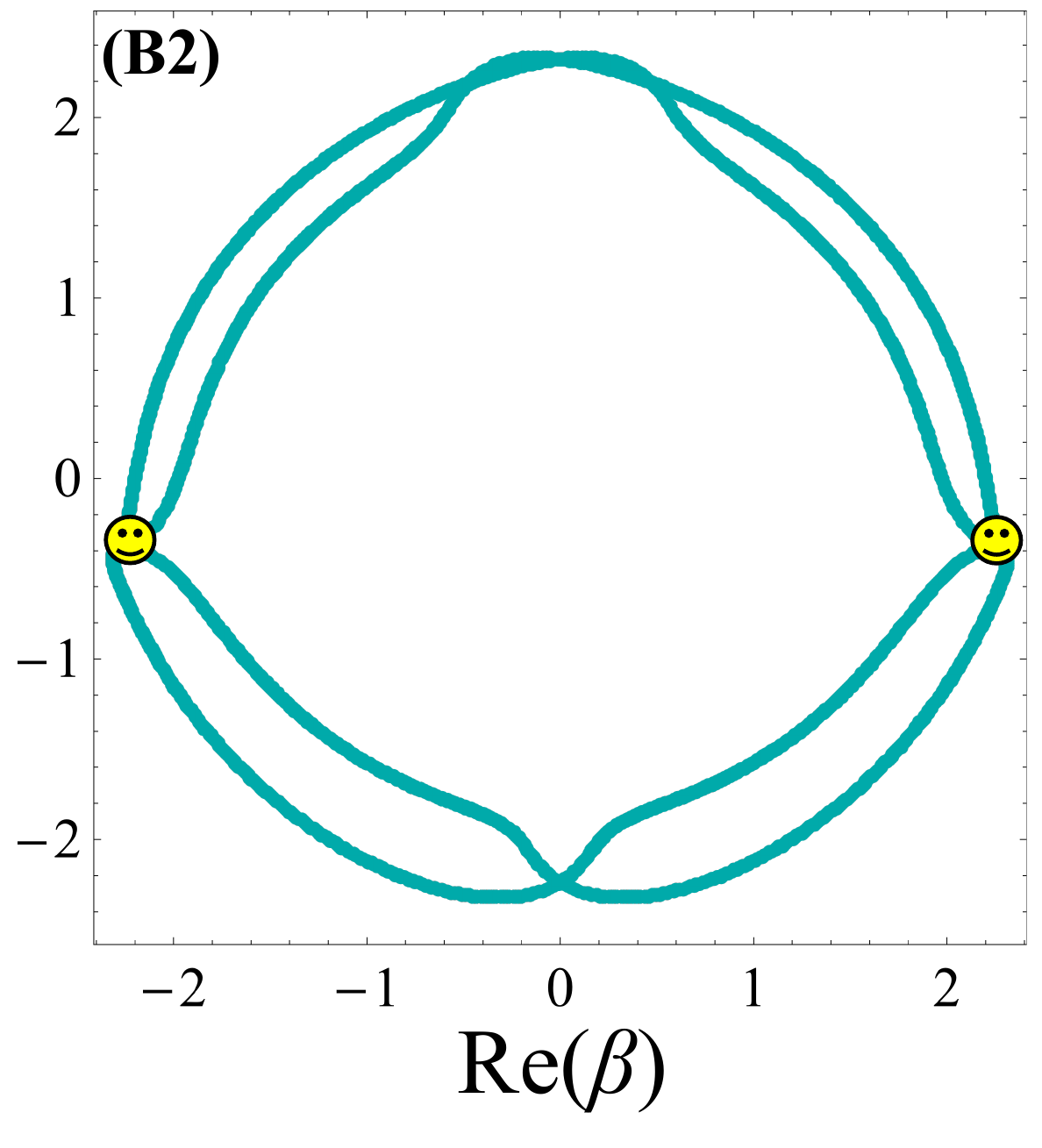}
    \includegraphics[width=0.232 \linewidth]{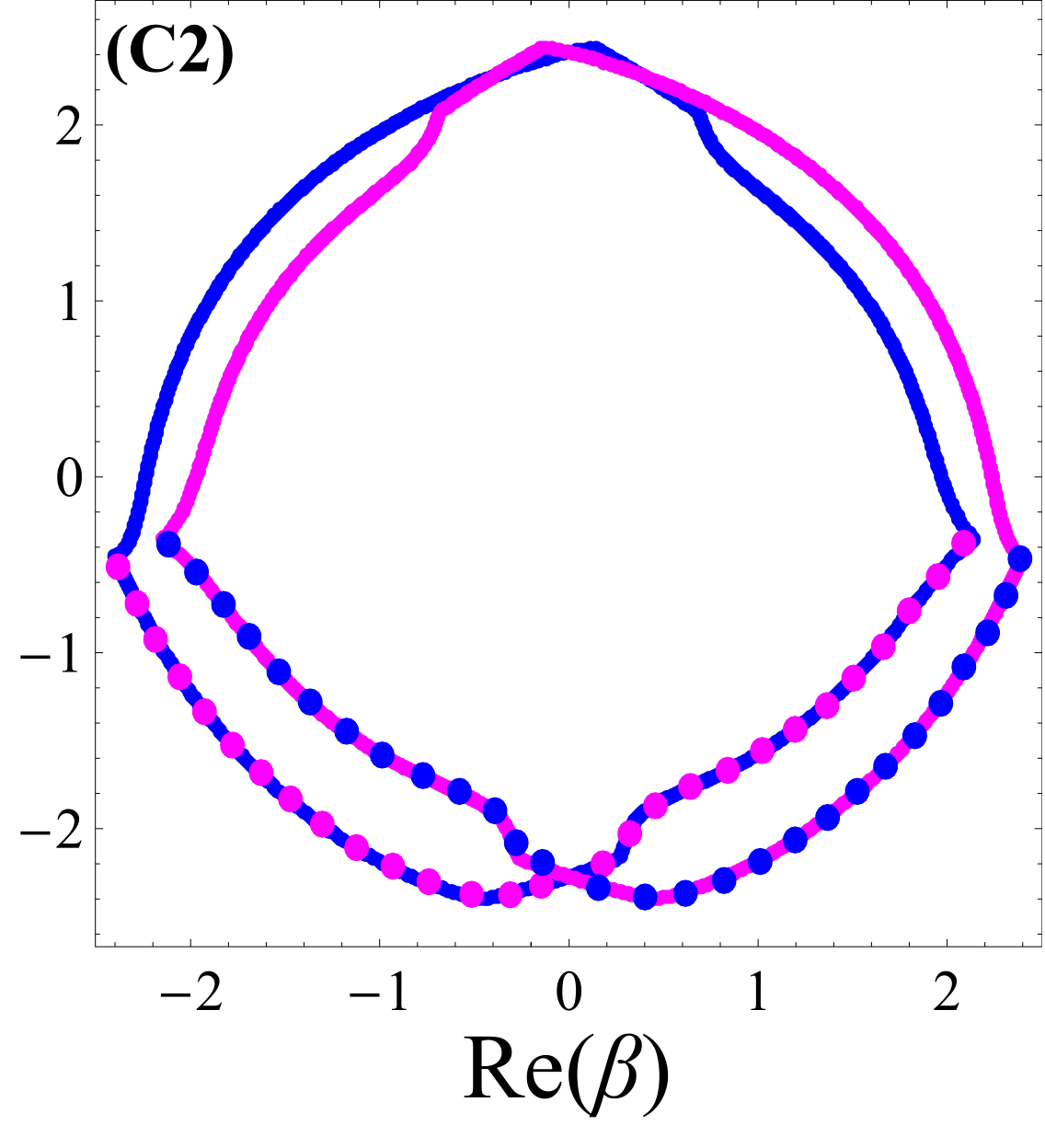}
    \includegraphics[width=0.23 \linewidth]{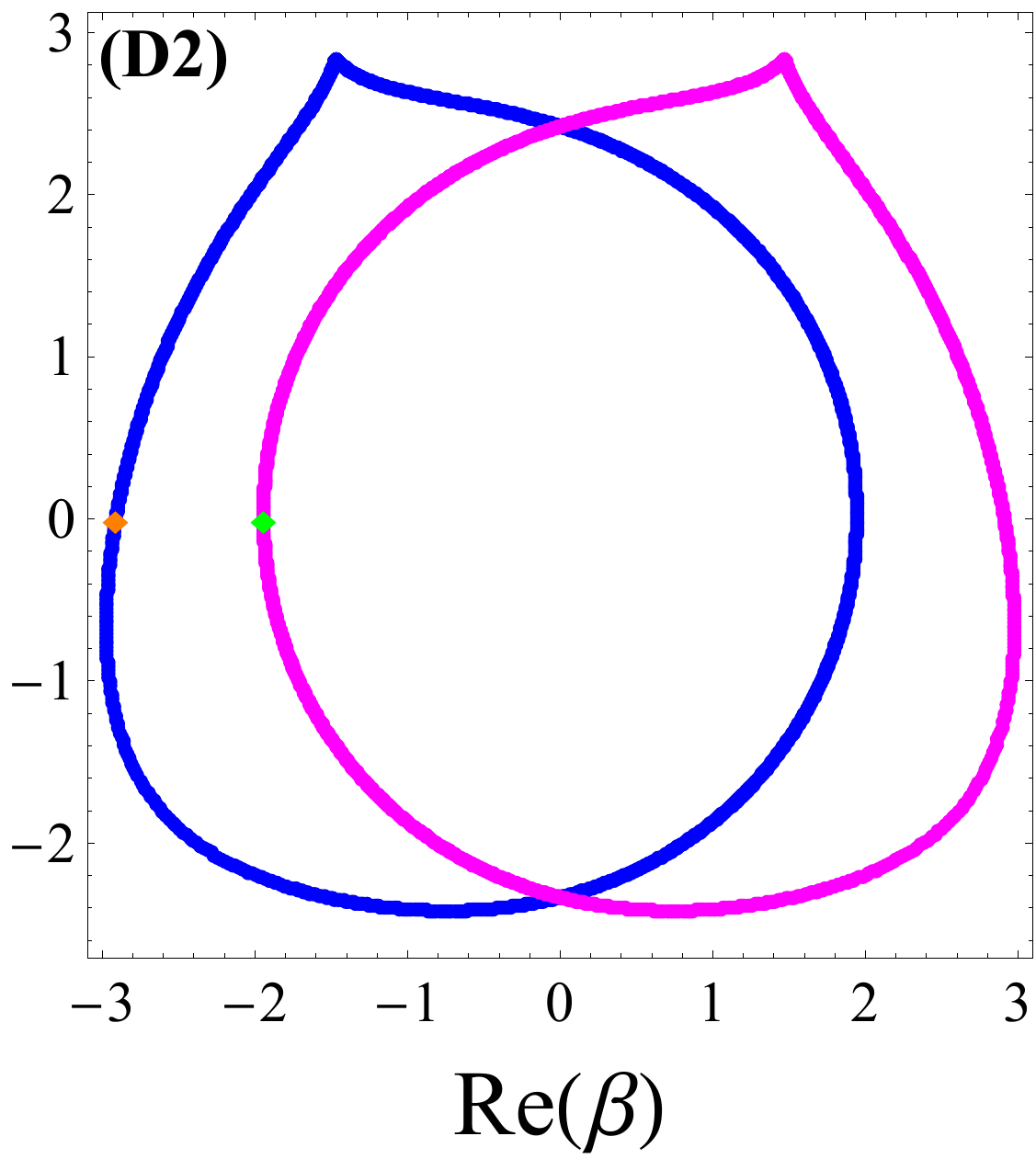}
    \caption{The OBC bands of the non-Hermitian model in Eq. (\ref{eqmainmodelreal}) demonstrate clear braiding topology in the $\left(\mathrm{Re}(E), \mathrm{Im}(E), \theta\right)$ space: (A) a Hopf link at $\lambda_{3}=0.36$; (B) a topological transition with two emergent EPs (yellow points) at $\lambda_{3}\approx 0.375$; (C) a separated unlink at $\lambda_{3}=0.4$; and (D) an isolated unlink at $\lambda_{3}=1$. $\lambda_1=\lambda_2=0$, $t_{p}=0.15$, $t_{m}=0.85$, $t_{a}=0.2$, and $m=0.1$. We also show (A1)-(D1) the corresponding bands $E_{1,2}$ in the complex $E$ plane and (A2)-(D2) the sub-GBZs $\mathcal{C}_{1,2}$ in the complex $\beta$ plane. The blue and magenta arrows show the spectral flows as $\theta$ evolves along the sub-GBZs, where the orange and green diamonds mark the $\theta=-\pi$ locations. Note that the $``\infty"$-shaped central loop in (C1) is visited once only by each band, which overlaps there. The magenta (blue) dots in (C) and (C2) denotes the parts of the bands and sub-GBZs exchanged after the transition.}
    \label{figknotmodel}
\end{figure*}

{\em An illustrated model.---}
We study the following 1D non-Hermitian tight-binding model under OBC as an example of the continuity criterion, the OBC band braiding, and the topological transitions: 
\begin{align}
    \label{eqmainmodelreal}  \hat{\mathcal{H}}_{br}=\sum_{j}\left[\mathbf{c}^{\dagger}_{j}\mathcal{M}\mathbf{c}_{j}+\mathbf{c}^{\dagger}_{j}\mathcal{T}_{p}\mathbf{c}_{j+1}+\mathbf{c}^{\dagger}_{j+1}\mathcal{T}_{m}\mathbf{c}_{j}\right],
\end{align}
where $\mathbf{c}_{j}$ is the annihilation operator at site $j$ with $n=2$ internal degrees of freedom, e.g., spins or sublattices, and we parametrize the hopping matrices as: 
\begin{align}
    \label{eqhopping}
    \mathcal{M}&=\left(\begin{matrix}
    \lambda_{3} & -im\\
    im & -\lambda_{3}
    \end{matrix}\right), \nonumber\\
    \mathcal{T}_{p}&=\left(\begin{matrix}
    t_{p}+\lambda_{1} & -t_{a}\\
    t_{a} & t_{p}-\lambda_{1}
    \end{matrix}\right),
    \mathcal{T}_{m}=\left(\begin{matrix}
    t_{m}+\lambda_{2} & t_{a}\\
    -t_{a} & t_{m}-\lambda_{2}
    \end{matrix}\right).
\end{align}
The corresponding Laurent polynomial reads:
\begin{align}
    \label{eqmainmodelham}
    \mathcal{H}_{br}(\beta)=\left(
        \begin{matrix}
            h_{1}(\beta)+h_{2}(\beta) & -h_{0}(\beta)\\
            h_{0}(\beta) & h_{1}(\beta)-h_{2}(\beta)
        \end{matrix}
    \right),
\end{align}
where $h_{0}(\beta)=t_{a}(\beta-\beta^{-1})+im$, $h_{1}(\beta)=t_{p}\beta+t_{m}\beta^{-1}$, and $h_{2}(\beta)=\lambda_{1}\beta+\lambda_{2}\beta^{-1}+\lambda_{3}$. Next, we explore its band braidings and topological transitions. For simplicity, we set $\lambda_{1}=\lambda_{2}=0$, $t_p=0.15$, $t_m=0.85$, $t_a=0.2$, $m=0.1$, and vary $\lambda_3$ in the main text and leave more general cases to the supplemental materials \cite{supp}. 

Applying our continuity criterion, we obtain the two OBC bands $E_{1}$ (blue) and $E_{2}$ (magenta) and their respective sub-GBZs as shown in Fig. \ref{figknotmodel}. The bands' braiding topology is visible and distinguishable in the $\left(\mathrm{Re}(E), \mathrm{Im}(E), \theta\right)$ 3D space: the Hopf link of the separated bands [Fig. \ref{figknotmodel}(A)] transforms into an unlink of the separated [Figs. \ref{figknotmodel}(C)] or isolated [Figs. \ref{figknotmodel}(D)] bands as $\lambda_3$ is gradually increased. The topological transition occurs with the emergence of two stable EPs where the two OBC bands meet [Fig. \ref{figknotmodel}(B)]. Each EP contributes to a $\Delta \nu_e=-1/2$ change to the vorticity, which alters from $\nu_{h}=1$ of the Hopf link to $\nu_{u}=0$ of the unlink and sufficiently distinguishes the braiding topology. 

Interestingly, as the red points on the two bands of the Hopf link approach [Fig. \ref{figknotmodel}(A1)] and touch at the EPs [yellow points in Fig. \ref{figknotmodel}(B1)], the bands switch partners---their parts between the EPs; see Fig. \ref{figknotmodel}(C1). This band exchange is also apparent from the sub-GBZ view, where the sub-GBZs $\mathcal{C}_{1}$ and $\mathcal{C}_{2}$ touches at the topological transition and switch their portions between the two EPs [Fig. \ref{figknotmodel}(A2) to \ref{figknotmodel}(C2)]. We note that such an exchange requires a unified starting point for the sub-GBZs, say, an identical convention for $\beta_1$ and $\beta_2$'s argument $\theta$, so that the sub-GBZs remain continuous and well-defined across the transition. For example, the green and orange points are $\theta=-\pi$ for the respective bands, regardless before or after the transition. 

One interesting property under OBC is that each point (except a few endpoints) on its spectrum has to be visited (at least) twice \cite{wu2022,hu2024geo}. Correspondingly, we mark the spectral flows along the OBC bands as $\theta\in[-\pi,\pi]$ traverses its range. Each point (exclusion of endpoints) on the Hopf link bands is visited twice. After the topological transition, however, the flow is intercepted and interchanged between the bands, each of which only visited the central loop once. It is their overlap that ensures the two-visit rule and unique for multiband OBC systems. Also, as the OBC bands touch at the EPs, the two-visit rule gives rise to two natural choices of flow directions and exchange partners; see supplemental materials for further discussions. 

When $\lambda_{3}$ equals zero, on the other hand, Eq. (\ref{eqmainmodelham}) reduces to the model in Ref. \cite{li2023nonblochbraiding}. The two sub-GBZs overlap, and as $m$ is varied, the Hopf link and unlink are separated by a topological transition through a UDP, which splits into two EPs upon a small perturbation, e.g., a small $\lambda_3$.

{\em Discussions and conclusions.---}
We have proposed a continuity criterion to separate the respective sub-GBZs and non-Bloch bands of non-Hermitian multiband quantum systems under OBC. This allowed us to build their homotopic characterizations---braiding topology, such as the total vorticity over the OBC bands, and the intermediate topological transitions signaled by emergent DPs. We have also demonstrated our conclusions in an example non-Hermitian two-band model. 

We have discovered that non-Hermitian open-boundary bands undergo part exchanges when the system transitions through specific EPs. Remarkably, this phenomenon also exists within the non-Hermitian Bloch bands, though it had remained unidentified until now. We have assumed closed sub-GBZs $\mathcal{C}_{i}$ homeomorphic to $S^{1}$ in the main text, yet our continuity criterion and braiding topology, homotopic characterization of $B_{n}$ instead of $PB_{n}$, also remain valid for more subtle scenarios with intertwined sub-GBZs, as we discuss in the supplemental materials \cite{supp}. Our analysis also does not depend on symmetries, whose participation in further topological classifications of the non-Bloch bands is an interesting open question for future studies. 

{\em Acknowledgments.---} We thank Kunkun Wang and Shu-Xuan Wang for the helpful discussions and comments. We acknowledge support from the National Key R\&D Program of China (No.2022YFA1403700) and the National Natural Science Foundation of China (No.12174008 \& No.92270102).

\bibliography{reference}

\onecolumngrid
\flushbottom
\clearpage
\appendix

\begin{center}
\textbf{\large Supplemental Online Material for ``Braiding Topology of Non-Hermitian Open-Boundary Bands''}
\end{center}

\begin{center}
{Yongxu Fu$^*$ and Yi Zhang$^\dagger$  }  
\end{center}

\begin{center}
{\em \footnotesize
International Center for Quantum Materials, School of Physics, Peking University, Beijing 100871, China
}
\end{center}

\section{S1. Detail and applicability of the algorithm of continuity criterion}

\begin{figure}
    \centering
    \includegraphics[width=1 \linewidth]{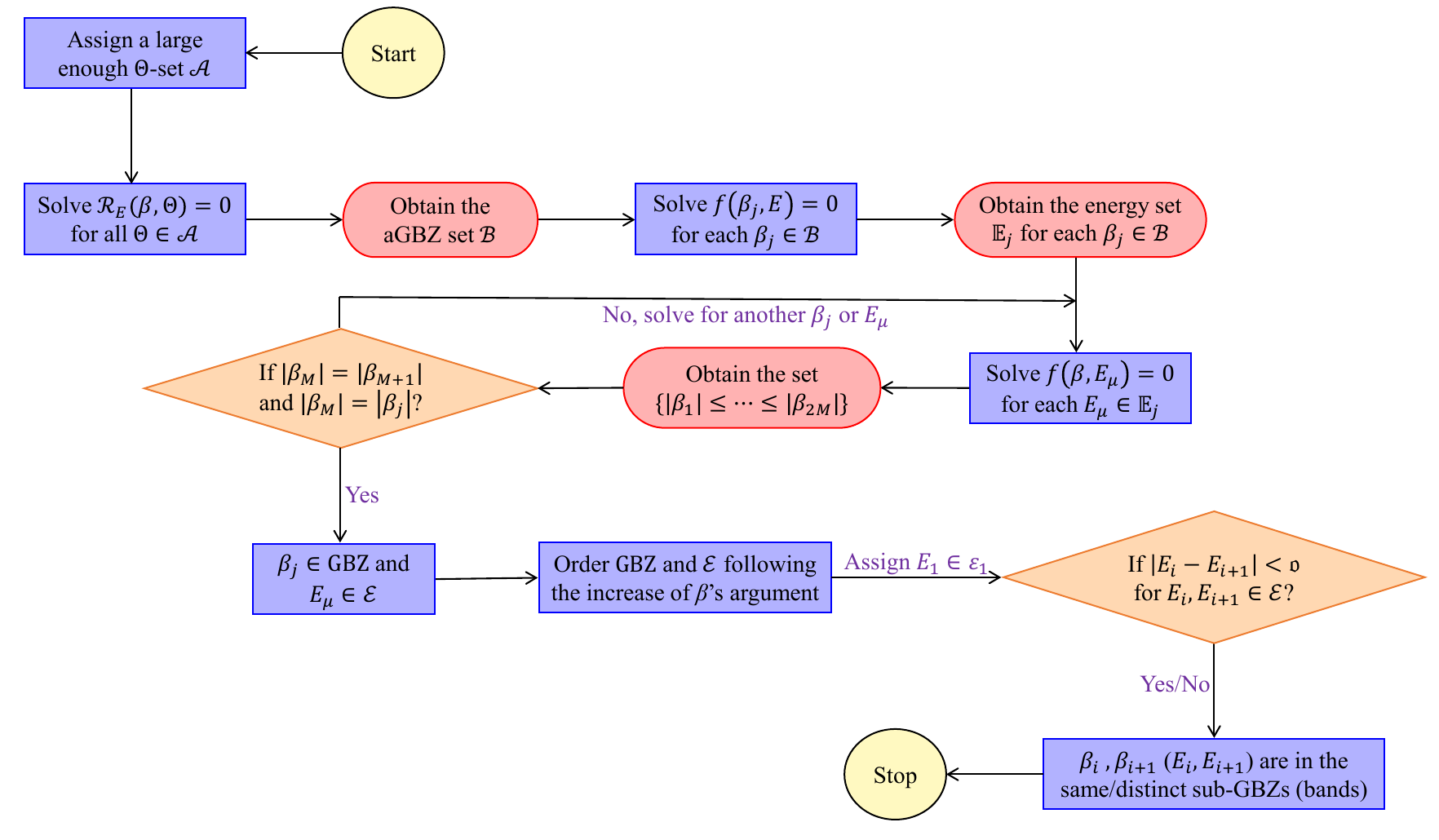}
    \caption{The flow chart of the algorithm to obtain the sub-GBZs (OBC bands) and enforce the continuity criterion.}
    \label{suppfigflowchart}
\end{figure}

In this section, we provide a detailed elucidation of the algorithm used to obtain the sub-generalized Brillouin zones (sub-GBZs) along with the related OBC bands and enforce the continuity criterion that compensates for the main text.
First, we solve the resultant equation $\mathcal{R}_{E}(\beta,\Theta)\equiv\mathcal{R}_{E}\left[f(\beta,E),f(\beta e^{i\Theta},E)\right]=0$,
over the range $\Theta\in [-\pi,\pi]$. In practice, we discretize a sufficiently large $\Theta$-set $\mathcal{A}$ in $[-\pi,\pi]$ such that the adjacent $\Theta$'s close to each other, thereby simulating the continuity. For a given $\Theta$, the characteristic equations (CHEs) $f(\beta, E)=0$ and $f(\beta e^{i\Theta}, E)=0$ share a common energy $E$ if and only if the resultant equation holds. Thus, we can obtain all $\beta_{p+1}=\beta_{p} e^{\pm i\Theta}$ solutions satisfying the CHE, so that $\left|\beta_{1}\right|\leq\ldots\leq\left|\beta_{p}\right|=\left|\beta_{p+1}\right|\leq\ldots \leq\left|\beta_{2M}\right|$, $p=1,2,\ldots,2M-1$, for a specific $\Theta$. The union of the solutions for all $\Theta$ constitutes the auxiliary GBZ (aGBZ) set $\mathcal{B}$. Next, we select the $\beta$ solutions with $p=M$, which coincide with the GBZ condition, and map out the GBZ as we vary $\Theta$. In practice, we solve the CHE $f(\beta_{j}, E)=0$ for each $\beta_{j}\in \mathcal{B}$, and obtain the corresponding energy set $\mathbb{E}_{j}$. By substituting each $E_{\mu}\in\mathbb{E}_{j}$ back into the CHE $f(\beta, E_{\mu})=0$, we obtain the corresponding set $\left\{\left|\beta_{1}\right|\leq\ldots\leq\left|\beta_{2M}\right|\right\}$. The GBZ condition is determined by whether $\left|\beta_{M}\right|=\left|\beta_{M+1}\right|$ and $\left|\beta_{M}\right|=\left|\beta_{j}\right|$ simultaneously hold. Here, the equality $\left|\beta_{M}\right|=\left|\beta_{M+1}\right|$ is realized through $\left|\left|\beta_{M}\right|-\left|\beta_{M+1}\right|\right|<\eta$ for a sufficiently small $\eta$ in the numerical calculations. If the GBZ condition holds, we judge that $\beta_{j}$ ($E_{\mu}$) lies in the GBZ (OBC-band set $\mathcal{E}$); if not, we repeat the above procedures for another $\beta_{j}$ and $E_{\mu}$ until the respective sets $\mathcal{B}$ and $\mathbb{E}_{j}$ are complete. Finally, after obtaining the GBZ set with the corresponding ensemble of OBC bands $\mathcal{E}$, we apply the continuity criterion: as the argument $\theta$ of $\beta$ increases, both $\beta_i$ and $\varepsilon_{i}$ within each band $i$ evolve continuously. In practice, it helps to separate this ensemble of energies into individual bands $\varepsilon_{i}$ one by one, following the order of $\beta$'s argument at a specific numerical precision. Each $\{\beta_i\}$ for $\theta\in[-\pi, \pi]$ constitutes a sub-GBZ $\mathcal{C}_{i}$. Specifically, after ordering the sets GBZ and $\mathcal{E}$ following the increase of $\beta$'s argument, we label $\beta_{i}$ and $E_{i}$ as the sequential elements of GBZ and $\mathcal{E}$, respectively. We artificially assign the first element $E_{1}$ ($\beta_{1}$) in $\mathcal{E}$ (GBZ) to the first OBC band $\varepsilon_{1}$ (sub-GBZ $\mathcal{C}_{1}$). If the condition $\left|E_{i}-E_{i+1}\right|<\mathfrak{o}$ holds (or does not hold) for a sufficiently small $\mathfrak{o}$, then $\beta_{i},\beta_{i+1}$ ($E_{i},E_{i+1}$) belong to the same (or distinct) sub-GBZs (OBC bands). We repeat this sorting until $\beta_{i}$ completes its $2\pi$ cycle in the set GBZ. This process allows us to enforce the continuity criterion numerically. The flow chart of the above algorithm is illustrated in Fig. \ref{suppfigflowchart}.

\begin{figure}
    \centering
    \includegraphics[width=0.42 \linewidth]{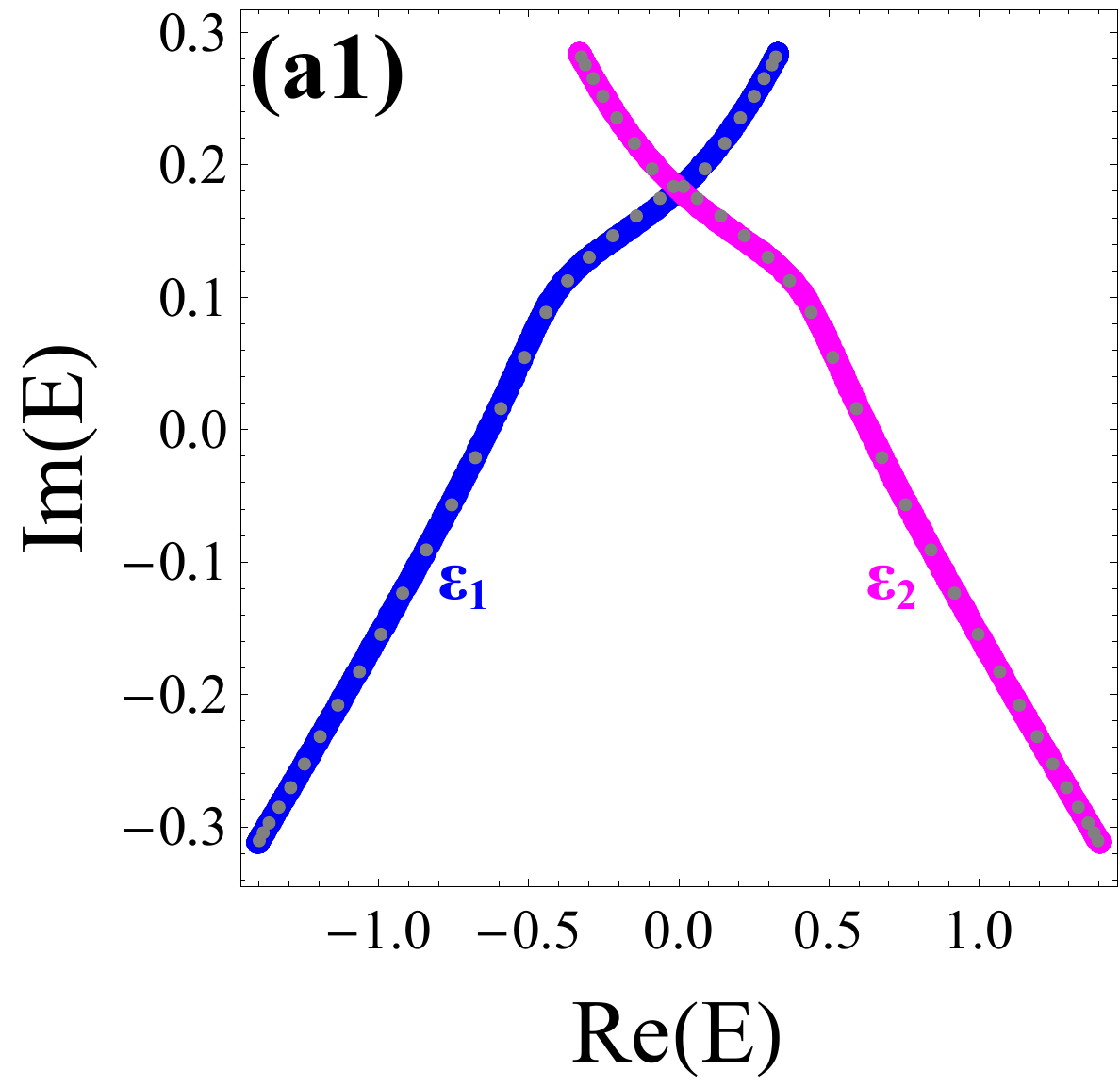}
    \includegraphics[width=0.385 \linewidth]{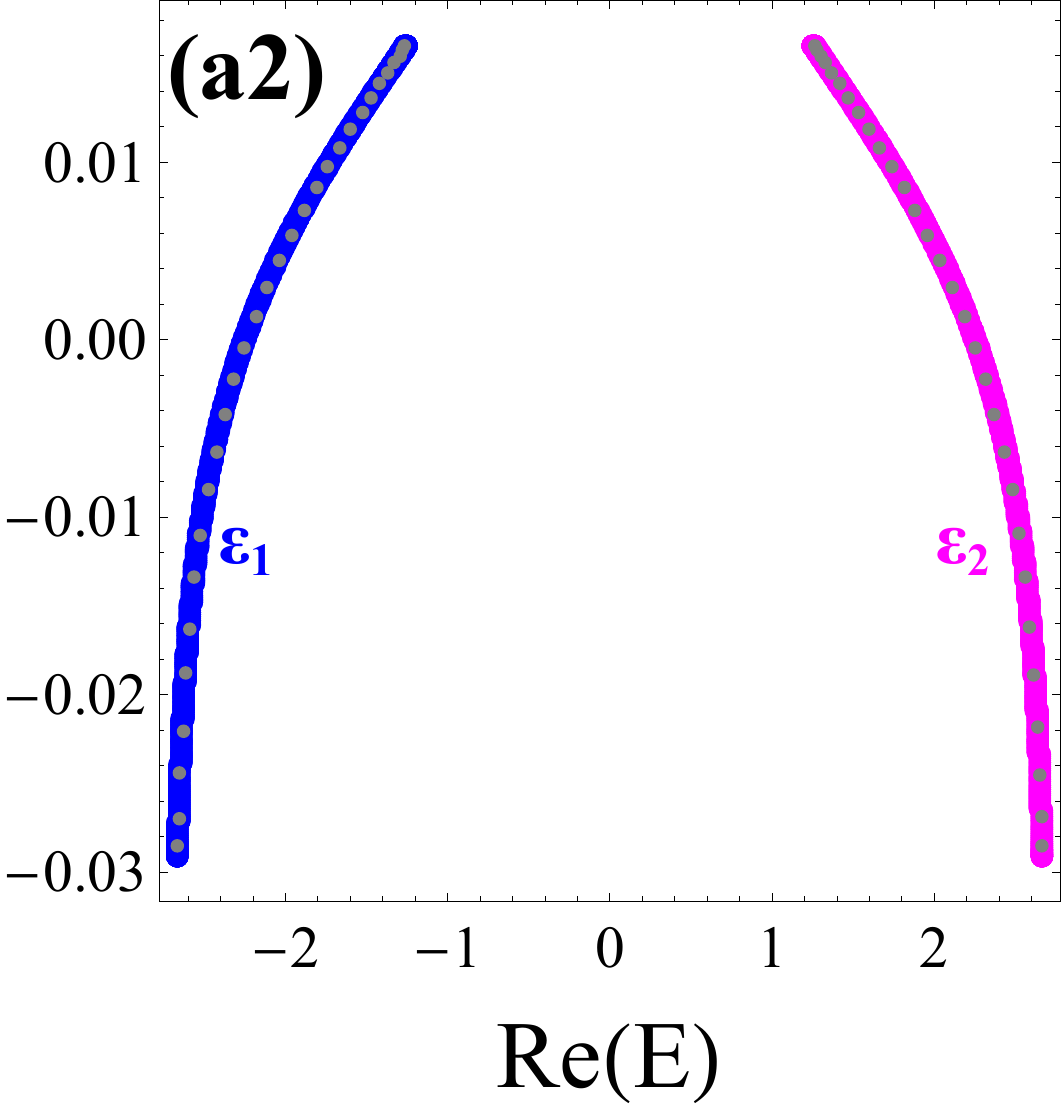}\\
    \includegraphics[width=0.415 \linewidth]{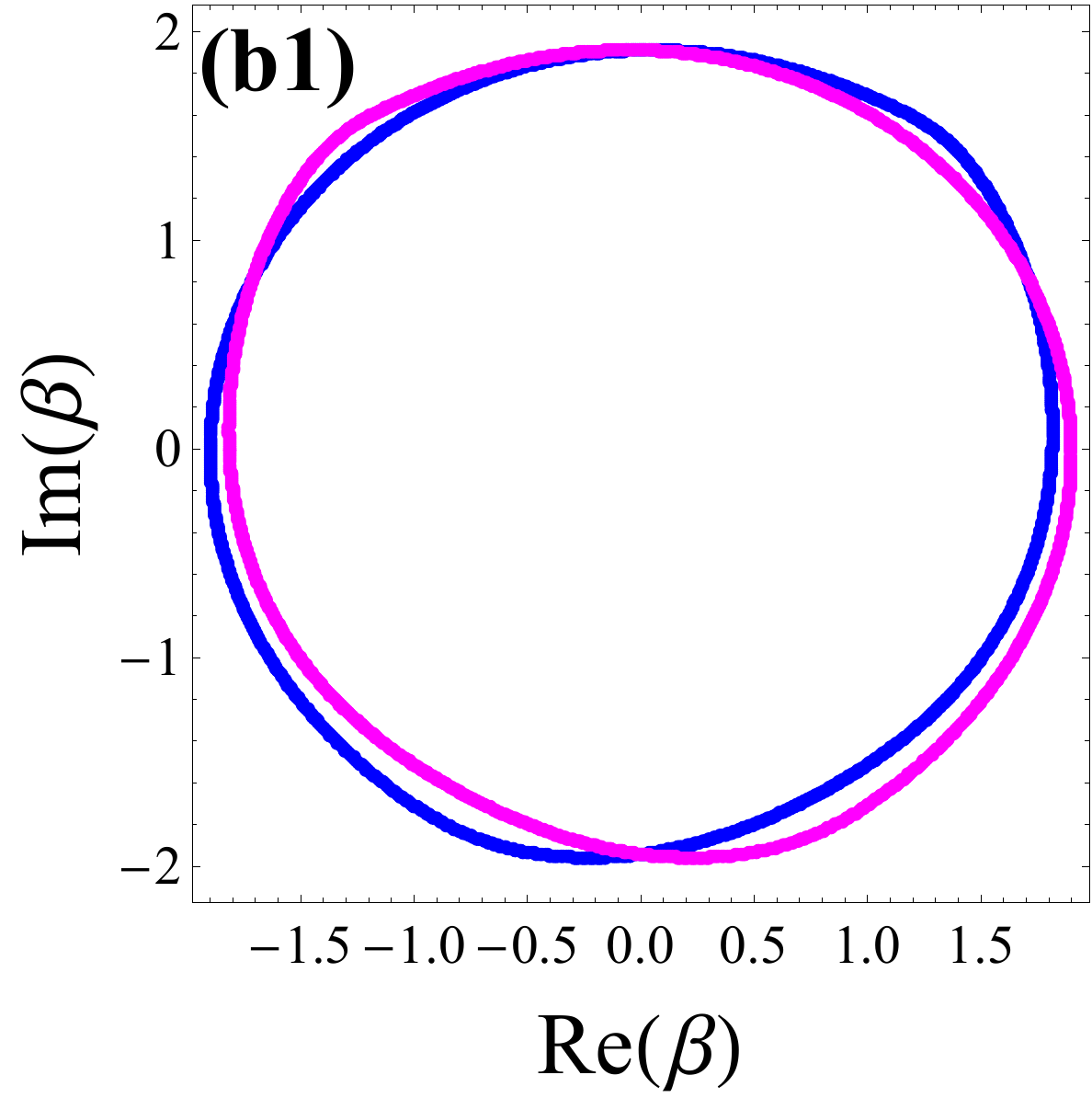}
    \includegraphics[width=0.37 \linewidth]{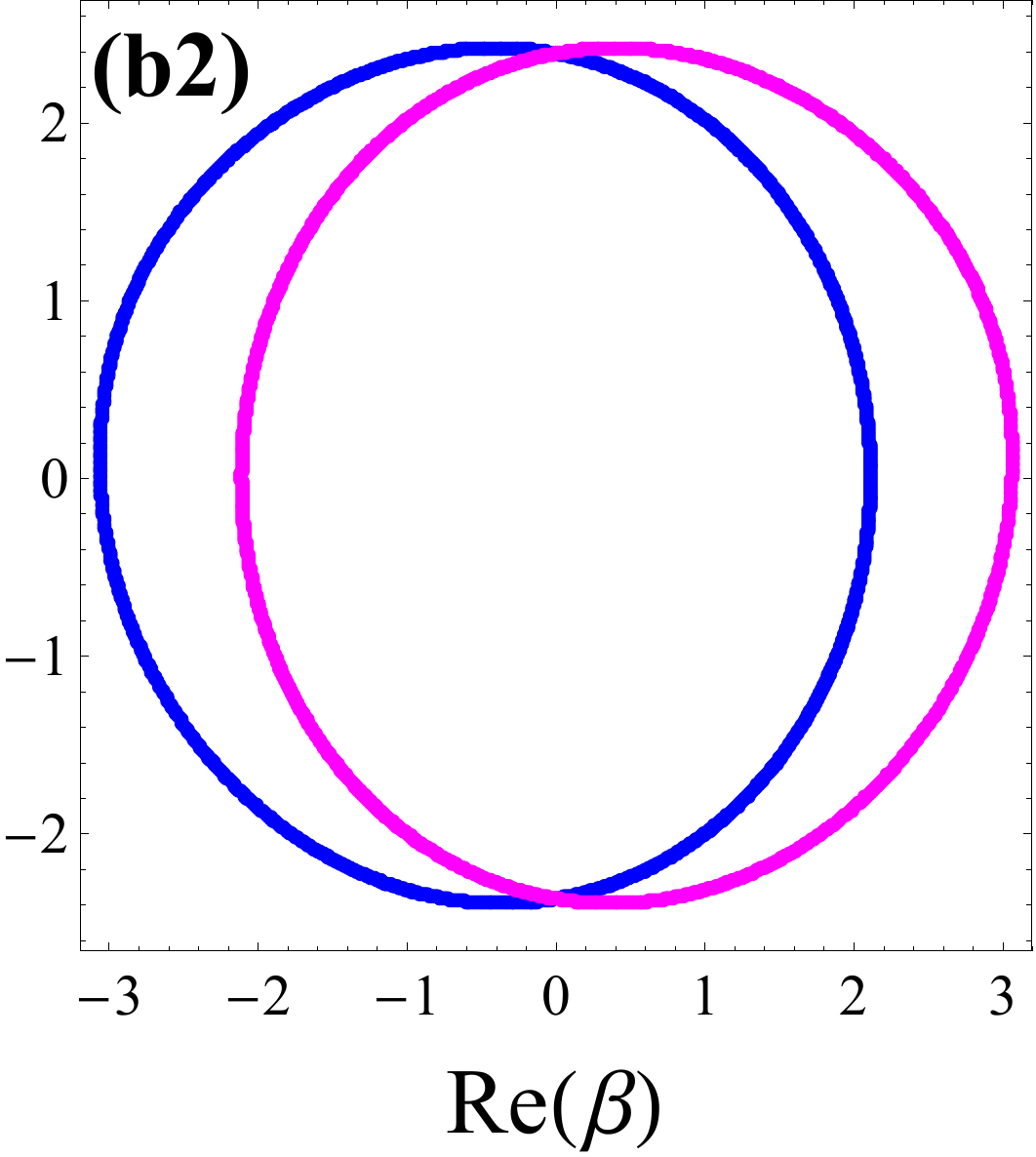}
    \caption{Illustration of the non-Bloch Hamiltonian $\mathcal{H}_{br}(\beta)$ in the main text utilizing the algorithm of the continuity criterion. The open-boundary bands $\varepsilon_{1,2}$ (blue and magenta lines) in (a1) and (a2) correspond to sub-GBZs with $(m, \lambda_{3})=(0.5, 0.2)$ and $(0.1, 2)$ in (b1) and (b2), respectively, which accord with the OBC spectra (grey points) through diagonalizing the Hamiltonian $\hat{\mathcal{H}}_{br}$ in the real space with $30$ sites. The other parameters are $t_{p}=0.15, t_{m}=0.85, t_{a}=0.2, \lambda_{1}=\lambda_{2}=0$.}
    \label{suppfigcontinu}
\end{figure}

As illustrated in Figs. \ref{suppfigcontinu}(a1) and \ref{suppfigcontinu}(a2), we test the algorithm of the continuity criterion with the given suitably small parameters $\eta=10^{-5}$ and $\mathfrak{o}=0.08$, which shows that the spectra (blue and magenta lines) utilizing the algorithm fit with the results through diagonalizing $\hat{\mathcal{H}}_{br}$ under OBC (gray dots) for $30$ sites perfectly. The related sub-GBZs are also shown in Figs. \ref{suppfigcontinu}(b1) and \ref{suppfigcontinu}(b2).  Nevertheless, the more sites we diagonalize the Hamiltonian with, the more deviations from the algorithm results emerge due to the numerical error, which needs more numerical precision but more time to erase. Our algorithm for the continuity criterion can effectively and unambiguously determine the multiple OBC bands and the associated sub-GBZs without encountering precision issues. This is achieved by relying on the analytical resultant equation and abiding by the continuity criterion incorporating with the GBZ condition, instead of the numerical diagonalization of the Hamiltonian matrix.

\begin{figure}
    \centering
    \includegraphics[width=0.35 \linewidth]{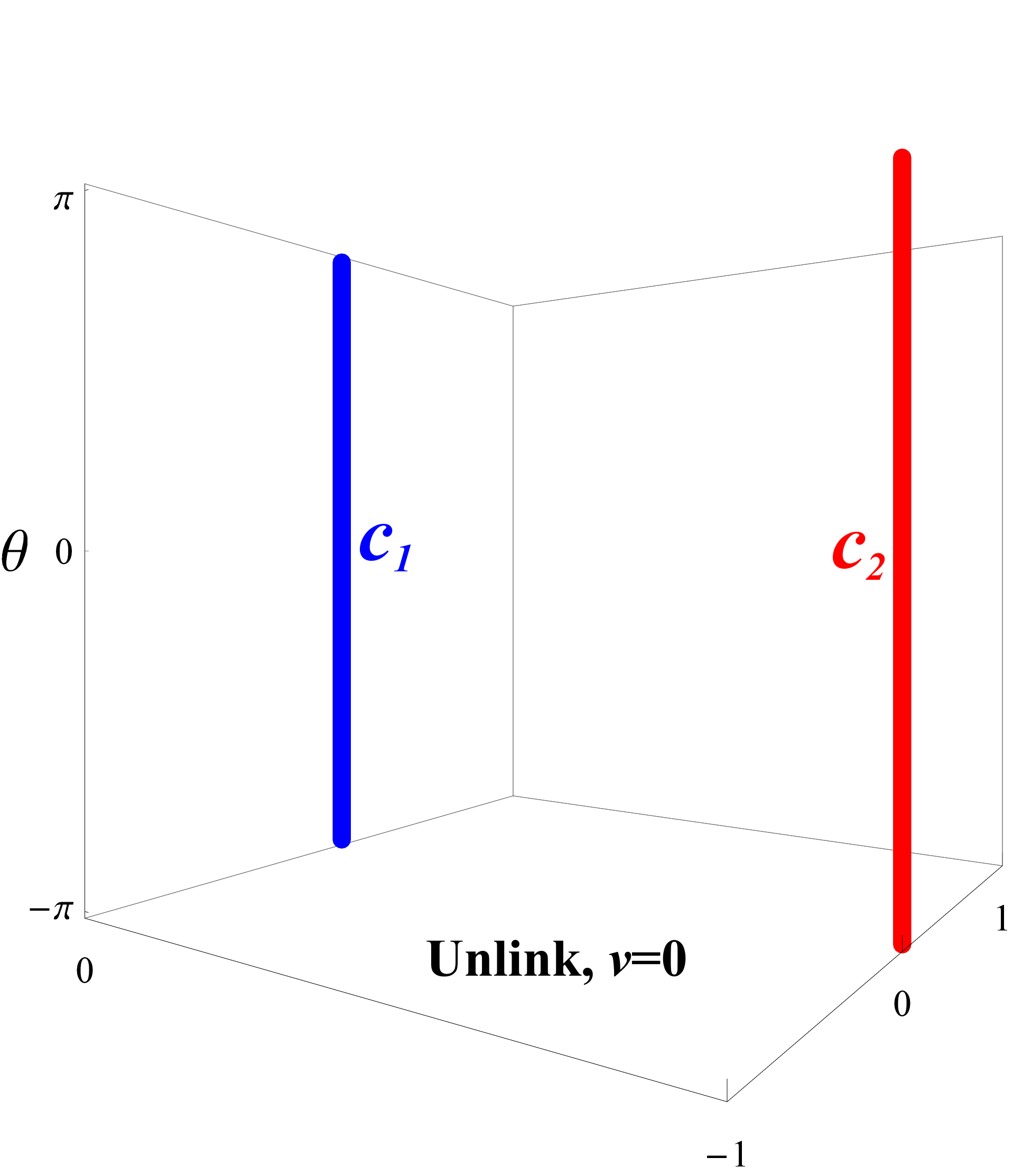}\qquad\qquad
    \includegraphics[width=0.35 \linewidth]{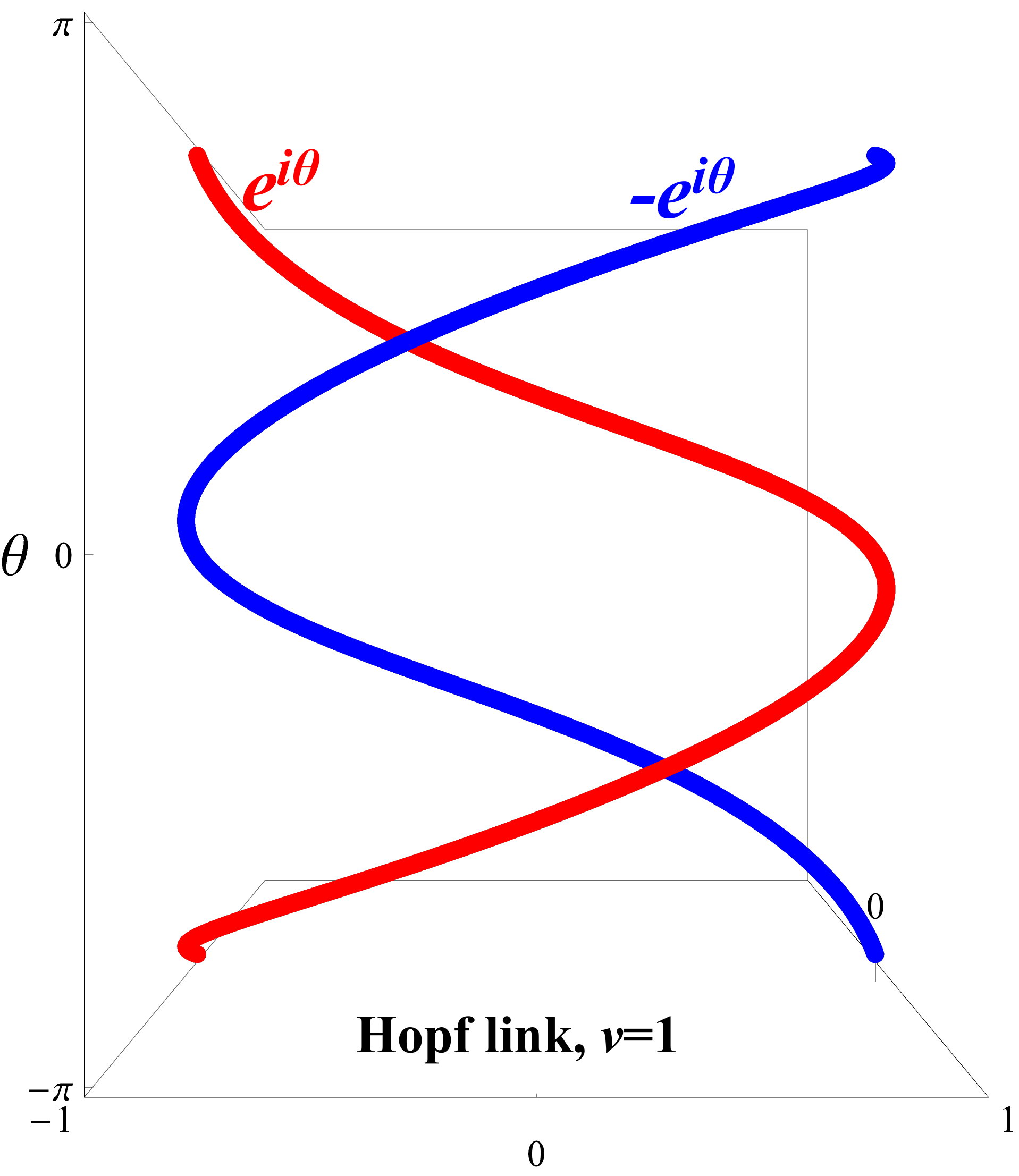}
    \includegraphics[width=0.35 \linewidth]{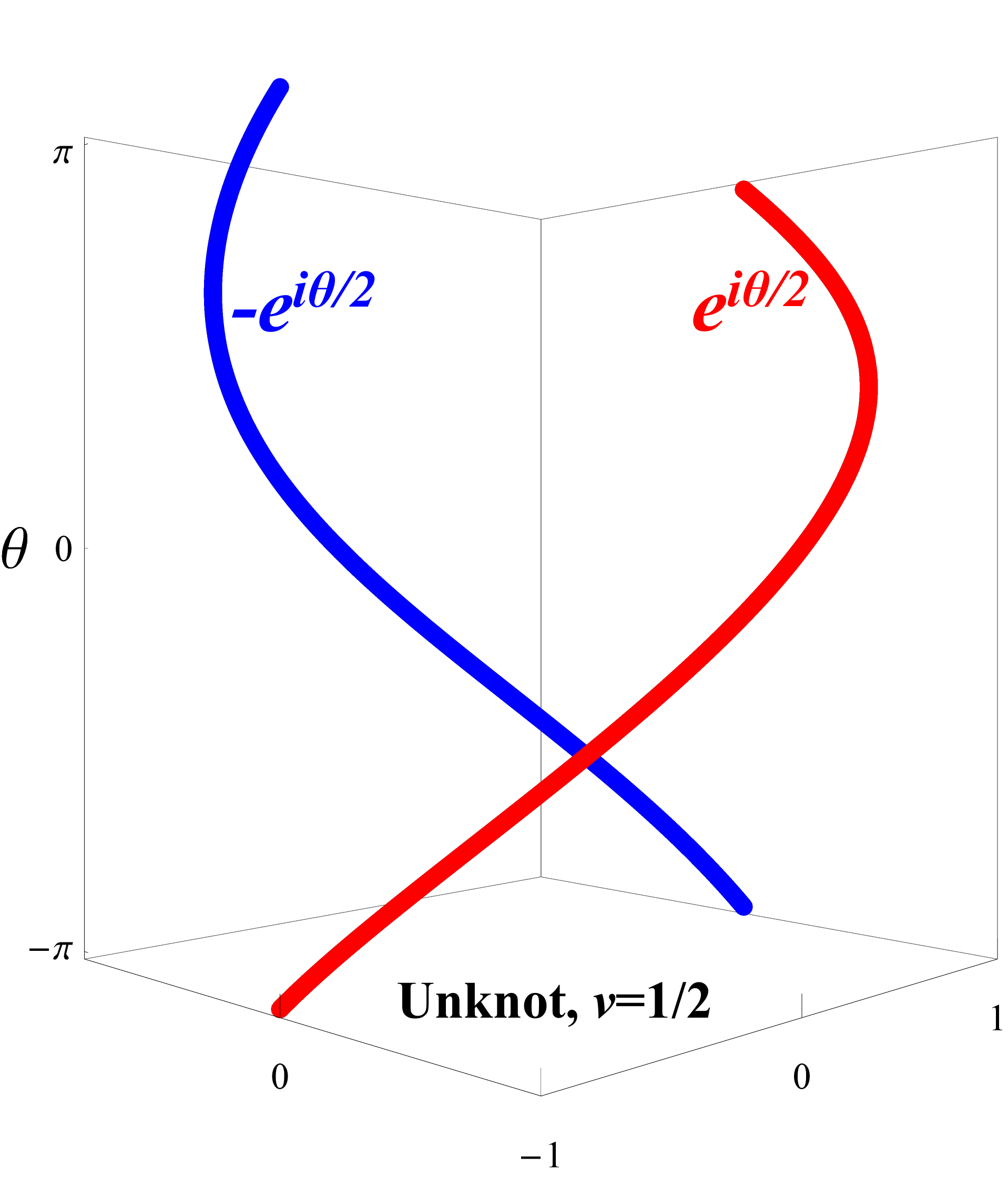}\qquad\qquad
    \includegraphics[width=0.35 \linewidth]{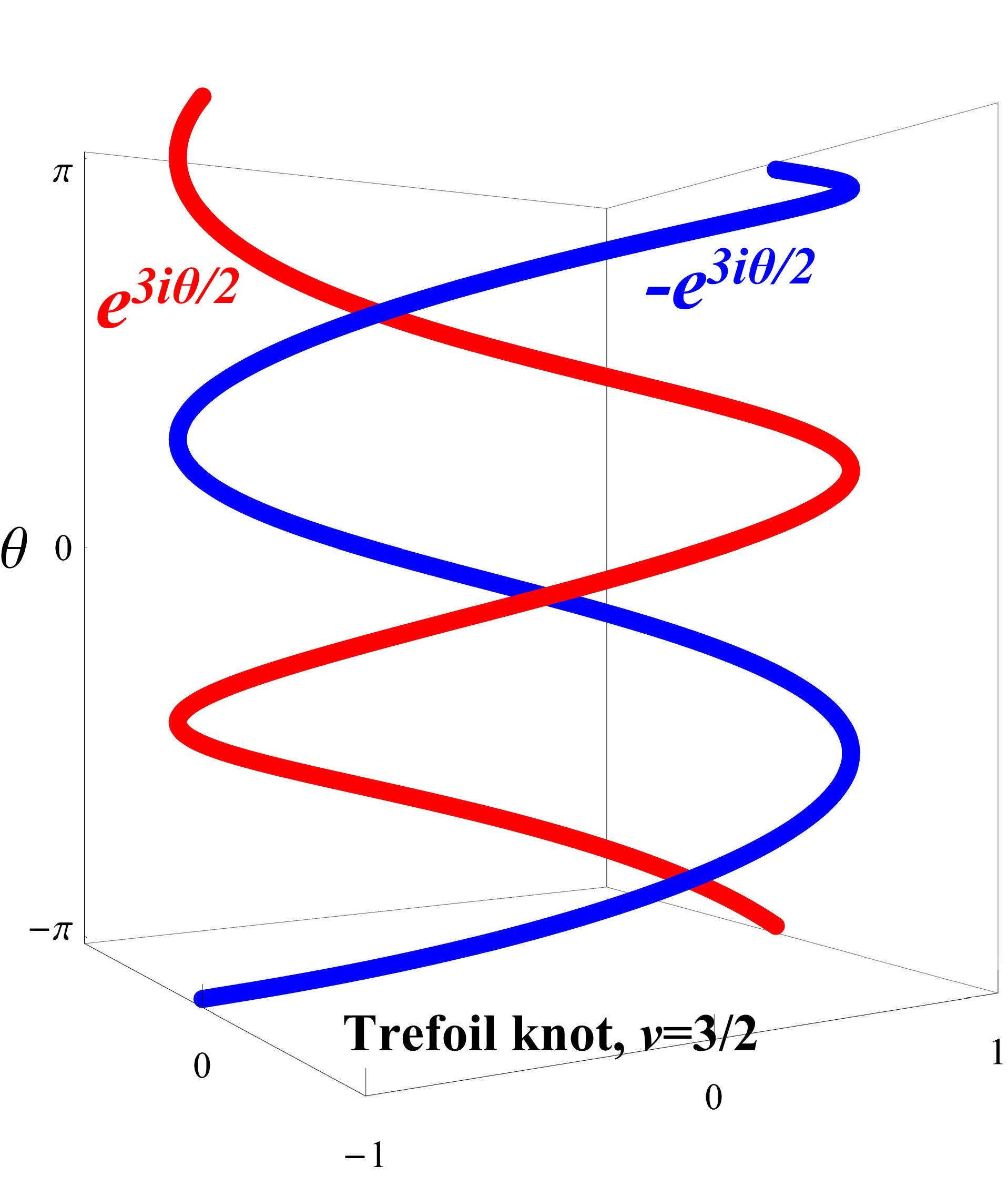}
    \caption{Illustration of the relation between the vorticity and braid crossings of two-bands, where $c_{1},c_{2}$ are constants.}
    \label{suppfigvor}
\end{figure}

\section{S2. Vorticity and braid crossings}
The vorticity charactering the braiding topology of open-boundary bands is defined in the main text,
\begin{align}
    \label{suppeqvortotal}
    \nu=\frac{1}{2}\sum_{i\neq j}\nu_{ij}=\frac{1}{2}\sum_{i\neq j}\frac{1}{2\pi}\oint_{S^{1}}\frac{d}{d\theta}\mathrm{arg}\left[\varepsilon_{i}(\theta)-\varepsilon_{j}(\theta)\right]d\theta.
\end{align}
As illustrated in Fig. \ref{suppfigvor}, we show the structures of unlink, Hopf link, unknot, and trefoil knot as examples of two-band braiding, concerning $\nu=0,1,1/2$, and $3/2$, respectively. From a specific perspective in three-dimensional (3D) space [e.g., the perspective in Fig. \ref{suppfigvor}], or by projecting the 3D braid of energy bands (strings) onto a designated two-dimensional (2D) plane, we can label instances where a left band crosses a right band from above or below as over or under crossings, respectively, as $\theta$ evolves along $[-\pi, \pi]$. The results from different perspectives (projection planes) are topologically equivalent up to the Reidemeister moves \cite{kassel2008braid}. By defining $W=2\nu$, an over/under crossing of the braid contributes to $W$ ($\nu$) a $\pm 1$ ($\pm 1/2$) as exhibited in Fig. \ref{suppfigvor}, where the band's expressions are specified. Noteworthily, the invariant $W$ is equivalent to the discriminant invariant in Ref. \cite{yang2021doubling,hu2022knot} and the non-Bloch winding number in Ref. \cite{li2023nonblochbraiding} mathematically. The braiding structures in Fig. \ref{suppfigvor} can topologically deform to various irregular braidings corresponding to open-boundary bands as long as the bands do not touch each other, such as the bands given in the main text. We emphasize that the unlink and Hopf link correspond to the bands with closed sub-GBZs, while the unknot and trefoil knot are related to the bands with non-closed sub-GBZs, as shown in the examples in the main text and below. The relation between the vorticity and braid crossings can be directly verified to more than two bands, that is, $\nu=(n_{+}-n_{-})/2$, where $n_{\pm}$ counts the total number of over/under crossings.

\section{S3. Stability of degenerate points}
In this section, we briefly review the stability argument of degenerate points (DPs) \cite{hu2021knots,shen2018,heiss_2012,lee2016anom,yang2021doubling} to support the mechanism of the braiding phase transition in the main text. 

The DPs, the band touching points, serve as the critical points of band crossing. The most common DPs occurrence is two-fold degeneracy, which are stable DPs since higher-fold degenerate DPs need more constraints and can split into several two-fold degenerate DPs in the presence of generic perturbations. Hence, it is enough to focus on the two-fold degenerate DPs to explore the braiding phase transition of non-Hermitian open-boundary bands \cite{yang2021doubling}. 

Consider a generic two-band Hamiltonian in the 2D parameter space, e.g., $\gamma_{1},\gamma_{2}$,
\begin{align}
    \label{suppeq2dham}
    H(\gamma_{1},\gamma_{2})=h_{0}(\gamma_{1},\gamma_{2})\sigma_{0}+h_{x}(\gamma_{1},\gamma_{2})\sigma_{x}+h_{y}(\gamma_{1},\gamma_{2})\sigma_{y}+h_{z}(\gamma_{1},\gamma_{2})\sigma_{z},
\end{align}
where $\sigma_{0}$ is 2D identity matrix, $\sigma_{\alpha},\alpha=x,y,z$ are Pauli matrices, and $h_{i},i=0,x,y,z$ are complex functions. The two bands of $H$ are 
\begin{align}
    \label{suppeq2dband}
    E_{\pm}=h_{0}\pm\sqrt{h_{x}^{2}+h_{y}^{2}+h_{z}^{2}}.
\end{align}
Without loss of generality, we can set $h_{0}=0$, which just shifts the energy level. At DPs, the condition $h_{x}^{2}+h_{y}^{2}+h_{z}^{2}=0$ gives the exceptional points (EPs), where the Hamiltonian is defective, i.e., only one eigenvector existing; in contrast, the condition $h_{x}=h_{y}=h_{z}=0$ gives the non-defective points (NDPs). The EP condition includes two equations, and the NDP condition contains six conditions. In the 2D space, only the EPs are stable, while the NDPs are unstable, requiring fine-tuning \cite{shen2018,hu2021knots}.  

The characteristic invariant of EPs is the vorticity around a small contour $\Gamma$ enclosing EPs, that is Eq. (\ref{suppeqvortotal}) with replacing $S^{1}$ to $\Gamma$ for two-bands. A statement given in Ref. \cite{yang2021doubling} is that only the EPs with $\left|\nu\right|=1/2$ are stable, while other EPs with $\left|\nu\right|>1/2$ can split into several EPs with $\left|\nu\right|=1/2$ by generic perturbations. Furthermore, the two-fold degenerate EP with $\left|\nu\right|=1/2$ always terminates a branch cut in the energy spectrum.

\section{S4. Examples of open-boundary band braiding}
We investigate the band braiding for the model $\mathcal{H}_{br}(\beta)$ with $\lambda_{1}=\lambda_{2}=0$ in the main text. In this section, we show some additional examples with nonzero $\lambda_{1},\lambda_{2}$. 

\begin{figure}
    \centering
    \includegraphics[width=0.3 \linewidth]{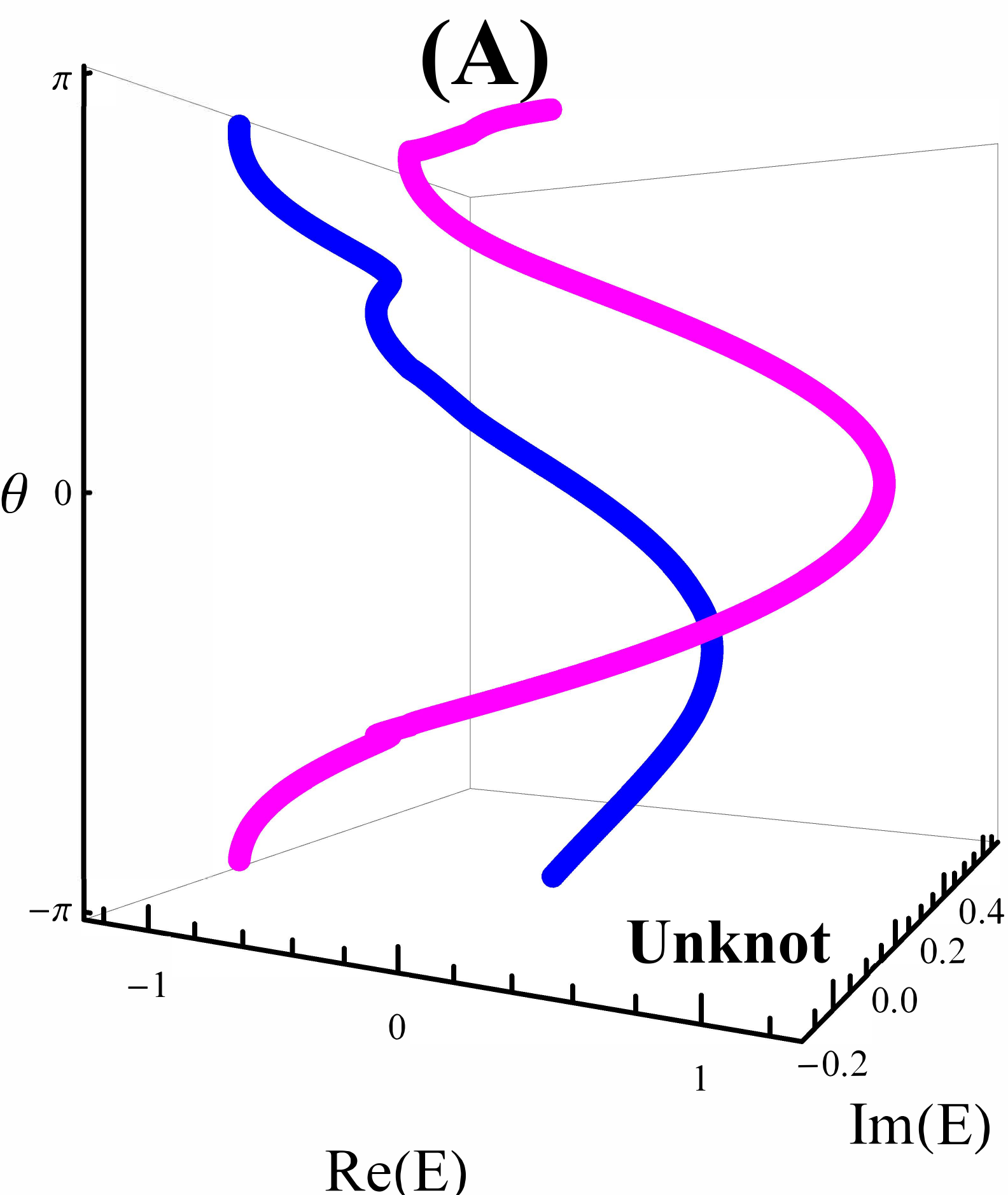}
    \includegraphics[width=0.3 \linewidth]{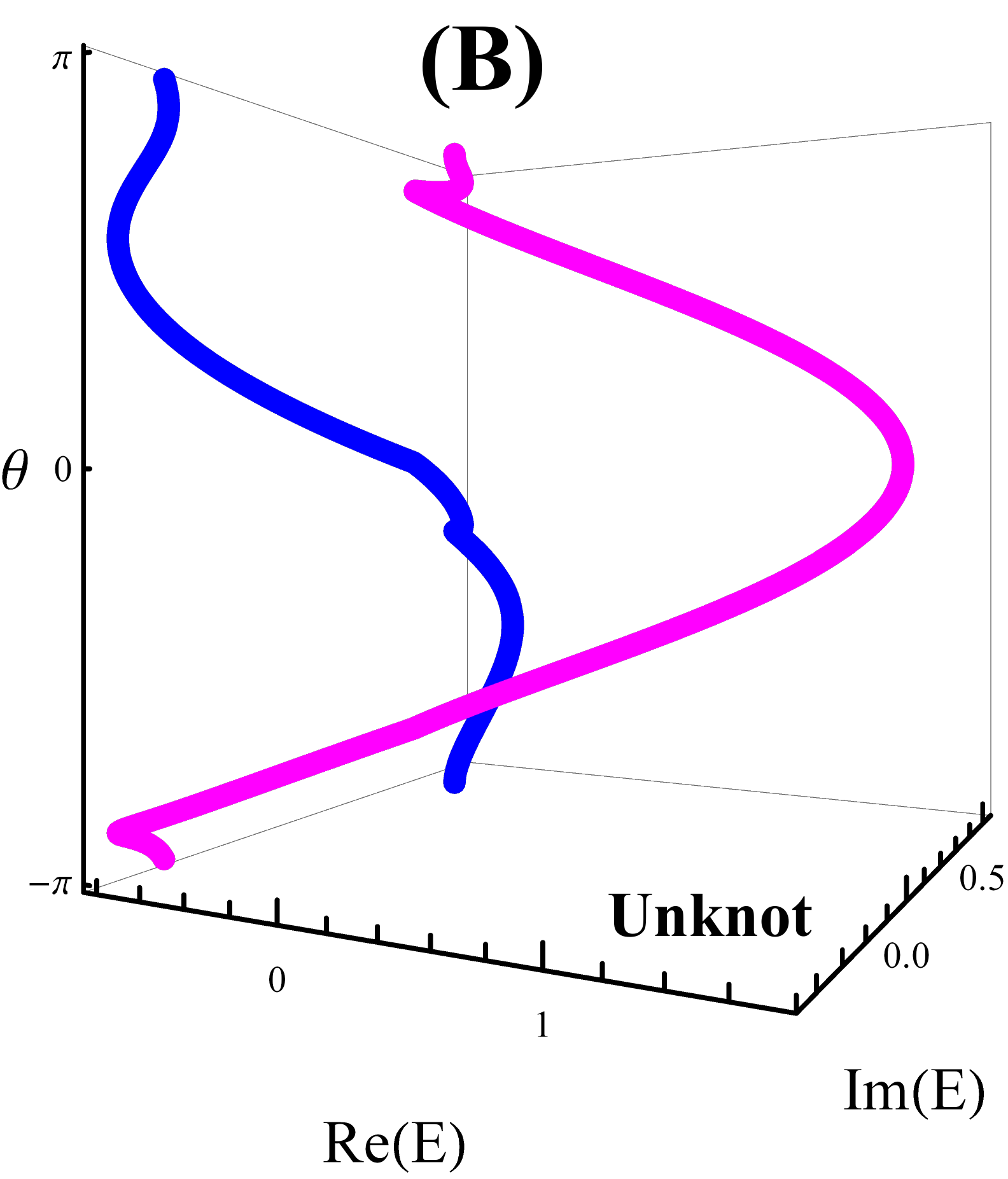}
    \includegraphics[width=0.3 \linewidth]{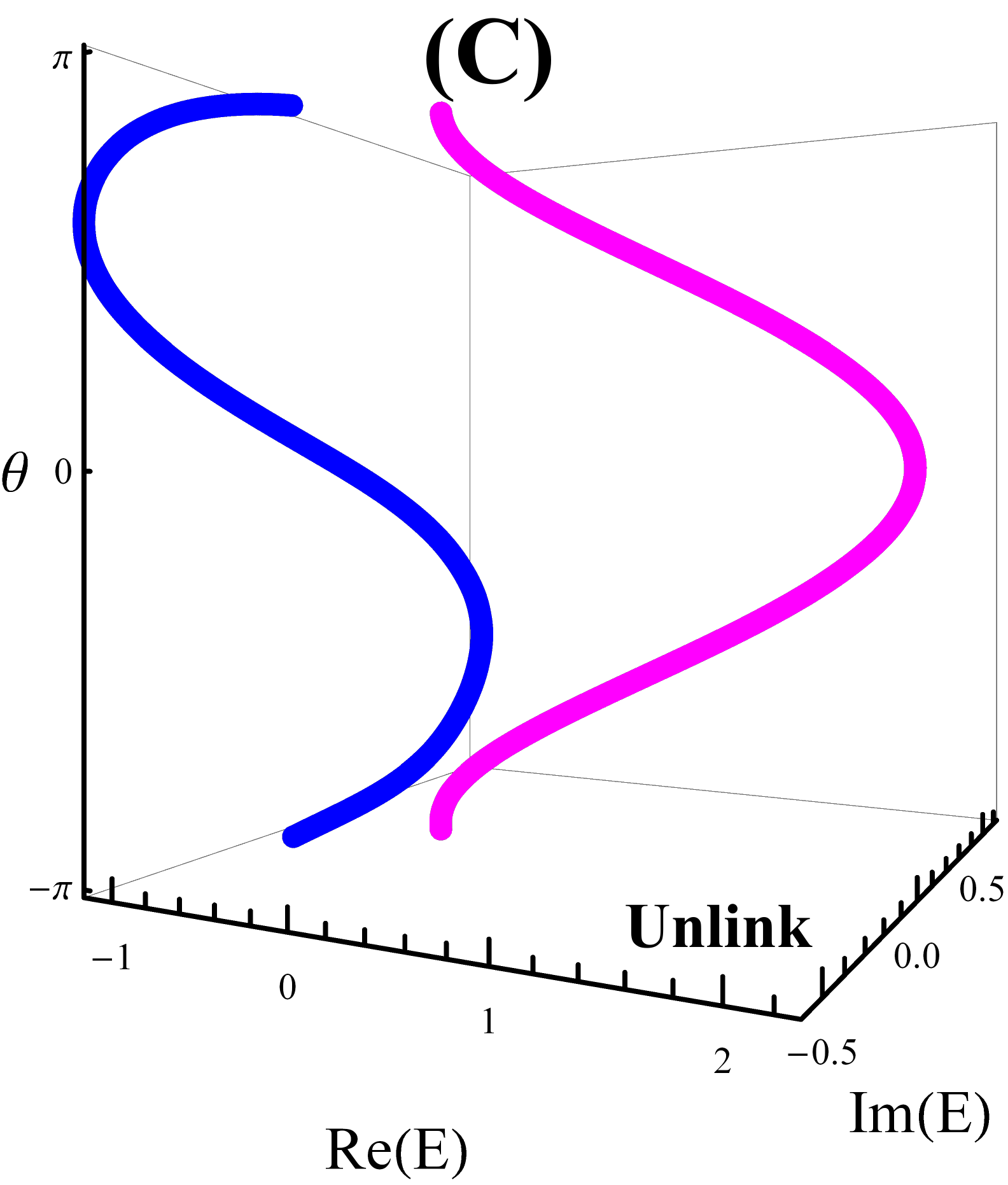}\\
    \includegraphics[width=0.33 \linewidth]{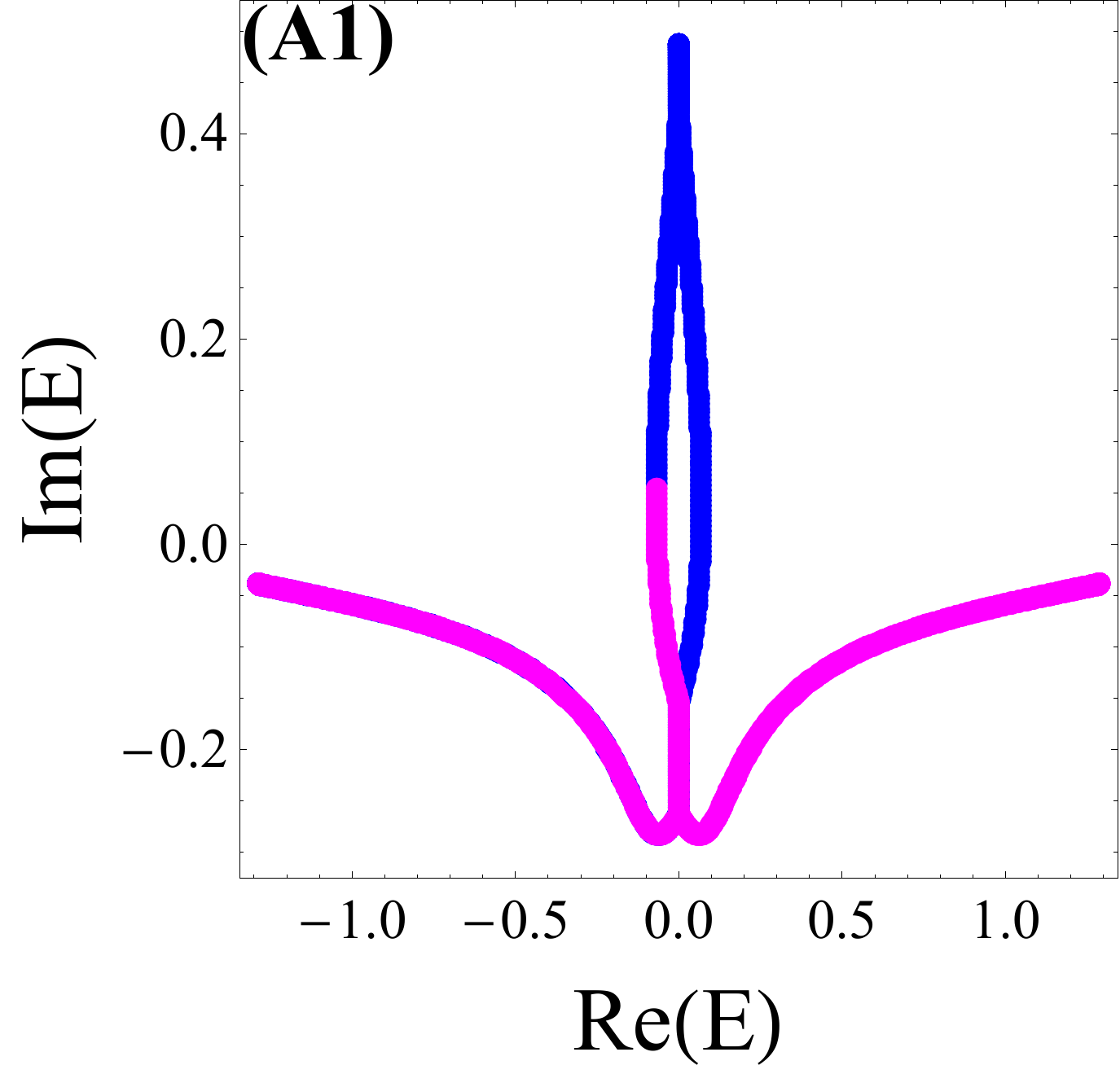}
    \includegraphics[width=0.3 \linewidth]{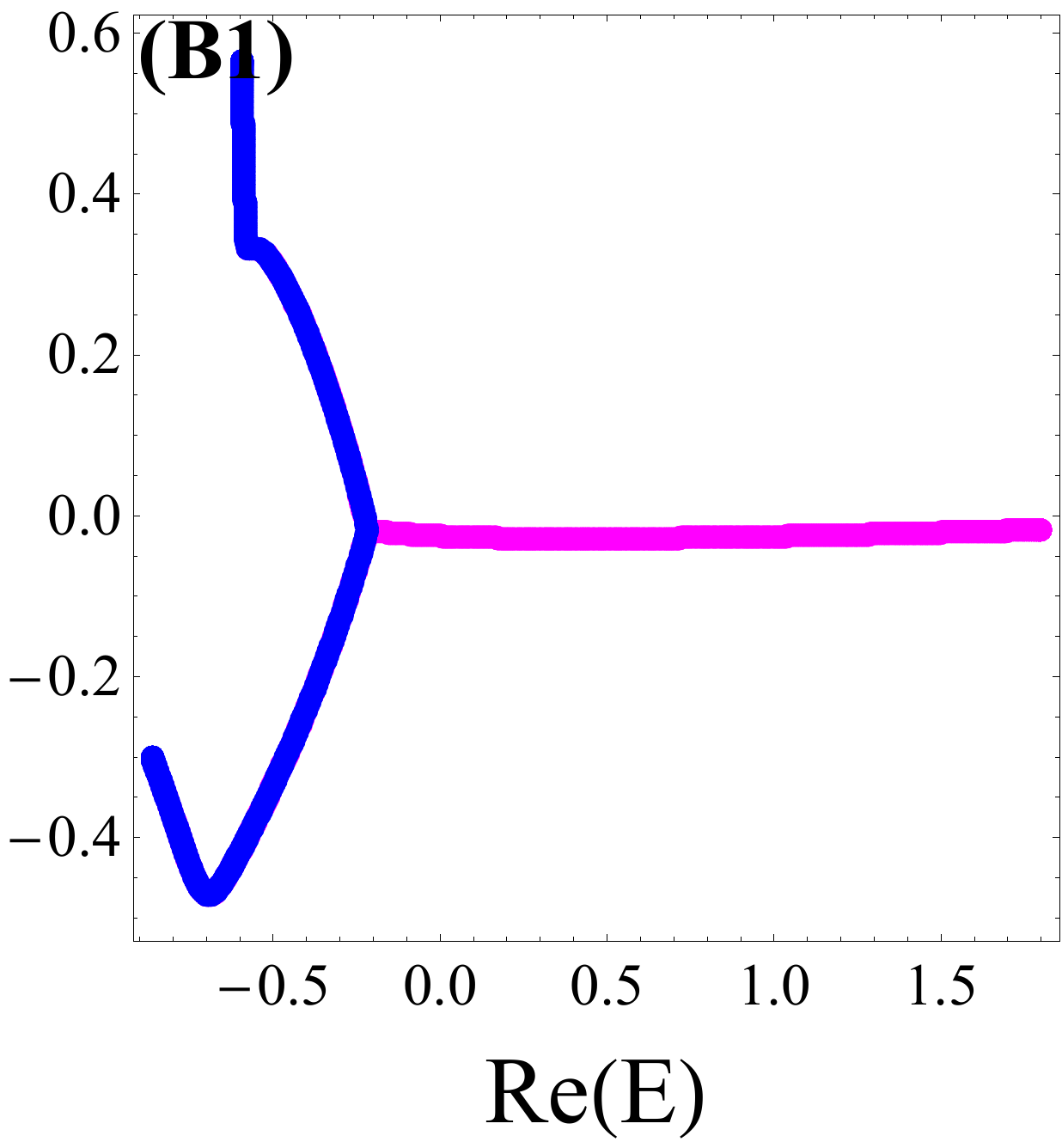}
    \includegraphics[width=0.3 \linewidth]{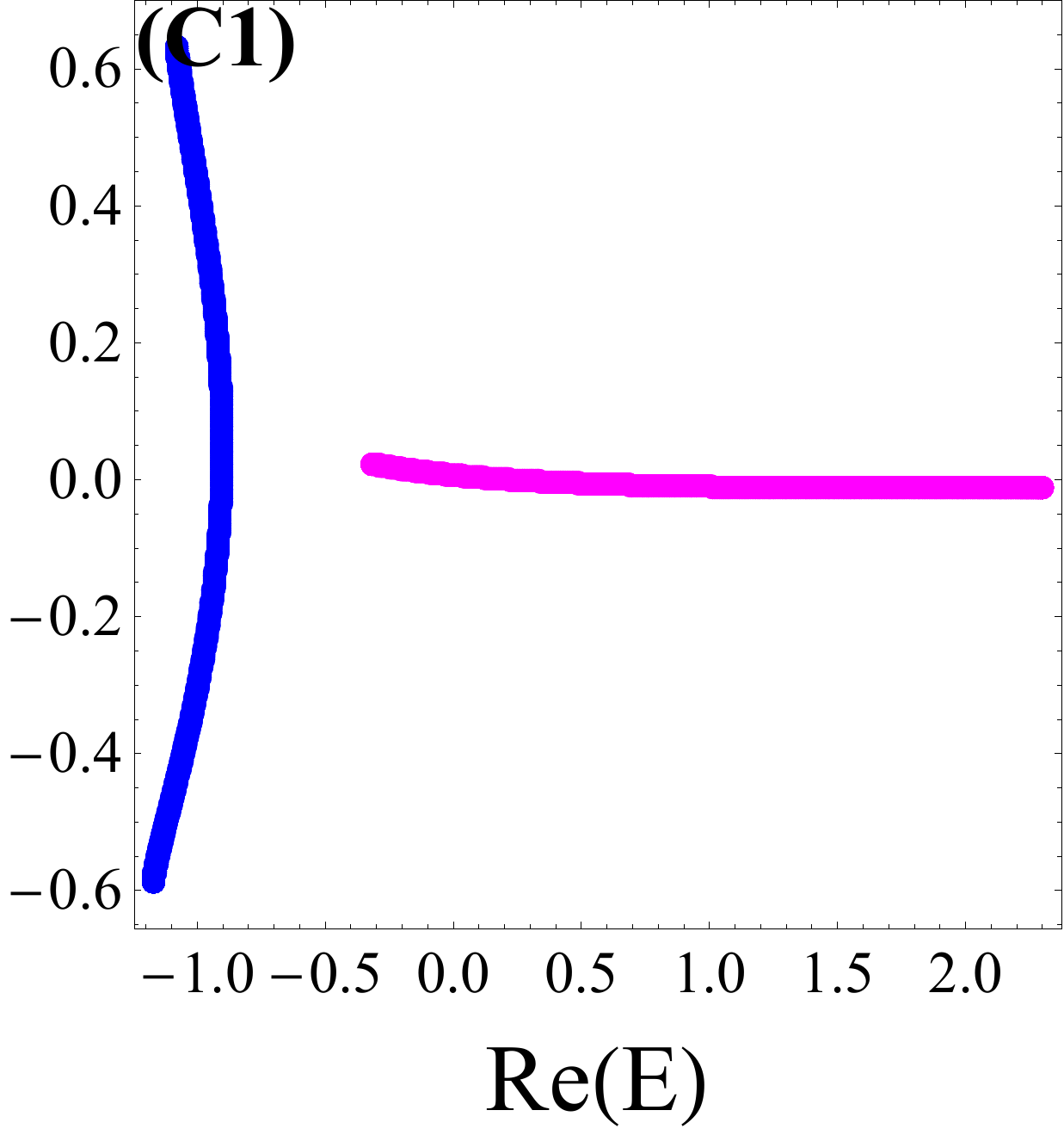}\\
    \includegraphics[width=0.33 \linewidth]{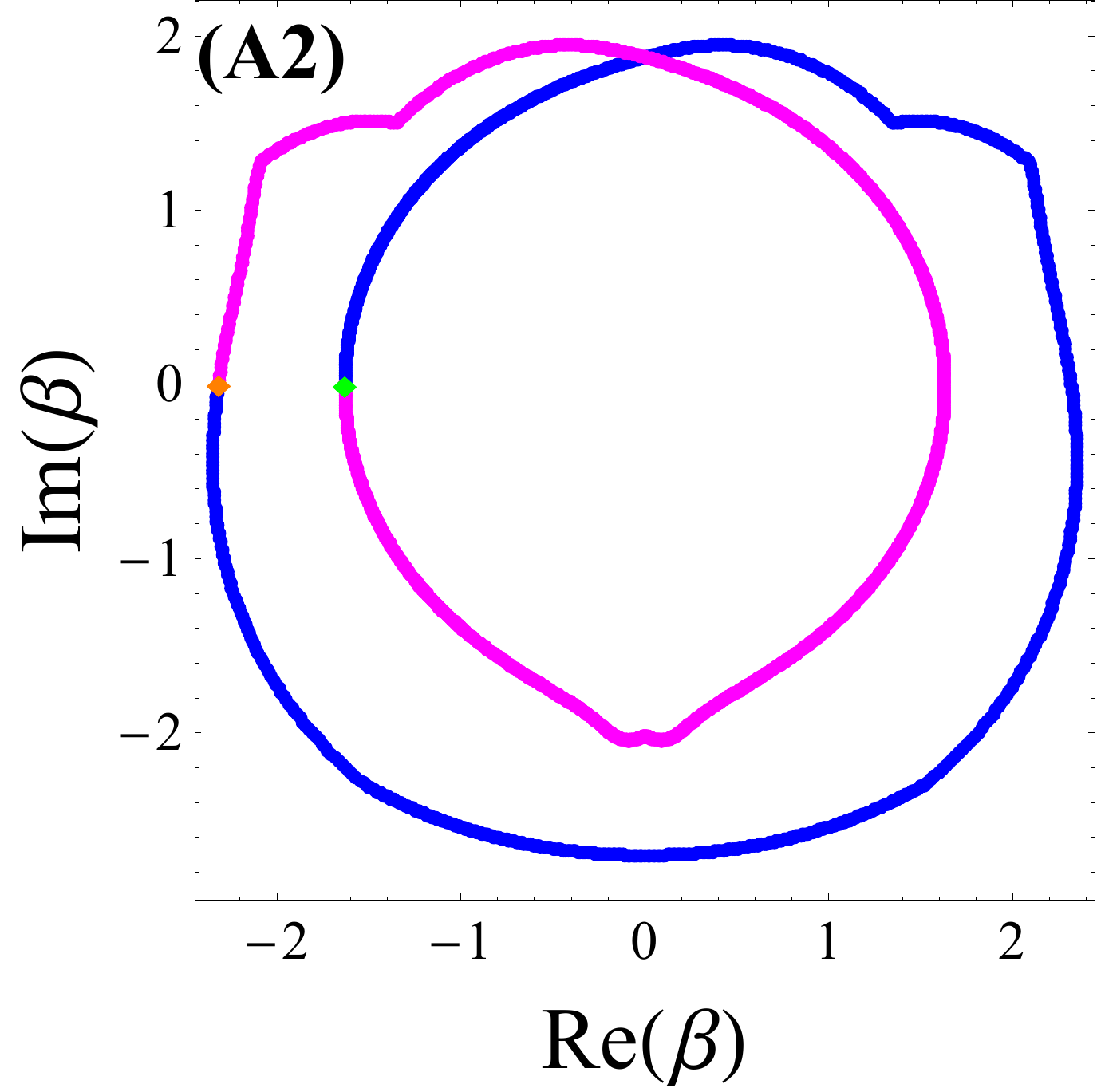}
    \includegraphics[width=0.3 \linewidth]{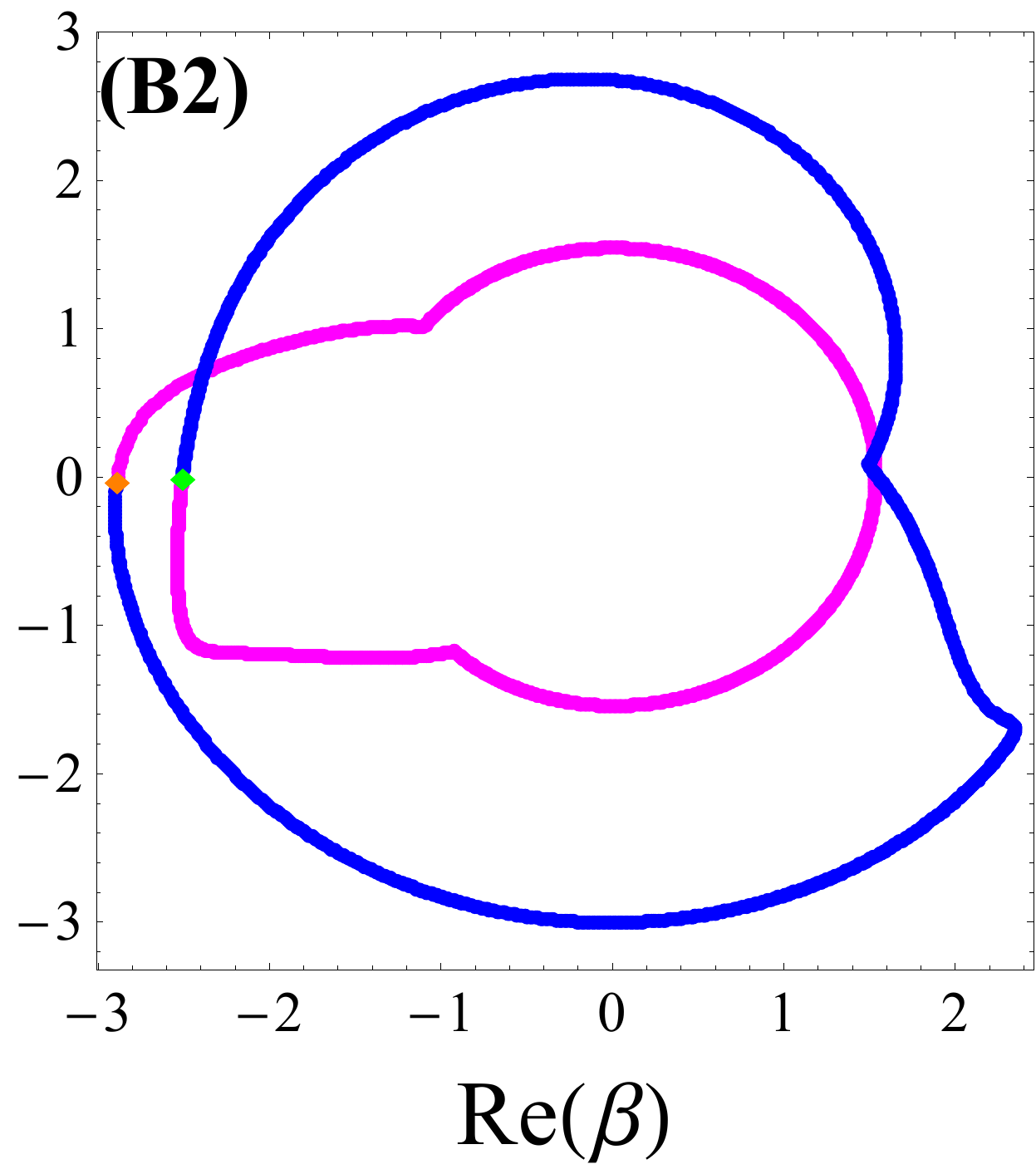}
    \includegraphics[width=0.3 \linewidth]{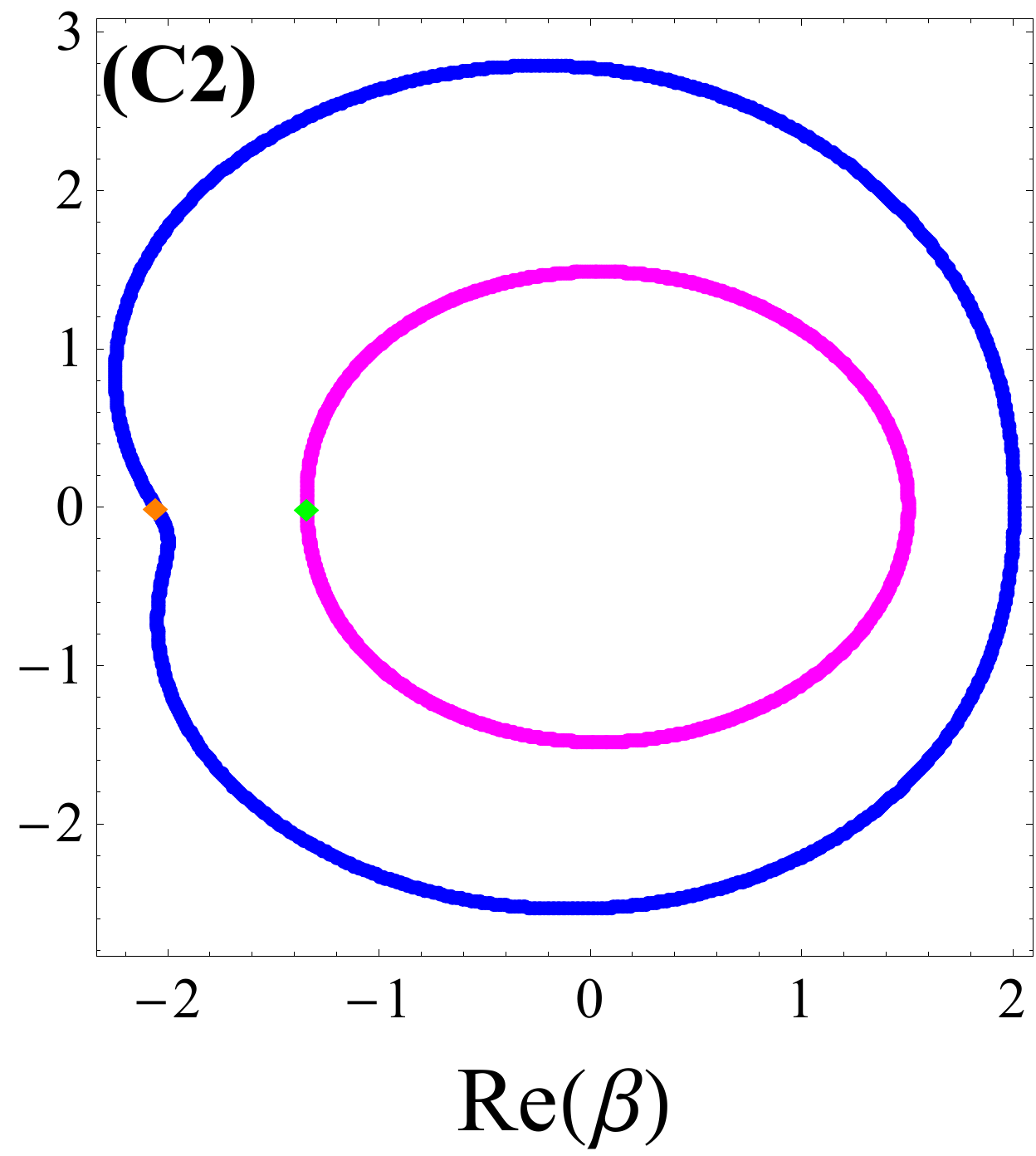}
    \caption{The model $\mathcal{H}_{br}(\beta)$ with the parameters $t_{p}=0.15,t_{m}=0.85,t_{a}=0.2,\lambda_{1}=0.3,\lambda_{2}=0.1,m=0.1$ in the main text. (A)-(C) The band braiding corresponding to $\lambda_{3}=0,0.5$, and $1$, projected into the complex energy plane shown in (A1)-(C1), respectively. (A2)-(C2) The sub-GBZs with respect to (A)-(C), where the orange and green dots label the starting points of blue and magenta sub-GBZs, respectively . The sub-GBZs in (A2) and (B2) are intertwined, leading to the emergence of the unknot braiding in (A) and (B).}
    \label{suppfigmainmodela}
\end{figure}

In Fig. \ref{suppfigmainmodela}, we show the open-boundary bands with intertwined (non-closed) sub-GBZs [Figs. \ref{suppfigmainmodela}(A2) and \ref{suppfigmainmodela}(B2)], which exhibit the starting point of one sub-GBZ (band) being the endpoint of another sub-GBZ (band) and vice versa, thus leading to the unknot braiding [Figs. \ref{suppfigmainmodela}(A) and \ref{suppfigmainmodela}(B)]. Note that the map from the argument $\theta$ of each sub-GBZ (instead of itself in the main text) to $S^{1}$ is homeomorphic, thus the homotopic characteristic of open-boundary bands $\varepsilon_{i}(\theta)$ is still valid. The unknot does not belong to the pure braid group $PB_{n}$ rather than the generic braid group $B_{n}$. The non-closed GBZs evolve into closed ones [Fig. \ref{suppfigmainmodela}(C2)] passing through the critical point of braiding topology, which must emerge as a stable EP in contrast to the critical point with two EPs between Hopf link and unlink phases in the main text, leading to the unlink braiding in Fig. \ref{suppfigmainmodela}(C). The bands corresponding to the unknot (unlink) phase are separated (isolated) [Figs. \ref{suppfigmainmodela}(A1)-\ref{suppfigmainmodela}(C1)]. 

\begin{figure}
    \centering
    \includegraphics[width=0.3 \linewidth]{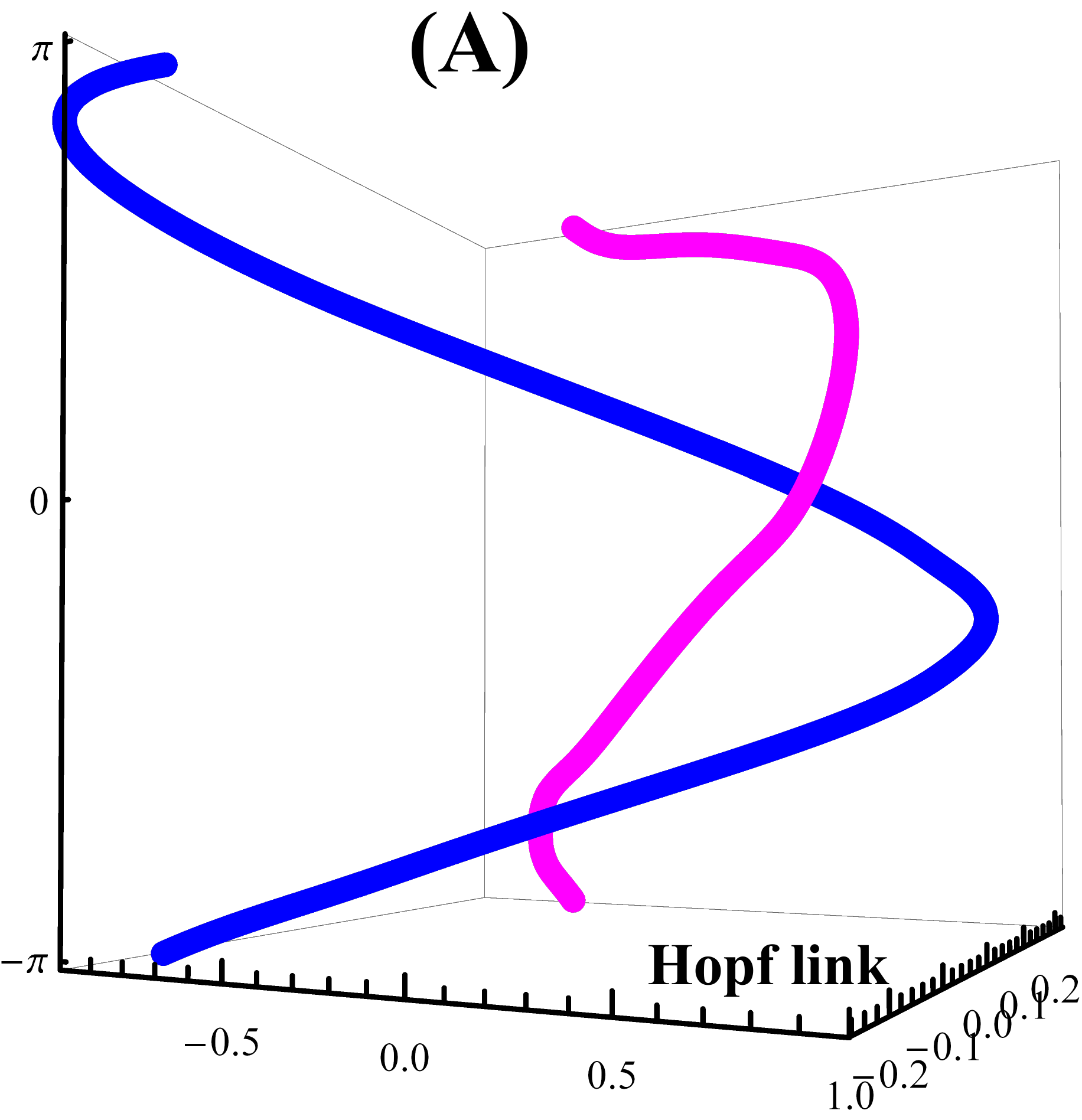}
    \includegraphics[width=0.3 \linewidth]{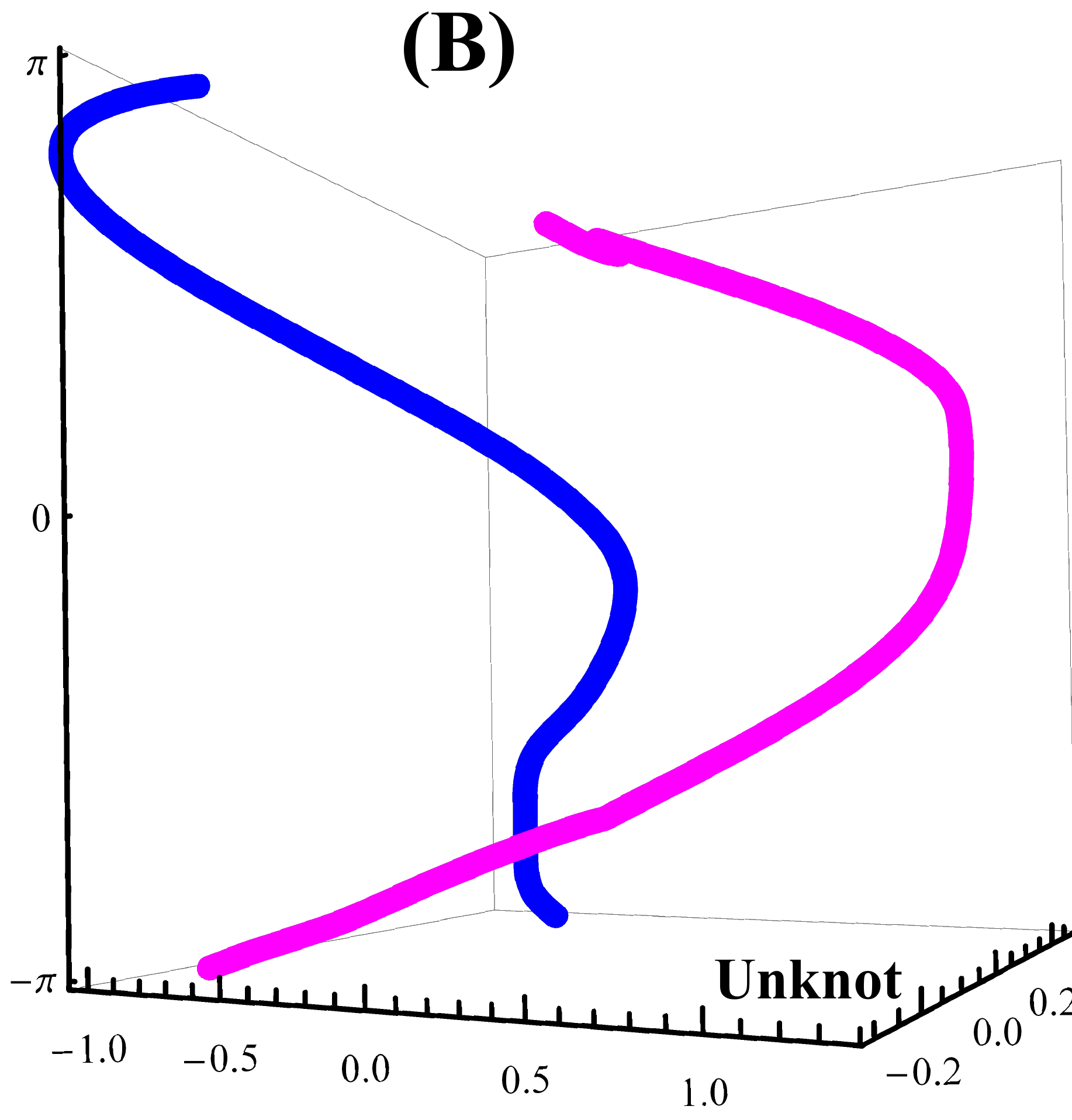}
    \includegraphics[width=0.3 \linewidth]{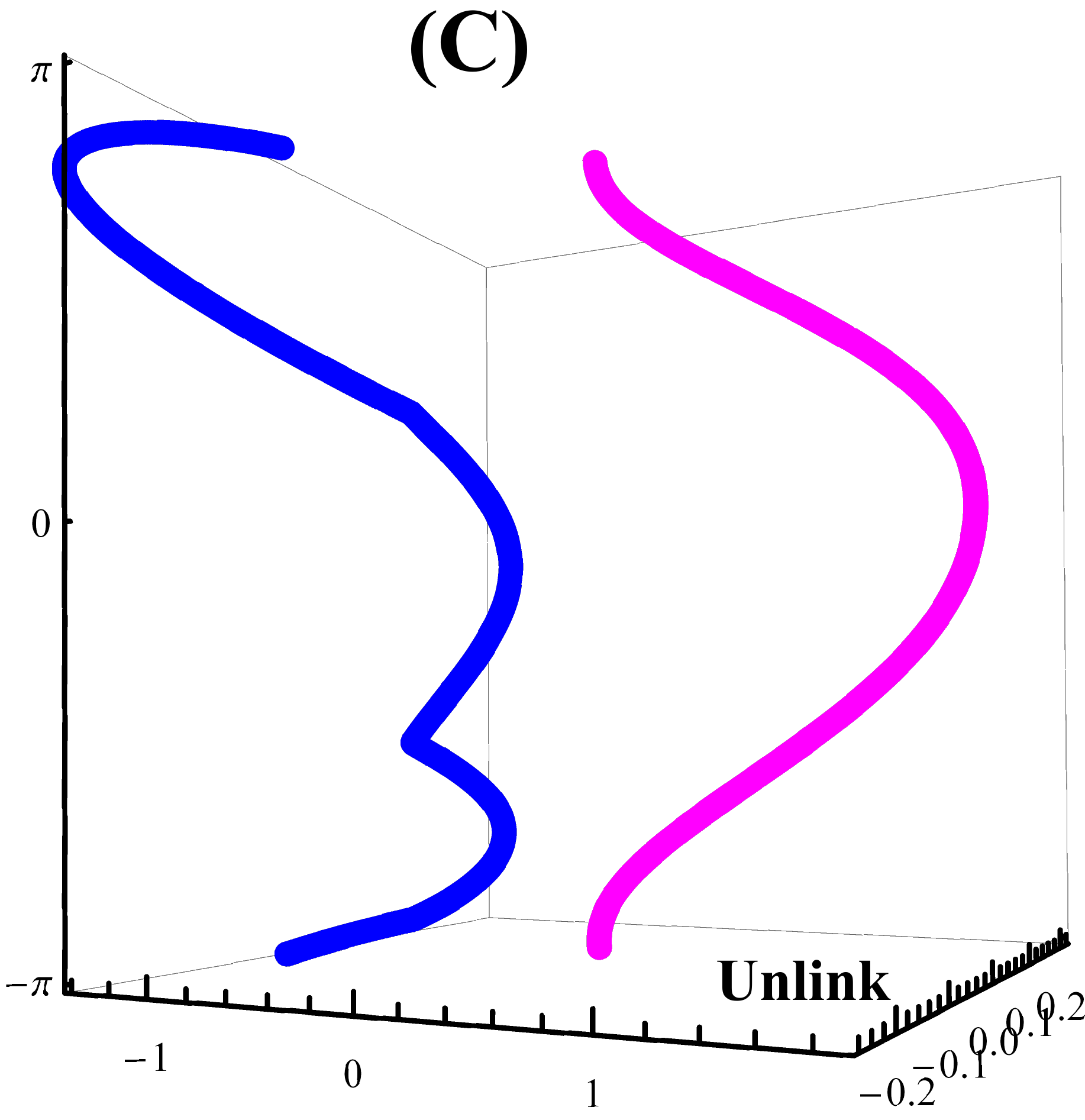}\\
    \includegraphics[width=0.33 \linewidth]{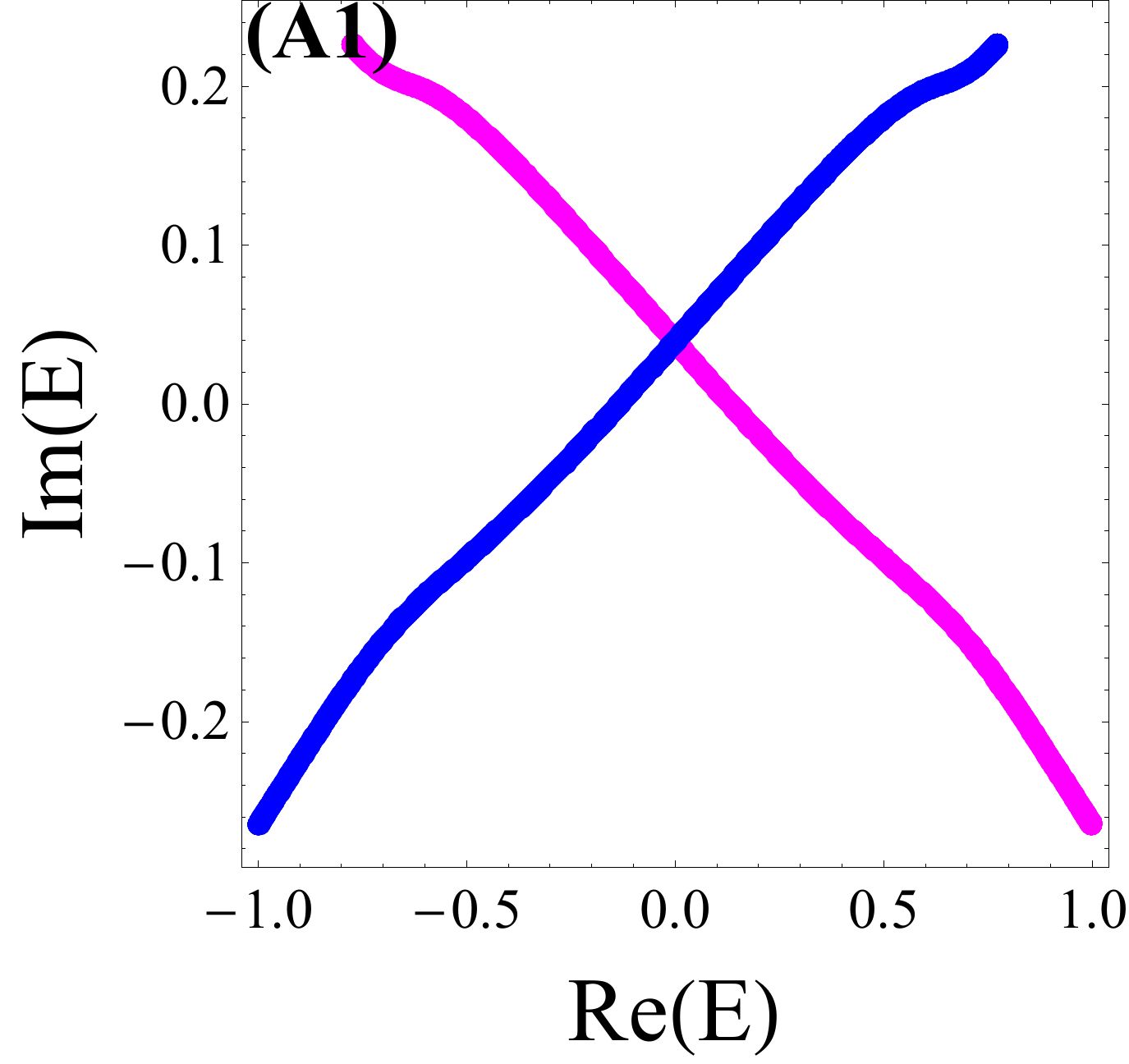}
    \includegraphics[width=0.29 \linewidth]{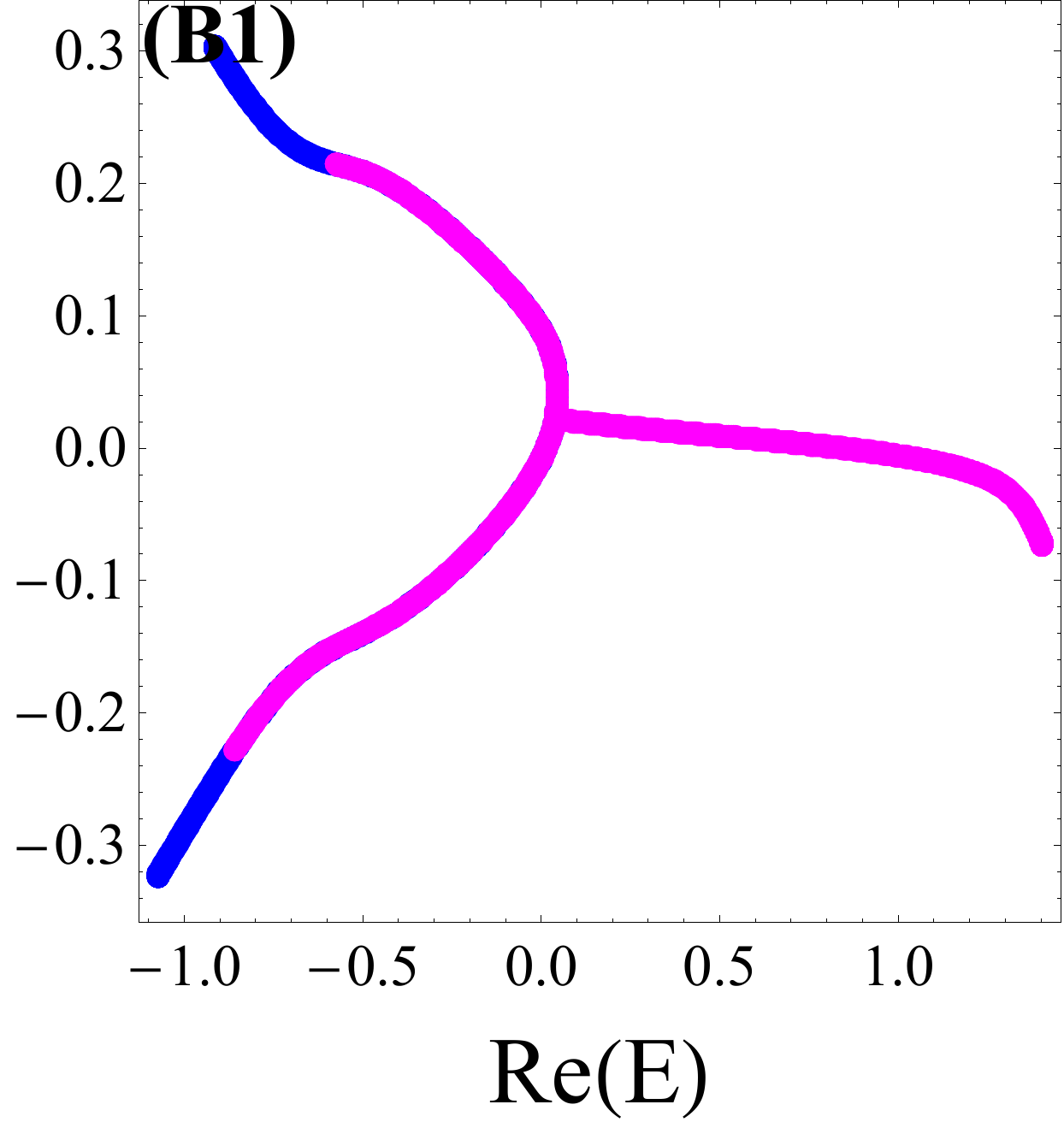}
    \includegraphics[width=0.3 \linewidth]{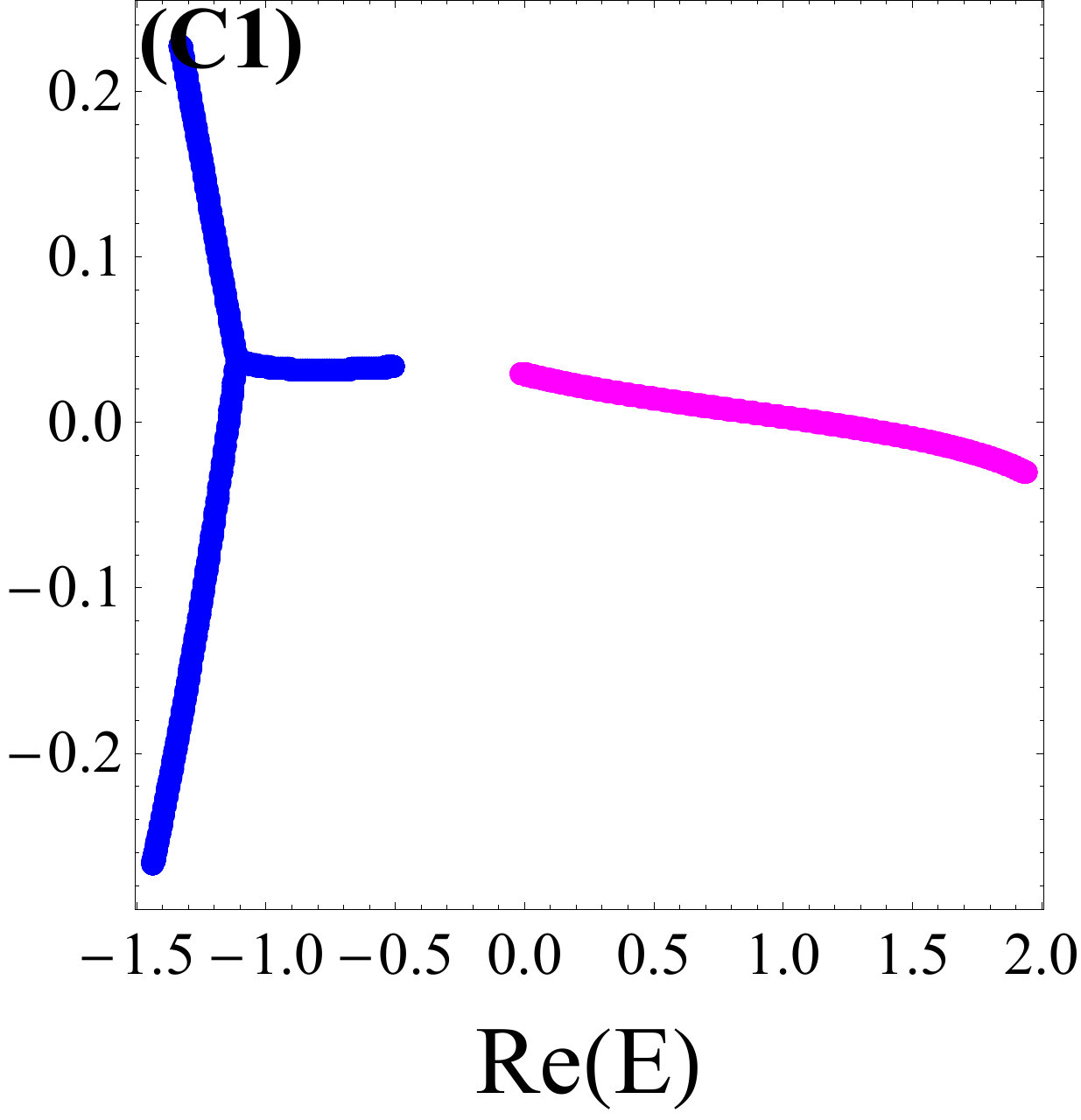}\\
    \includegraphics[width=0.332 \linewidth]{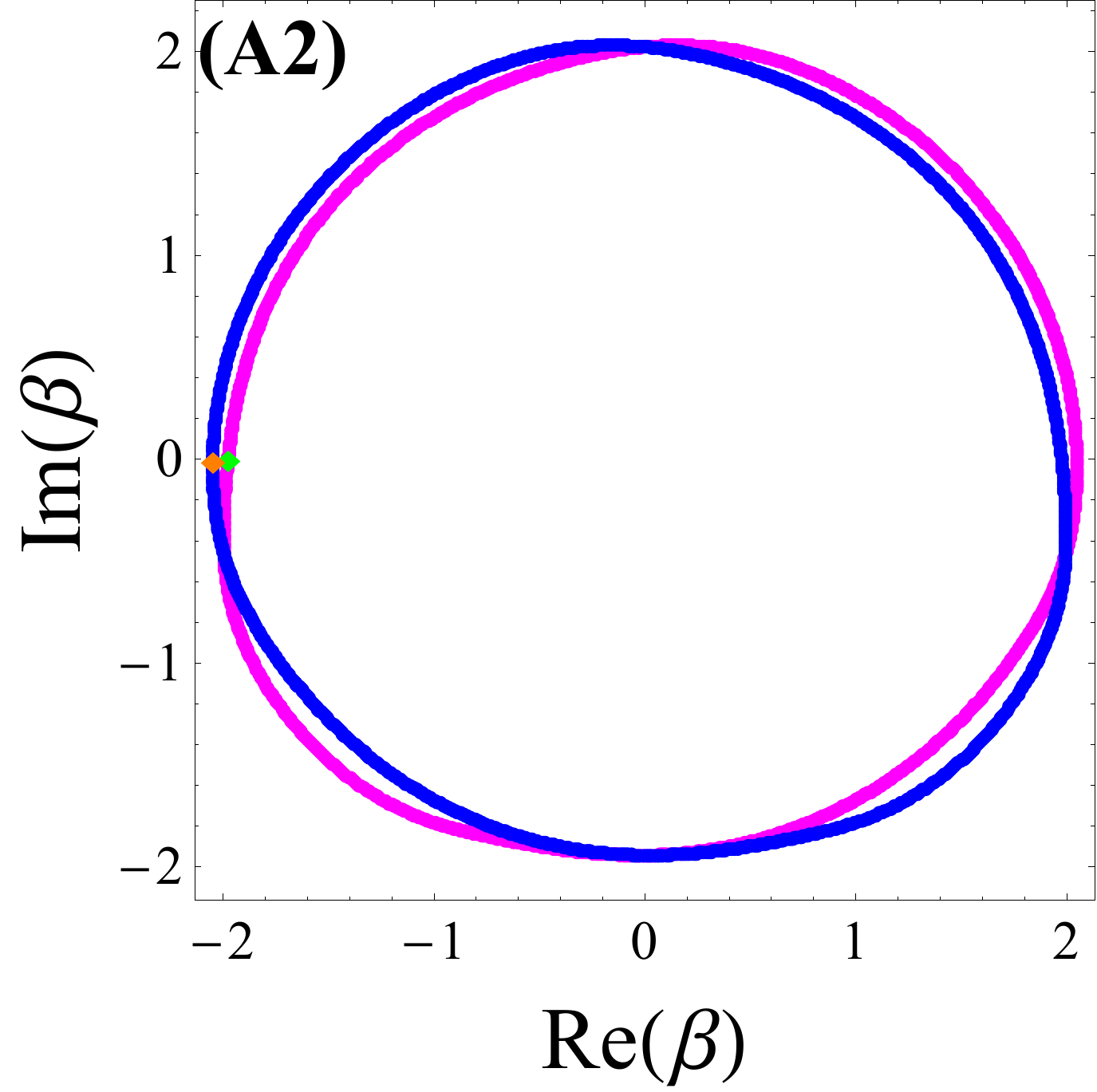}
    \includegraphics[width=0.297 \linewidth]{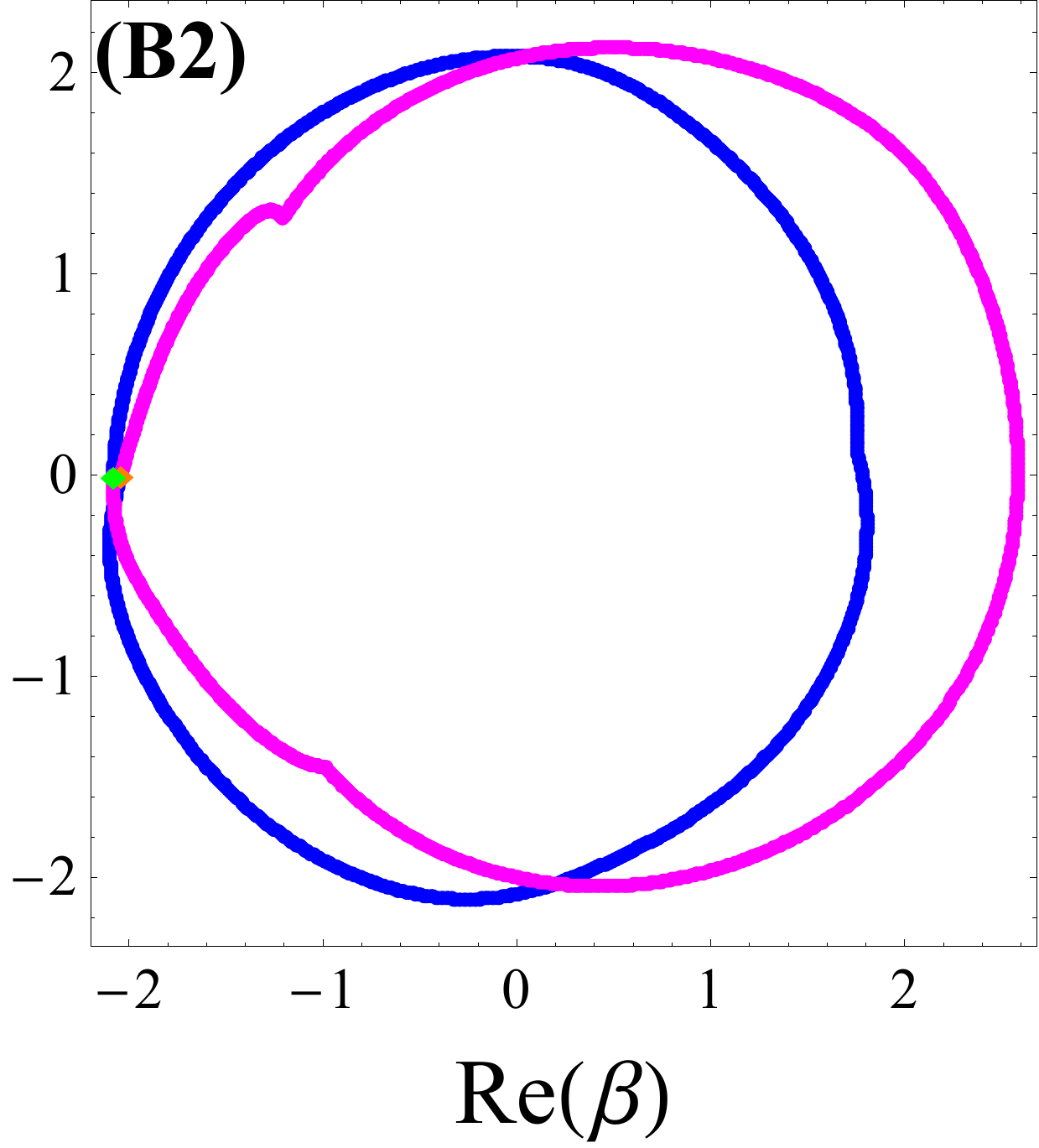}
    \includegraphics[width=0.295 \linewidth]{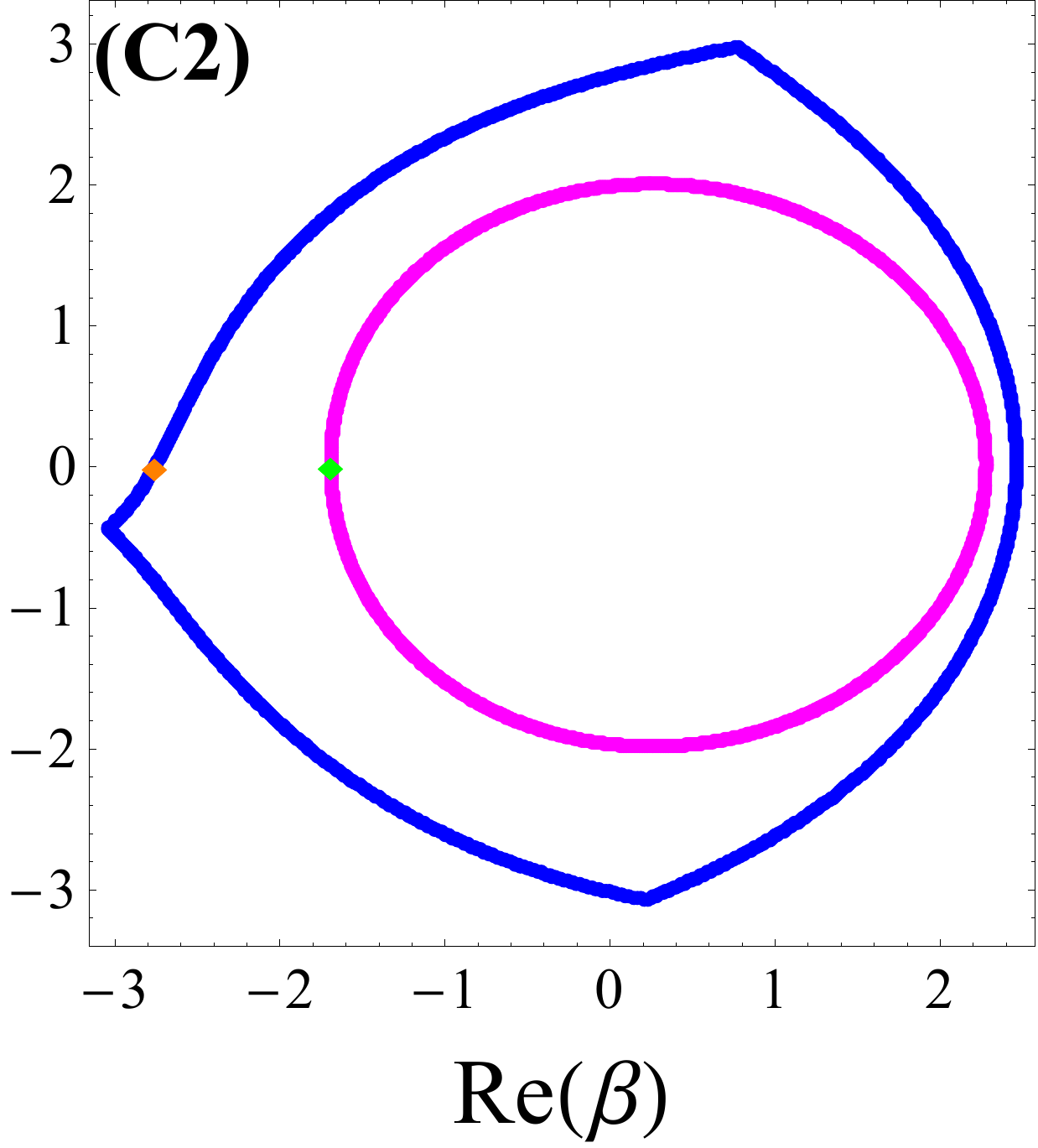}
    \caption{The model $\mathcal{H}_{br}(\beta)$ with the parameters $t_{p}=0.15,t_{m}=0.85,t_{a}=0.2,\lambda_{1}=\lambda_{2}=0.1,m=0.1$ in the main text. (A)-(C) The band braiding corresponding to $\lambda_{3}=0,0.5,$ and $1$, projected into the complex energy plane shown in (A1)-(C1), respectively. (A2)-(C2) The sub-GBZs with respect to (A)-(C), where the orange and green dots label the starting points of blue and magenta sub-GBZs, respectively. The sub-GBZs in (B2) are intertwined, leading to the emergence of the unknot braiding in (B).}
    \label{suppfigmainmodelb}
\end{figure}

In Fig. \ref{suppfigmainmodelb}, we show the open-boundary bands with three braiding phases, i.e., Hopf link, unknot, and unlink [Figs. \ref{suppfigmainmodelb}(A)-\ref{suppfigmainmodela}(C)]. Of particular significance is the critical point, which we identify as a stable EP, between the phase of Hopf link (unknot) and unknot (unlink). The bands corresponding to the Hopf link and unknot (unlink) phases are separated (isolated) [Figs. \ref{suppfigmainmodelb}(A1)-\ref{suppfigmainmodelb}(C1)], and the sub-GBZs corresponding to the Hopf link and unlink (unknot) phases are (non-)closed [Figs. \ref{suppfigmainmodelb}(A2)-\ref{suppfigmainmodelb}(C2)].  

\begin{figure}
    \centering
    \includegraphics[width=0.35 \linewidth]{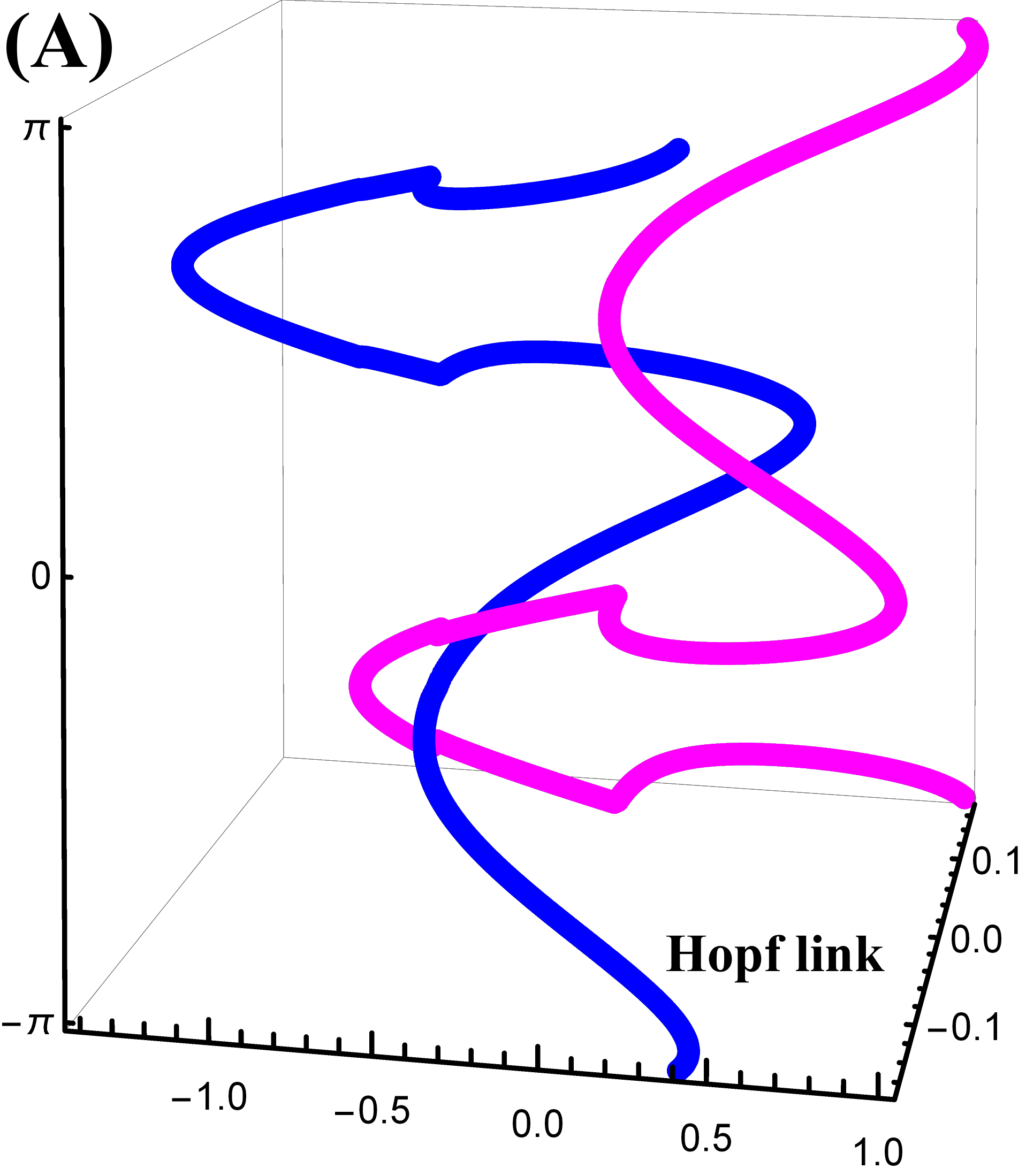}
    \includegraphics[width=0.35 \linewidth]{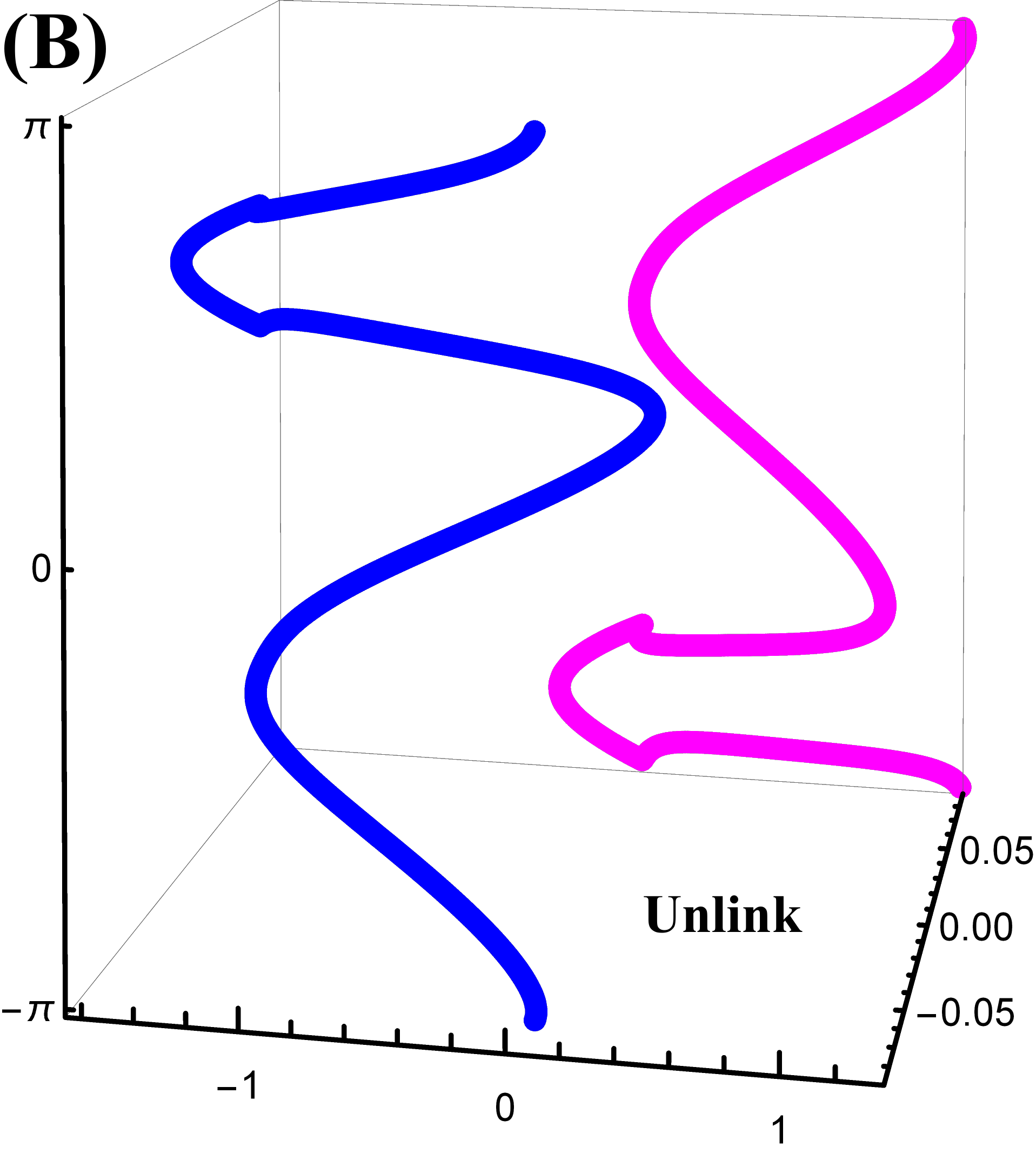}\\
    \includegraphics[width=0.4 \linewidth]{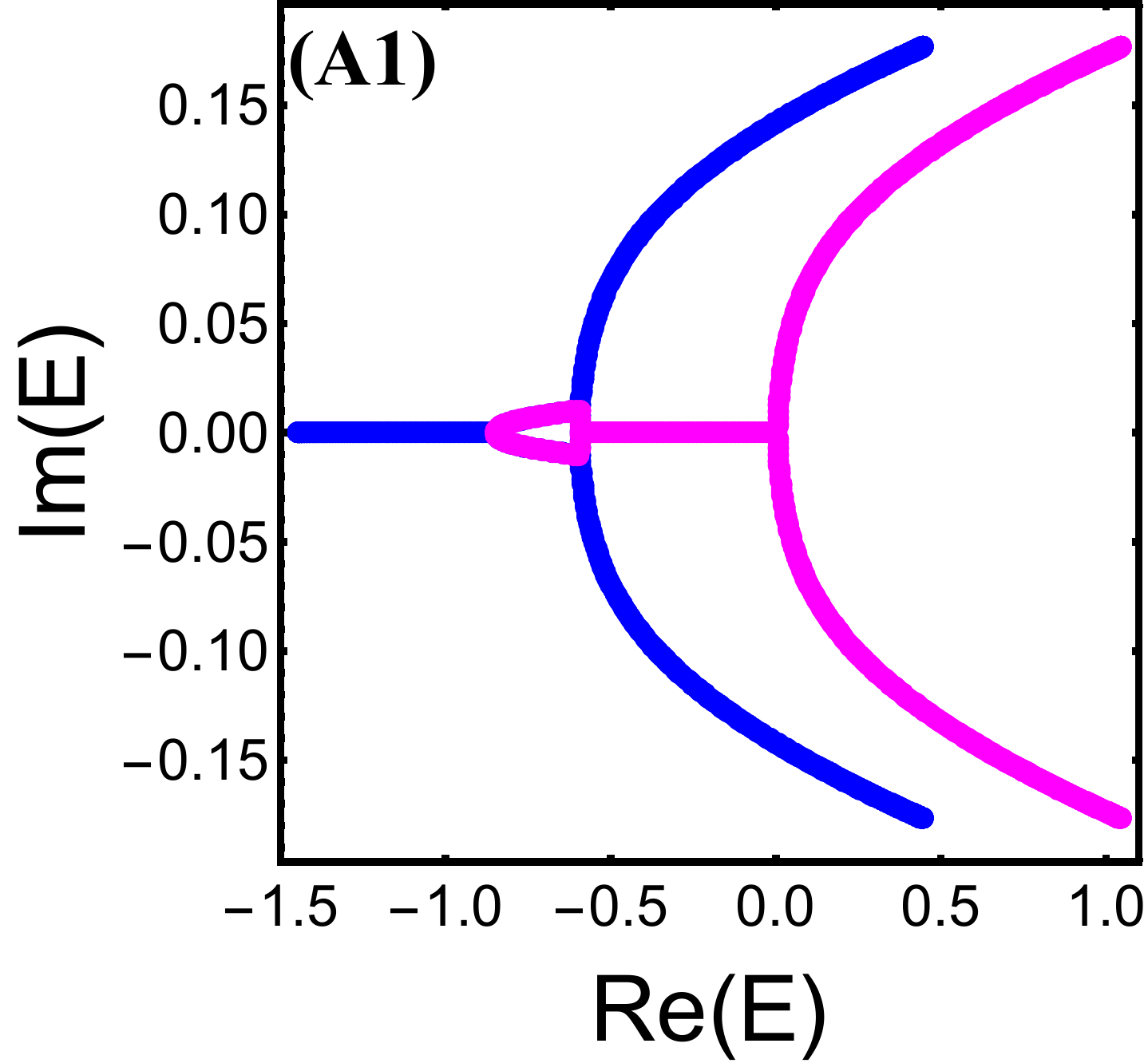}
    \includegraphics[width=0.36 \linewidth]{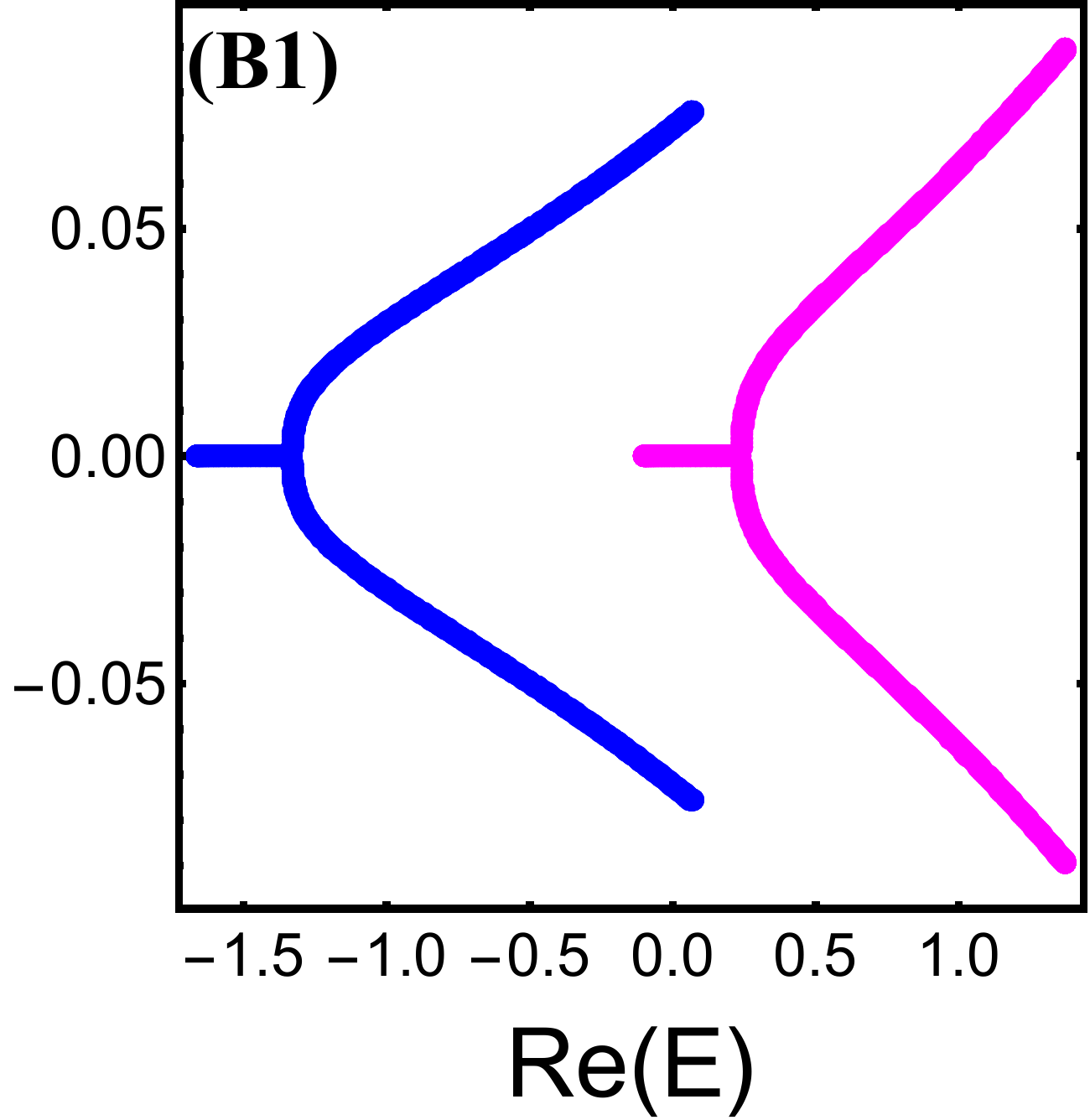}\\
    \includegraphics[width=0.4 \linewidth]{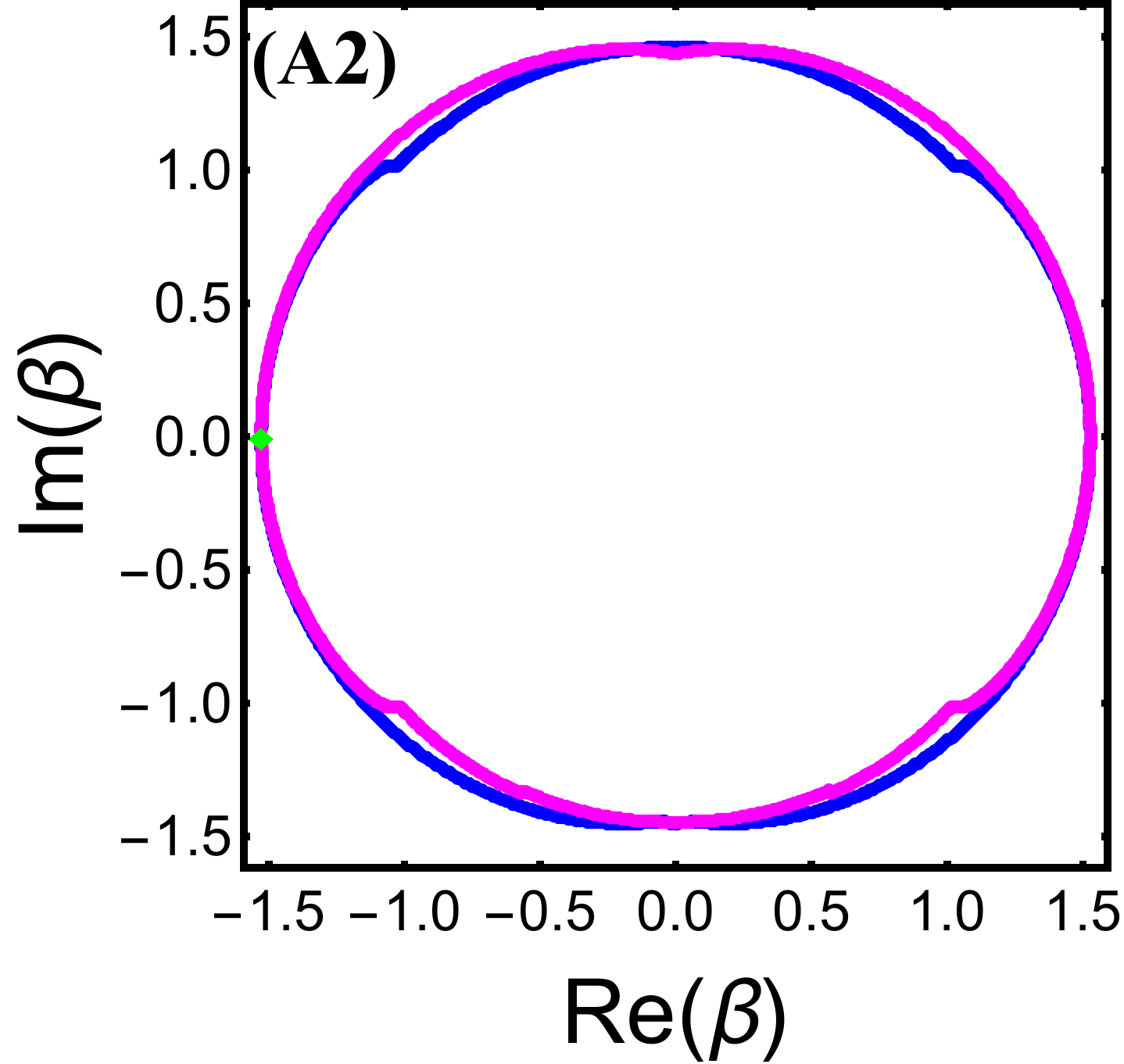}
    \includegraphics[width=0.355 \linewidth]{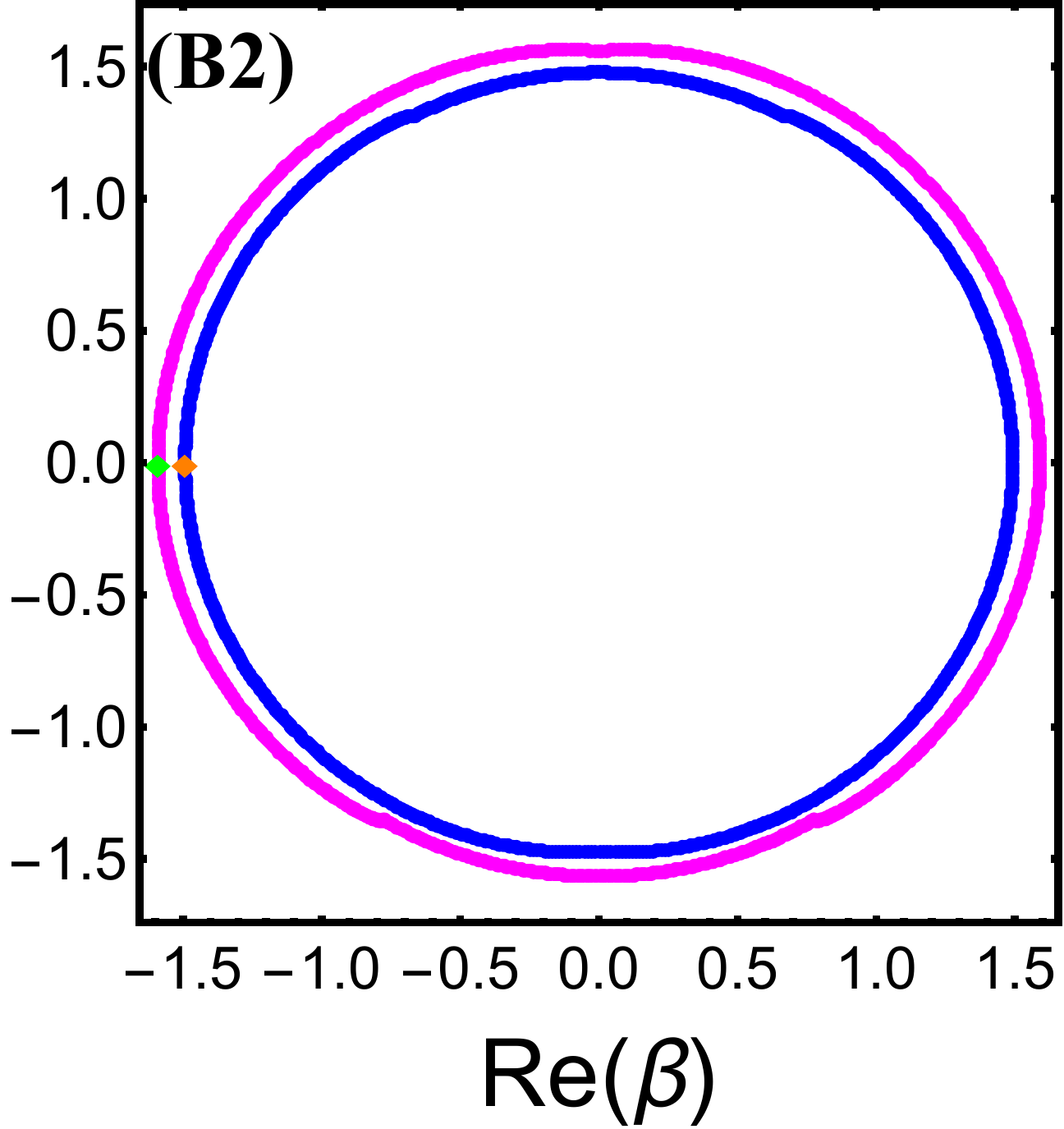}
    \caption{The model (\ref{suppeqcomplexmodel}) with the parameters $t_{p}=0.15,t_{m}=0.85,t_{a}=0.2,m=0.3$. (A)(B) The band braiding corresponding to $\lambda_{3}=0$ and $0.6$, projected into the complex energy plane shown in (A1)(B1), respectively. (A2)(B2) The sub-GBZs with respect to (A)(B), where the orange and green dots label the starting points of blue and magenta sub-GBZs, respectively.}
    \label{suppfiggenemodel}
\end{figure}

We can also apply our formalism to other models, such as the following non-Bloch Hamiltonian with next-nearest neighbor hoppings,
\begin{align}
    \label{suppeqcomplexmodel}
    \mathcal{H}_{c}(\beta)=\left(\begin{matrix}
        t_{p}\beta^{2}+t_{m}\beta^{-2}+\lambda_{3} & -t_{a}(\beta-\beta^{-1})-im\\
        t_{a}(\beta-\beta^{-1})+im & t_{p}\beta^{2}+t_{m}\beta^{-2}-\lambda_{3}
    \end{matrix}\right),
\end{align}
which is the generalization of the model $\mathcal{H}(\beta)$ in the main text. As illustrated in Fig. \ref{suppfiggenemodel}, there exists the Hopf link and unlink phases [Figs. \ref{suppfiggenemodel}(A) and \ref{suppfiggenemodel}(B)], which are related to the separated and isolated bands [Figs. \ref{suppfiggenemodel}(A1) and \ref{suppfiggenemodel}(B1)], as well as the closed sub-GBZs [Figs. \ref{suppfiggenemodel}(A2) and \ref{suppfiggenemodel}(B2)].

\section{S5. Disambiguation of Brillouin zone and generalized Brillouin zone}

In the context of the non-Bloch band theory for one-dimensional (1D) non-Hermitian systems, the Brillouin zone (BZ) is considered as the Hermitian reduction from the GBZ \cite{yao2018,yokomizo2019}. However, this viewpoint of BZ and GBZ is ambiguous---BZ is the result of the combination of translation invariance and periodic boundary condition (PBC) in the crystals, i.e., the standard Bloch band theory in solid state physics \cite{quinn2018,Phillips_2012}, while GBZ is the consequence of the solutions under open boundary condition (OBC) in both Hermitian and non-Hermitian systems. Hence, the unambiguous terminology BZ and GBZ used in non-Hermitian systems should be clarified.

In the Bloch band theory, the eigenstates of the crystals, valid for both Hermitian and non-Hermitian systems under PBC, display the Bloch wavefunctions based on the (first) BZ in the reciprocal space. Consider the 1D Hatano-Nelson (HN) model for an example; the tight-binding Hamiltonian is 
\begin{align}
    \label{supphatanoham}
    \hat{H}_{hn}=\sum_{j}\left[(t+\gamma)\hat{c}^{\dagger}_{j}\hat{c}_{j+1}+(t-\gamma)\hat{c}^{\dagger}_{j+1}\hat{c}_{j}\right],
\end{align}
where $\hat{c}_{j}$ ($\hat{c}^{\dagger}_{j}$) is the fermion annihilation (creation) operator at the $j$-th site, and the non-Hermiticity attributes to the nonreciprocal hoppings $t\pm\gamma$ with the Hermitian limit $\gamma\rightarrow 0$. The Bloch theorem suggests that the Bloch Hamiltonian
\begin{align}
    \label{supphatanohambloch}
    H_{hn}(k)=2t\cos k+2i\gamma\sin k,
\end{align}
gives the eigenstates  
\begin{align}
    \label{supphatanotbastate}
    \ket{k}=\frac{1}{\sqrt{N}}\sum_{j=1}^{N}e^{ikR_{j}}\ket{j},
\end{align}
corresponding to the eigenvalues $H_{hn}(k)$ with the wave vectors $k=\frac{2\pi q}{N}\in [0,2\pi]\equiv \textrm{BZ}, q=1,2,\ldots,N$, where $N$ and $R_{j}$ are the number of sites and coordinate of the $j$-th site in the Bravais lattice, respectively \cite{alase2016generalized}. Here, we have set the lattice constant $a=1$; thus, $R_{j}$ is integer value. Through the Fourier transformation, we obtain the basis of the Wannier representation
\begin{align}
    \label{supphatanowannier}
    \ket{j}=\frac{1}{\sqrt{N}}\sum_{k\in\mathrm{BZ}}e^{-ikR_{j}}\ket{k},
\end{align}
while $\left\{\ket{k}, k\in\mathrm{BZ}\right\}$ is the basis of the Bloch representation. The Bloch wavefunction is the projection of $\ket{k}$ into the continuous real space
\begin{align}
    \label{supphatanoblaochwave}
    \psi_{k}(x)=\braket{x|k}=\frac{1}{\sqrt{N}}\sum_{j=1}^{N}e^{ikR_{j}}\braket{x|j}\equiv \frac{1}{\sqrt{N}}\sum_{j=1}^{N}e^{ikR_{j}}\phi(x-R_{j})=e^{ikx}\frac{1}{\sqrt{N}}\sum_{j=1}^{N}e^{-ik(x-R_{j})}\phi(x-R_{j})=e^{ikx}u_{k}(x),
\end{align} 
where $u_{k}(x)=\frac{1}{\sqrt{N}}\sum_{j=1}^{N}e^{-ik \left(x-R_{j}\right)}\phi(x-R_{j})$ satisfying 

\begin{align}
   u_{k}(x+R_{j'})=\frac{1}{\sqrt{N}}\sum_{j=1}^{N}e^{-ik\left[x-(R_{j}-R_{j'})\right]}\phi\left(x-(R_{j}-R_{j'})\right)=\frac{1}{\sqrt{N}}\sum_{l=1}^{N}e^{-ik\left(x-R_{l}\right)}\phi\left(x-R_{l}\right)=u_{k}(x)
\end{align}
due to $R_{l}\equiv R_{j}-R_{j'}$ is still a lattice vector.

\begin{figure}
    \centering
    \includegraphics[width=0.4 \linewidth]{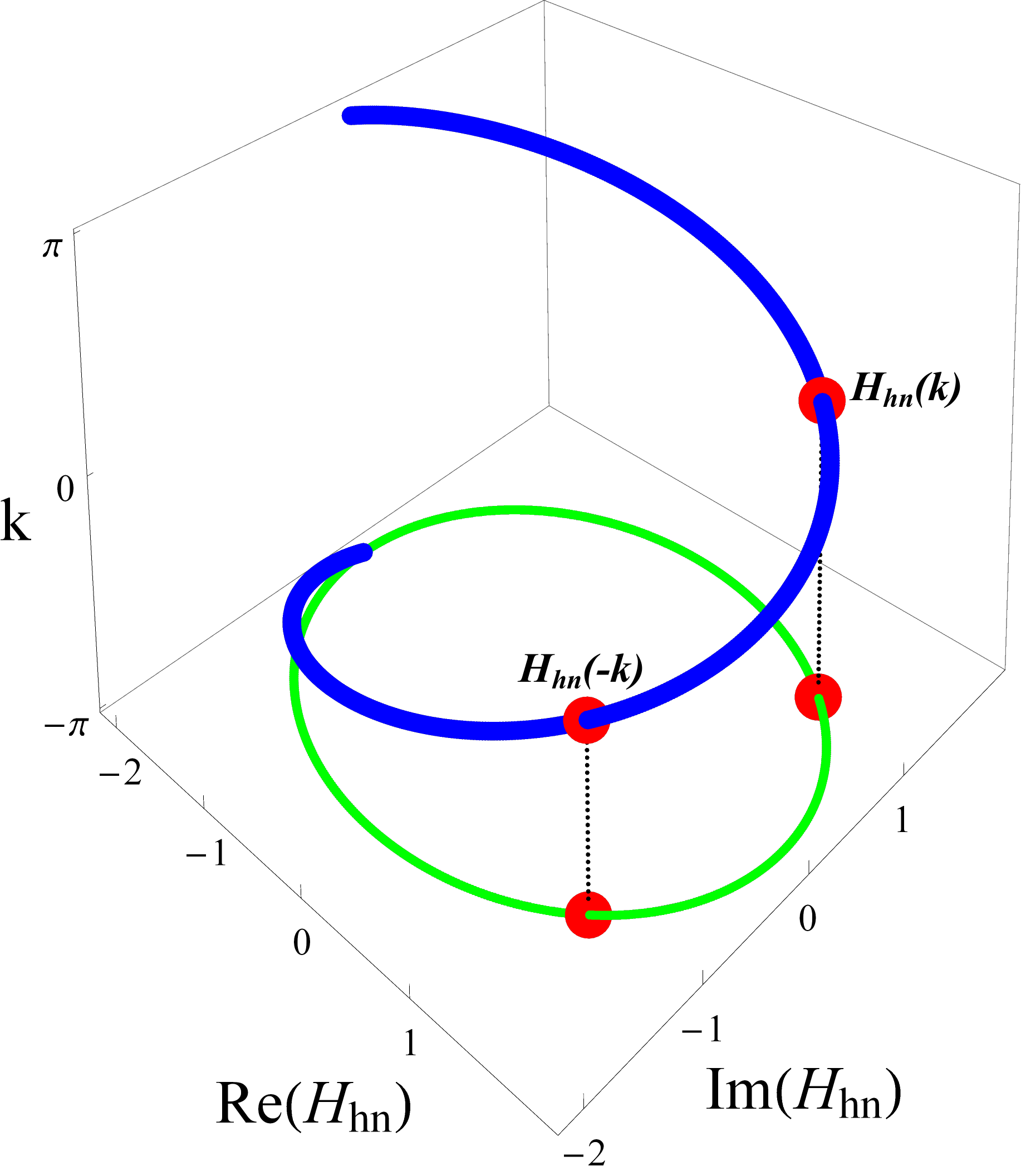}
    \includegraphics[width=0.4 \linewidth]{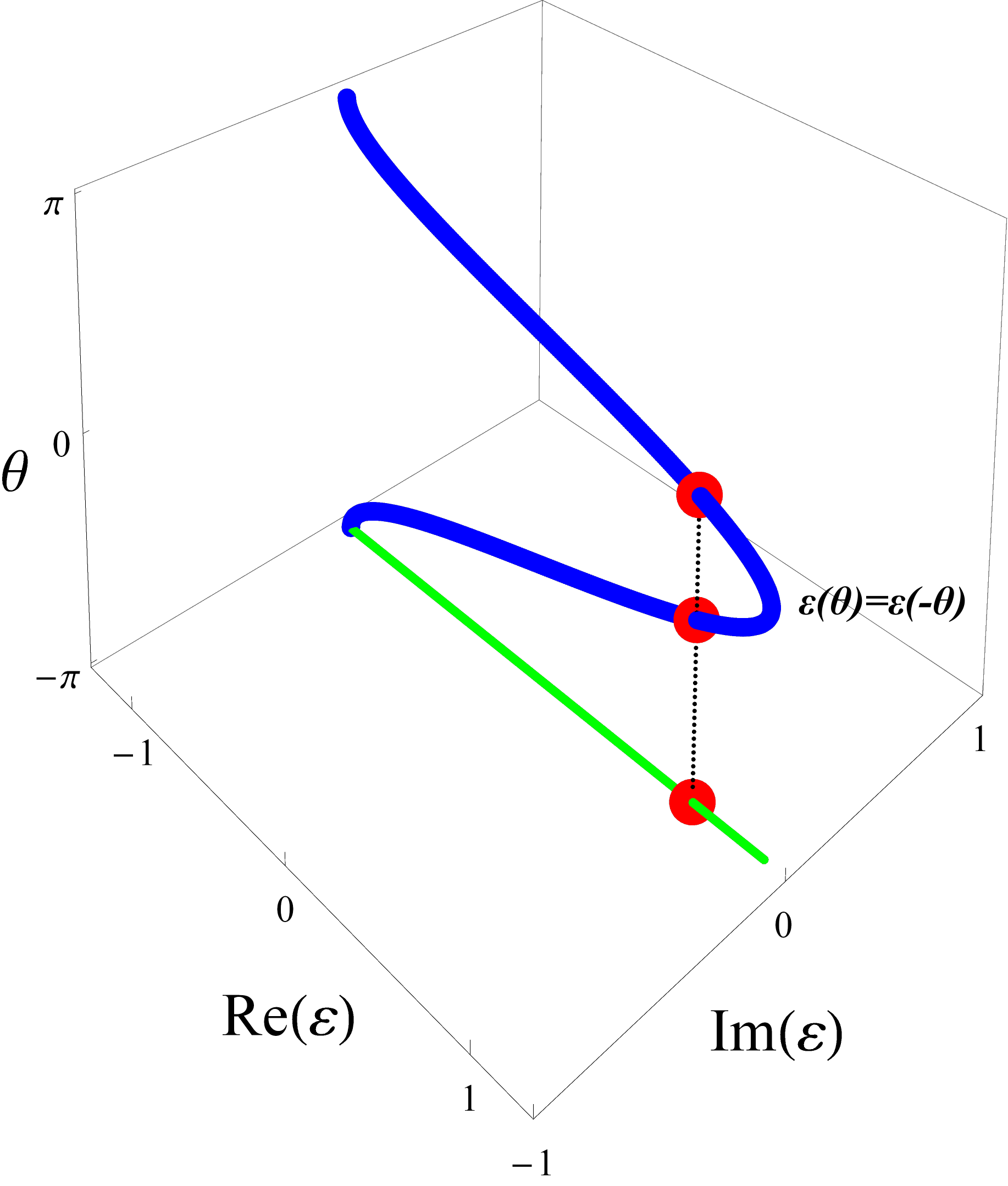}
    \caption{The spectrum $H_{hn}(k)$ and $\varepsilon(\theta)$ of the HN model in Eq. (\ref{supphatanoham}) under PBC (left) and OBC (right), respectively. The parameters are $t_{1}=1,\gamma=0.8$.}
    \label{fighnbraid}
\end{figure}

When we switch off the hoppings connecting the left and right boundaries, the system intuitively exhibits standing-wave solutions originating from the reflection of the Bloch plane wave between the two (hard) boundaries, which is exactly the results of the non-Bloch band theory under OBC \cite{yokomizo2019}. The analytical solutions of the HN model under OBC are
\begin{align}
    \varepsilon(\theta)&=2\sqrt{(t+\gamma)(t-\gamma)}\cos \theta,\label{suppeqhnsoluen}\\
    \ket{\theta}&=\mathcal{N}_{\theta}\sum_{j}r^{j}\sin (\theta j)\ket{j}\label{suppeqhnsoluvec},
\end{align}
where $\mathcal{N}_{\theta}$ is the normalized coefficient and $r=\sqrt{\frac{t-\gamma}{t+\gamma}}, \theta=\frac{m\pi}{N+1}, m\in \mathbbm{Z}$. The GBZ is a circle with the radius $r$, which coincides with BZ in the Hermitian limit ($r=1$), and $\theta$ acts in the same role as the wave vector $k$ in the Bloch band theory. However, the non-Bloch band theory exhibits distinct eigenstates with $\theta$ in contrast to the Bloch band theory with $k$. The fixed $\pm k$ correspond to two distinct eigenstates $\ket{\pm k}$ in the Bloch band [left panel of Fig. \ref{fighnbraid}], while the fixed $\pm \theta$ correspond to only one eigenstate [Eq. (\ref{suppeqhnsoluvec})] in the non-Bloch band [right panel of Fig. \ref{fighnbraid}], that is the OBC eigenstate is the standing wave $\sin (\theta j)$ originating from the superposition of the two opposite plane waves $\ket{\pm \theta}$ modulated by the exponential factor $r^{j}$, which is illustrated more clearly in the $(\mathrm{Re}(H_{hn}),\mathrm{Im}(H_{hn}),k)$ [or $(\mathrm{Re}(\varepsilon),\mathrm{Im}(\varepsilon),\theta)$] space [Fig. \ref{fighnbraid}]. The GBZ demonstrates numerous results that differ from the BZ for more general (multiband) systems, which is the focus of our discussion on the braiding topology of open-boundary bands in the main text.

\begin{figure}
    \centering
    \includegraphics[width=0.39 \linewidth]{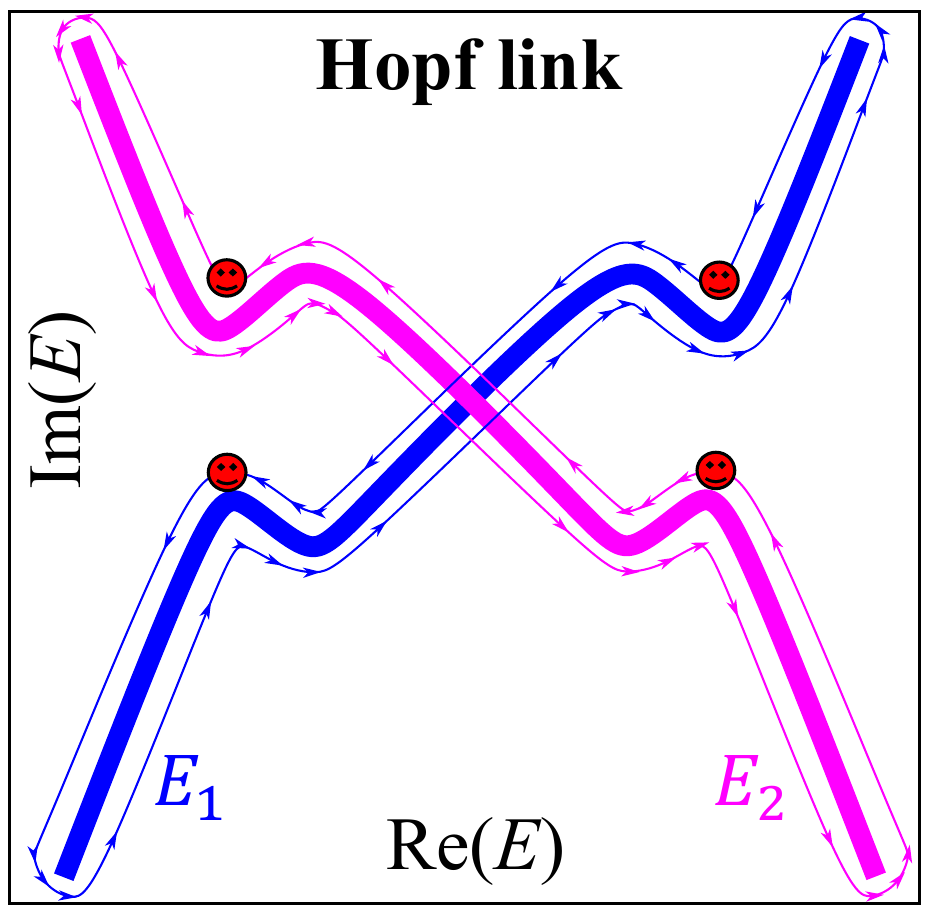}\qquad\qquad
    \includegraphics[width=0.402 \linewidth]{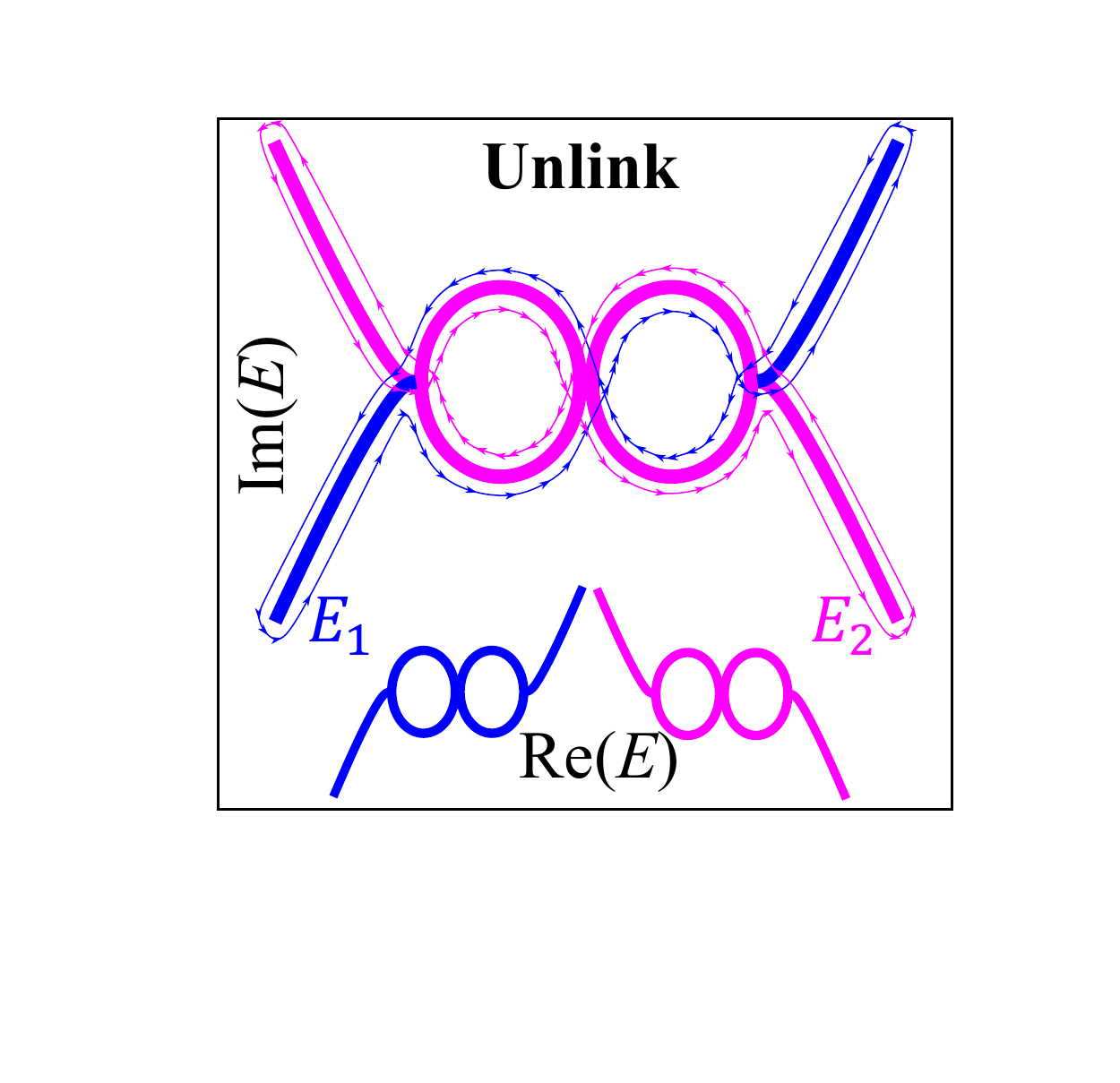}
    \caption{A schematic illustration for the complementary choice of flow directions and exchange partners to the OBC bands $E_{1,2}$ in the main text. The blue and magenta arrows label the spectral flows. The right panel inset shows the shape of the two bands that unlink individually and partially overlap, as shown in the main figure.}
    \label{suppfigtran}
\end{figure}

\section{S6. Choices of flow directions at EPs}
As elucidated in the main text, the two-visit rule has profound implications, leading to two natural choices of flow directions and exchange partners. These choices, represented by the OBC bands $E_{1,2}$, mirror those in the main text and intersect at the EPs. One of these choices is detailed in the main text. The other choice, as shown in the left panel of Fig. \ref{suppfigtran}, reveals that the red points on the two bands of the Hopf link approach and touch at the EPs are distinct from the first choice along the flow directions. After the topological braiding transition and band partners switching, the two bands form the unlink phase shown in the right panel of Fig. \ref{suppfigtran}. We disregard the starting points of the spectral flows due to their irrelevance to the braiding topology of the two bands.

\end{document}